\documentclass[11pt, reqno]{amsart}
\usepackage{amsfonts,latexsym}
\usepackage{amsmath}
\usepackage{amscd}
\usepackage{float,amssymb,mathrsfs,bm,multirow,graphics}
\usepackage[dvips]{graphicx}
\usepackage{amsbsy}
\usepackage[abs]{overpic}

\newcommand{\nequation}{\setcounter{equation}{0}}
\renewcommand{\theequation}{\mbox{\arabic{section}.\arabic{equation}}}
\newcommand{\R}{{\Bbb R}}

\newcommand{\C}{{\Bbb C}}
\newcommand{\Z}{{\Bbb Z}}

\newcommand{\proofbegin}{\noindent{\quad \it Proof.\,\,}}
\newcommand{\proofend}{\hfill$\Box$\bigskip}

\newtheorem{theorem}{Theorem}[section]
\newtheorem{proposition}[theorem]{Proposition}
\newtheorem{lemma}[theorem]{Lemma}

\newtheorem{remark}[theorem]{Remark}

\newtheorem{figuretext}[theorem]{Figure}

\input epsf
\title[Boundary value problems for the Einstein equations]{\sc Boundary value problems for the stationary axisymmetric Einstein equations: a disk rotating around a black hole}
\author{Jonatan Lenells}
\address{Institut f\"ur Angewandte Mathematik, Leibniz Universit\"at Hannover, Welfengarten 1, 30167 Hannover, Germany}
\email{lenells@ifam.uni-hannover.de}

\begin{document}

\begin{abstract} 
\noindent
We solve a class of boundary value problems for the stationary axisymmetric Einstein equations corresponding to a disk of dust rotating uniformly around a central black hole. The solutions are given explicitly in terms of theta functions on a family of hyperelliptic Riemann surfaces of genus $4$. In the absence of a disk, they reduce to the Kerr black hole. In the absence of a black hole, they reduce to the Neugebauer-Meinel disk. 
\end{abstract}

\maketitle



\tableofcontents
\section{Introduction}\label{introsec}\nequation
Two of the most famous solutions of the stationary axisymmetric Einstein equations are the Kerr black hole and the Neugebauer-Meinel disk. The former was discovered by Kerr in 1963 \cite{Kerr} and the latter by Neugebauer and Meinel in the 1990s \cite{NM1993, NM1994, NM1995}. In this paper, we construct analytic solutions of a class of BVPs for the stationary axisymmetric Einstein equations which combine the Kerr and Neugebauer-Meinel spacetimes. Thus, the BVPs considered involve a finite disk of dust rotating uniformly around a central black hole. In the limit of a vanishing disk, the solutions tend to the Kerr black hole. In the absence of a black hole, they reduce to the Neugebauer-Meinel disk. The constructed disk/black-hole systems are given explicitly in terms of theta functions on a family of hyperelliptic Riemann surfaces of genus $4$. Given the importance of the Kerr and Neugebauer-Meinel solutions, we believe that the class of solutions presented here could also be of interest. 

The general analysis of rotating relativistic bodies is exceedingly complicated because it involves the study of free boundary value problems (BVPs) for the Einstein equations, which are nonlinear partial differential equations in four dimensions. However, in cases where the surface of the body is known and the motion is stationary and axisymmetric (a reasonable assumption in many astrophysical situations), the physical problem gives rise to a BVP for a single {\it integrable} equation in two dimensions---the celebrated Ernst equation. The integrability of the Ernst equation implies that powerful solution-generating techniques are at hand. Thus, through the application of suitable nonlinear transformations, new stationary axisymmetric spacetimes can be generated from already known ones. Furthermore, a large class of solutions of the Ernst equation can be given explicitly in terms of theta functions on Riemann surfaces \cite{K}. In this way, it is possible to write down a large number of exact analytic solutions to the stationary axisymmetric Einstein equations and to study them using the methods of algebraic geometry cf. \cite{KR2005}.

Nevertheless, for the solution of a concrete BVP, the power of this approach is often limited. Indeed, although a large class of exact solutions can be produced, the problem of determining which particular solution within this class that satisfies the given BVP is in general a highly nonlinear problem.
It is therefore remarkable that Neugebauer and Meinel in the 1990s were able to solve explicitly, using constructive methods, the BVP corresponding to the physically relevant situation of a rotating dust disk. The structure of an infinitesimally thin, relativistic disk of dust particles which rotate uniformly around a common center was first explored numerically by Bardeen and Wagoner 40 years ago \cite{BW1971}, who also pointed out that ``there may be some hope of finding an analytic solution'' \cite{BW1969}. Bi\v{c}\'ak notes in the comprehensive review \cite{Bicak} that the subsequent construction of such an analytic solution by Neugebauer and Meinel represents ``the first example of solving the BVP for a rotating object in Einstein's theory by analytic methods.'' Let us point out that since the Neugebauer-Meinel dust disk can be written in terms of theta functions on Riemann surfaces of genus 2, it belongs to the general class of solutions introduced in \cite{K} and can therefore be analyzed by means of algebro-geometric methods. On the other hand, the Kerr black hole is the most famous example of a stationary axisymmetric spacetime and has had an immense impact on the development of general relativity and astrophysics (see e.g. \cite{Chandra}). 

The approach in this paper is primarily inspired by the work of Neugebauer, Meinel, and collaborators \cite{MAKNP}, but also by a novel method for the analysis of BVPs for integrable PDEs which has been developed by Fokas and his collaborators within the past 15 years. A central development in the theory of nonlinear PDEs in the second half of the 20th century, and continuing to the present, has been the introduction of the Inverse Scattering Transform (IST). This technique was put forward in the famous 1967 paper \cite{GGKM} in connection with the Korteweg-de Vries (KdV) equation and the range of its applicability began to unfold with the investigation of the nonlinear Schr\"odinger (NLS) equation [ZS]. One of the most important later developments in this area has been the generalization of the IST formalism from initial-value to initial-boundary value problems introduced by Fokas \cite{F1997, F2002} and subsequently developed further by several authors cf. \cite{Fbook}. 
The Fokas method consists of two steps: (a) Construct an integral representation of the solution characterized via a matrix Riemann-Hilbert (RH) problem formulated in the complex $k$-plane, where $k$ denotes the spectral parameter of the associated Lax pair. Since this representation involves, in general, some unknown boundary values, the solution formula is {\it not} yet effective. (b) Characterize the unknown boundary values by analyzing a certain equation called the {\it global relation}. In general, this characterization involves the solution of a nonlinear problem; however, for certain so-called {\it linearizable} boundary conditions, step (b) can be solved in closed form.

In a recent work \cite{LF} steps (a) and (b) were implemented for the class of BVPs of the Ernst equation corresponding to a thin rotating disk of finite radius. In particular, it was found that the boundary conditions of the particular BVP corresponding to the uniformly rotating Neugebauer-Meinel disk are linearizable. The present paper is to some extent a continuation of \cite{LF}: The main novel observation is that the BVP which combines the Kerr black hole boundary conditions with those of a uniformly rotating disk is also linearizable. 

Physically, disk/black-hole systems are important as models for black hole accretion disks in the context of active galactic nuclei and X-ray binaries \cite{ABP}. Accretion disks are flattened astronomical objects made of rapidly rotating gas which slowly spirals onto a central gravitating body. Accretion onto a black hole is generally assumed to be thin and axisymmetric (see e.g. \cite{Pringle}) and many of the most energetic phenomena in the universe have been attributed to the accretion of matter onto black holes. In particular, active galactic nuclei and quasars are believed to be the accretion disks of supermassive black holes. We refer to \cite{ABP} and references therein for more information on the physical aspects of black hole accretion disks. Our approach here is mathematical and we do not investigate any possible physical relevance of the presented solutions.

The manuscript is organized as follows. In section \ref{diskblackholesec}, we formulate our disk/black-hole BVP and write down its full solution in terms of theta functions. 
In section \ref{examplesec}, we consider a particular example.
In sections \ref{specsec}-\ref{axishorizonsec}, the derivation of the solution is presented in several steps.
In section \ref{paramsec}, we consider the singularity structure of the solution and its dependence on various parameters.
In the appendix, we consider the relationship between the solution derived here and the general class of solutions of the Ernst equation studied in \cite{KM, KKS}.

\section{Disk/black-hole systems}\nequation\label{diskblackholesec}
In this section we introduce the Ernst equation, formulate the BVP corresponding to a dust disk rotating uniformly around a central black hole, and present its solution in terms of theta functions.

\subsection{The Ernst equation}
The metric of a stationary axisymmetric vacuum spacetime can be written in the Weyl-Lewis-Papapetrou form 
\begin{equation}\label{lineelement} 
 ds^2 = e^{-2U}\left[e^{2\kappa}(d\rho^2 + d\zeta^2) + \rho^2d\varphi^2\right] -
e^{2U}(dt + ad\varphi)^2,
\end{equation}
where $\rho \geq 0$ and $\zeta \in \R$ are Weyl's canonical coordinates and $t,\varphi$ are chosen so that $\partial_t$ and $\partial_\varphi$ are the two commuting asymptotically timelike and spacelike Killing vectors, respectively, cf. \cite{SKMHH}.
The metric functions $e^{2U}$, $a$, and $e^{2\kappa}$ are functions of $\rho$ and $\zeta$ alone. Introducing a real-valued potential $b$ by
\begin{equation}\label{bdef}  
  a_\rho = \rho e^{-4U}b_\zeta, \qquad a_\zeta = -\rho e^{-4U}b_\rho,
\end{equation}
it can be shown that the Einstein field equations for the metric (\ref{lineelement}) reduce to the following single nonlinear PDE in two dimensions for the complex-valued Ernst potential $f = e^{2U} + ib$:
\begin{equation}\label{ernst}  
  \frac{f + \bar{f}}{2}\left(f_{\rho\rho} + f_{\zeta\zeta} + \frac{1}{\rho} f_\rho\right) = f_\rho^2 + f_\zeta^2, \qquad \rho > 0, \quad \zeta \in \R.
\end{equation}  
Following standard practice, the real part of $f$ will be denoted by $e^{2U}$ despite the fact that it may take on negative values.

We also need the concept of a corotating frame. Given $\Omega \in \R$, we define the coordinates $(\rho', \zeta', \varphi', t')$ corotating with the angular velocity $\Omega$ by
$$\rho' = \rho, \qquad \zeta' = \zeta, \qquad \varphi' = \varphi - \Omega t, \qquad t' = t.$$
In these new coordinates, the metric (\ref{lineelement}) retains its form and the corotating metric functions $U_\Omega, a_\Omega, \kappa_\Omega$ are related to $U, a, \kappa$ via
\begin{subequations}
\begin{align} \label{UOmegaUrelation}
& e^{2U_\Omega} = e^{2U}\left[(1 + \Omega a)^2 - \Omega^2\rho^2e^{-4U}\right],
	\\
& (1 - \Omega a_\Omega)e^{2U_\Omega} = (1 + \Omega a)e^{2U},\qquad \kappa_\Omega - U_\Omega = \kappa - U.
\end{align}
\end{subequations}
The Ernst equation retains its form in the corotating system and we denote the corotating Ernst potential by $f_\Omega := e^{2U_\Omega} + i b_{\Omega}$. 

\subsection{The boundary value problem}
We now formulate the BVP for the Ernst equation (\ref{ernst}) which corresponds to a finite dust disk rotating uniformly around a central black hole. The formulation involves five parameters: the radius $\rho_0 > 0$ and angular velocity	 $\Omega \in \R$ of the disk; the `radius' $r_1 > 0$ andÊ angular velocity $\Omega_h \in \R$ of the black hole horizon; and the (necessarily constant) value of the corotating metric function $e^{2U_\Omega}$ on the disk. 
\begin{figure}
\begin{center}
    \includegraphics[width=.3\textwidth]{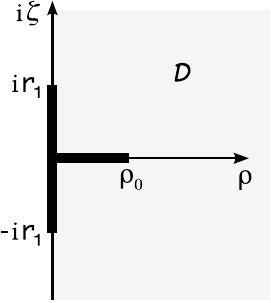} \quad
     \begin{figuretext}\label{holediskdomain.pdf}
       The exterior domain $\mathcal{D}$ of a disk of radius $\rho_0$ with the black hole horizon stretching along the imaginary axis from $-ir_1$ to $ir_1$.
     \end{figuretext}
 \end{center}
\end{figure}    
We will henceforth work with a complex variable $z = \rho + i\zeta$ and write $f(z)$ for $f(\rho, \zeta)$.\footnote{In general, given a function $h(\rho, \zeta)$, we will suppress the dependence on $\bar{z}$ and write $h(z)$ for $h(\rho, \zeta)$ even when $h$ is not analytic.}

Let $\mathcal{D}$ denote the exterior of a finite disk of radius $\rho_0 > 0$, i.e. $\mathcal{D}$ consists of all $z \in \C$ with strictly positive real part which do not belong to the interval $[0, \rho_0]$, see Figure \ref{holediskdomain.pdf}. The horizon of the black hole stretches in the $z$-plane along the imaginary axis from $-ir_1$ to $ir_1$.
We consider the problem of finding a function $f$ such that:
\addtocounter{equation}{1}
\begin{subequations}\label{BVP}
\begin{align}
&\bullet \ \text{$f$ satisfies (\ref{ernst}) in $\mathcal{D}$.}
	\\ \label{BVPeqsymm}
& \bullet \ \text{$f(z) = \overline{f(\bar{z})}$ (equatorial symmetry).}
	\\
& \bullet\ \text{$f(z) \to 1$ as $|z|^2 \to \infty$ (asymptotic flatness).}
	\\ \nonumber
& \bullet \ \text{$\frac{\partial f}{\partial \rho}(i\zeta) = 0$ for all $|\zeta| > r_1$  (regularity on the rotation axis)}
	\\ \label{BVPdisk}
& \bullet \ \text{$f_{\Omega}(\rho \pm i 0) = e^{2U_{\Omega}(+ i 0)}$ for $0 < \rho < \rho_0$ (boundary condition on the disk).}
	\\ \label{BVPhorizon}
& \bullet \ \text{$e^{2U_{\Omega_h}(i\zeta)} = 0$ for $0 < |\zeta| < r_1$ (boundary condition on the horizon).}
\end{align}
\end{subequations}
The boundary conditions (\ref{BVPdisk}) and (\ref{BVPhorizon}) are the boundary conditions corresponding physically to a uniformly rotating dust disk and a rotating black hole, respectively, cf. \cite{MAKNP}. If one sets $r_1 = 0$ in (\ref{BVP}) (i.e. one removes the black hole), then the solution of the obtained BVP is the Neugebauer-Meinel disk rotating with angular velocity $\Omega$. If one sets $\rho_0 = 0$ in (\ref{BVP}) (i.e. one removes the disk), then the solution of the obtained BVP is the Kerr black hole rotating with angular velocity $\Omega_h$.

\subsection{The solution}\label{solutionsubsec}
\begin{figure}
\begin{center}
    \includegraphics[width=.5\textwidth]{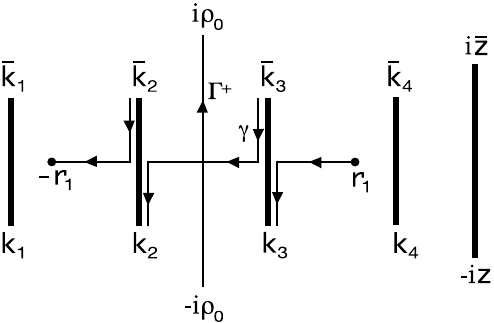}
     \begin{figuretext}\label{contours1.pdf}
       The Riemann surface $\Sigma_z$ presented as a two-sheeted cover of the complex $k$-plane with five branch cuts when $\text{Re}\, k_4 < \zeta$. The contours $\gamma$ and $\Gamma^+$ are also shown.
     \end{figuretext}
 \end{center}
\end{figure}

The formulation of the BVP (\ref{BVP}) involves the five independent parameters $\rho_0$, $\Omega$, $r_1$, $\Omega_h$, and the constant value of $e^{2U_\Omega}$ on the disk. However, it turns out that the condition that the solution be nonsingular at the rim of the disk imposes one relation among these parameters, so that the class of solutions is parametrized by only four parameters. It is convenient to adopt a parametrization in terms of the four parameters $\rho_0$, $r_1$, $w_2$, and $w_4$, where $w_2$ and $w_4$ are two real quantities related to the other parameters via equations (\ref{w0w2w4expressions}) below.
For a given choice of these parameters, the corresponding solution $f$ of the BVP (\ref{BVP}) can be written in terms of theta functions on the family of Riemann surfaces $\{\Sigma_z\}_{z \in \mathcal{D}}$ defined as follows:
Let $k_1, \bar{k}_1, \dots, k_4, \bar{k}_4 \in \C$ denote the eight zeros of $w^2(k) + 1$, where
\begin{equation}\label{wdef}
  w(k) = \frac{w_4 k^4 + w_2 k^2 + \rho_0^2(w_2 - w_4 \rho_0^2)}{(k^2 - r_1^2)},
\end{equation}
ordered so that $k_j$, $j =1, \dots, 4$, have negative imaginary parts and so that
$$\text{Re}\, k_1 \leq \text{Re}\, k_2 \leq \text{Re}\, k_3 \leq \text{Re}\, k_4.$$
Since these eight zeros are symmetrically distributed with respect to the origin, we have $-k_1 = \bar{k}_4$ and $-k_2 = \bar{k}_3$. 
For each $z = \rho + i\zeta$, $\Sigma_z$ is defined as the hyperelliptic Riemann surface of genus $4$ consisting of all points $(k, y) \in \C^2$ such that
\begin{equation}\label{Sigmazdef}  
  y^2 = (k + iz) (k - i\bar{z}) \prod_{j = 1}^4 (k - k_j) (k - \bar{k}_j),
\end{equation}
together with two points at infinity required to make the surface compact. We introduce branch cuts in the complex $k$-plane from $k_j$ to $\bar{k}_j$, $j = 1,\dots, 4$, and from $-iz$ to $i\bar{z}$, see Figure \ref{contours1.pdf}. For $k \in \hat{\C} = \C \cup \{\infty\}$, we let $k^+$ and $k^-$ denote the points which project onto $k$ and which lie in the upper and lower sheet of $\Sigma_z$, respectively. By definition, the upper (lower) sheet is characterized by $y/k^5 \to 1$ ($y/k^5 \to -1$) as $k \to \infty$. 

In view of the assumption of equatorial symmetry (\ref{BVPeqsymm}), we will in the sequel always assume that $\zeta \geq 0$. We will also, for the sake of definiteness, assume that
\begin{equation}\label{r1betweencuts}
  0 < \text{Re}\, k_3 < r_1 < \text{Re}\, k_4.
\end{equation}
Moreover, we assume that $\zeta \neq \text{Re}\, k_3$ andÊ $\zeta \neq \text{Re}\, k_4$, so that no branch cuts overlap (the solution for $\zeta = \text{Re}\, k_j$, $j = 3,4$, can be obtained by continuity). 

For $n$ complex numbers $\{a_j\}_1^n$, we let $[a_1, \dots, a_n]$ denote the directed contour $\cup_{j = 1}^{n-1} [a_j, a_{j+1}]$.
We let $\gamma$ denote the contour on $\Sigma_z$ which projects to the contour
\begin{equation}\label{gammaprojection}
[r_1, \text{Re}\,k_3 + \epsilon, k_3 + \epsilon]
\cup [\bar{k}_3 - \epsilon, \text{Re}\,k_3 - \epsilon, \text{Re}\,k_2 + \epsilon, k_2 + \epsilon]
\cup [\bar{k}_2 - \epsilon, \text{Re}\, k_2 - \epsilon, -r_1] 
\end{equation}
in the complex $k$-plane, where $\epsilon > 0$ is an infinitesimally small positive number, and which lies in the upper sheet for $\text{Re}\, k < \zeta$ and in the lower sheet for $\text{Re}\,k > \zeta$. We define $\Gamma^+$ as the contour in the upper sheet of $\Sigma_z$ which lies above the segment 
$\Gamma = [-i\rho_0, i\rho_0]$. The contours $\Gamma^+$ and $\gamma$ are shown in Figure \ref{contours1.pdf} in the case when $\text{Re}\, k_4 < \zeta$.

\begin{figure}
\begin{center}
    \includegraphics[width=.5\textwidth]{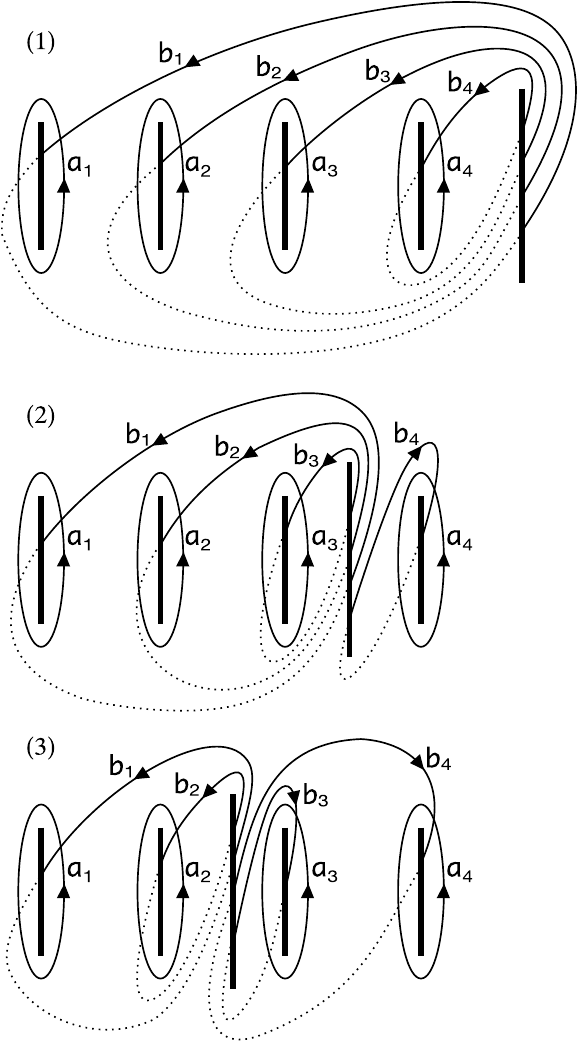} \qquad \qquad
     \begin{figuretext}\label{standardcutsall.pdf}
       The homology basis $\{a_j, b_j\}_1^4$ on the hyperelliptic Riemann surface $\Sigma_z$ of genus $g = 4$ in the case of (1) $\text{\upshape Re}\, k_4 < \zeta$, (2) $\text{\upshape Re} \, k_3 < \zeta < \text{\upshape Re} \,k_4$, and (3) $0 \leq \zeta < \text{\upshape Re}\, k_3$.
     \end{figuretext}
 \end{center}
\end{figure}

\begin{figure}
\begin{center}
    \includegraphics[width=.9\textwidth]{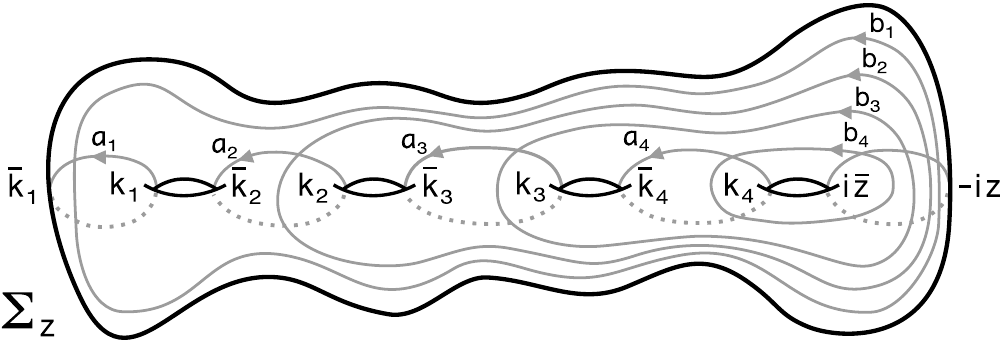} \quad
     \begin{figuretext}\label{Sigma3D.pdf}
       Three-dimensional picture of the hyperelliptic Riemann surface $\Sigma_z$ in the case when $\text{\upshape Re}\, k_4 < \zeta$.
     \end{figuretext}
 \end{center}
\end{figure}    

In order to define theta functions associated with $\Sigma_z$, we need to introduce a basis of the first homology group $H_1(\Sigma_z, \Z)$ of $\Sigma_z$. Since the Riemann surface $\Sigma_z$ depends on $z$, there are three qualitatively different cases to consider: 
\begin{enumerate}
  \setlength{\itemsep}{0pt}
\item[(1)] The cut $[-iz, i\bar{z}]$ lies to the right of $[k_4, \bar{k}_4]$ (i.e. $\text{Re}\, k_4 < \zeta$).

\item[(2)] The cut $[-iz, i\bar{z}]$ lies between $[k_3, \bar{k}_3]$ and $[k_4, \bar{k}_4]$ (i.e. $\text{Re} \, k_3 < \zeta < \text{Re} \,k_4$).

\item[(3)] The cut $[-iz, i\bar{z}]$ lies to the left of $[k_3, \bar{k}_3]$ (i.e. $0 \leq \zeta < \text{Re}\, k_3$).
\end{enumerate}
For these three cases, we let $\{a_j, b_j\}_{j=1}^4$ be the canonical basis of $H_1(\Sigma_z, \Z)$ shown in (1), (2), and (3) of Figure \ref{standardcutsall.pdf}, respectively. Thus, for $j = 1, \dots, 4$, $a_j$ surrounds the cut $[k_j, \bar{k}_j]$, whereas $b_j$ enters the upper sheet on the right side of $[-iz, i\bar{z}]$ and exits again on the right side of $[k_j, \bar{k}_j]$.

We define $\{\omega_j\}_1^4$ as the canonical basis of the space of holomorphic one-forms on $\Sigma_z$ dual to $\{a_j, b_j\}$. Then
$$\int_{a_j} \omega_i = \delta_{ij}, \qquad \int_{b_j} \omega_i =  B_{ij} , \qquad i,j = 1, \dots, 4,$$
where $B$ is the period matrix associated with the cut system $\{a_j, b_j\}$. We let $\omega = (\omega_1, \omega_2, \omega_3, \omega_4)^T$. The $4 \times 4$ matrix $B$ is symmetric and has a positively definite imaginary part. The associated theta function $\Theta(v |B)$ is defined by
\begin{equation}\label{Thetadef} 
  \Theta(v | B) = \sum_{N \in \Z^4} e^{2 \pi i\left(\frac{1}{2} N^T B N + N^T v\right)}, \qquad v \in \C^4.
\end{equation}
We let $\omega_{PQ}$ denote the Abelian differential of the third kind on $\Sigma_z$, which has two simple poles at the points $P$ and $Q$ with residues $+1$ and $-1$, respectively, and whose $a$-periods vanish, i.e. $\int_{a_j} \omega_{PQ} = 0$ for $j = 1,\dots, 4$.

We can now state our main result.

\begin{theorem}[Solution of the disk/black-hole BVP]\label{mainth}
Let $\rho_0, r_1, w_2, w_4$ be strictly positive numbers such that (\ref{r1betweencuts}) holds. [The requirement that the solution be singularity-free imposes further restrictions on these parameters, see section \ref{paramsec}.]
Let the function $h(k)$ be defined by
\begin{equation}\label{hdef}
  h(k) = \frac{1}{\pi i}\ln\left(\sqrt{w(k)^2 + 1} - w(k)\right), \qquad k \in \Gamma = [-i\rho_0, i\rho_0].
\end{equation}  
Define the $z$-dependent quantities $u \in \C^4$ and $I \in \R$ by
\begin{equation}\label{uIdef}
u = \int_{\Gamma^+} h \omega + \int_\gamma \omega,
\qquad I = \int_{\Gamma^+} h \omega_{\infty^+\infty^-} + \int_\gamma \omega_{\infty^+\infty^-}.
\end{equation}
Then the function 
\begin{equation}\label{ernstsolution}
f(z) = \frac{\Theta\left(u - \int_{-iz}^{\infty^-} \omega \bigl | B\right)}{\Theta\left (u + \int_{-iz}^{\infty^-} \omega \bigl | B\right)}e^{I},
\end{equation}
satisfies the BVP (\ref{BVP}) with the prescribed values of $\rho_0$ and $r_1$, and with the values of $\Omega_h$, $\Omega$, and $e^{2U_\Omega(+i0)}$ given by
\begin{align}\label{OmegahOmegae2U}
\Omega_h =  -\frac{1}{a_{hor}}, \qquad 
\Omega =  \frac{w_4 \Omega_h e^{2U_0} +  \sqrt{-2 w_4 \Omega_h^4 e^{2U_0}}}{w_4e^{2U_0}+2 \Omega_h^2}, \qquad  	
e^{2U_\Omega(+i0)} =  e^{2U_0}\left(1 - \frac{\Omega}{\Omega_h}\right)^2,
\end{align}
where $a_{hor}\in \R$ denotes the (necessarily constant) value of the metric function $a$ on the horizon and $e^{2U_0}\in \R$ denotes the real part of $f(+i0)$. Explicit expressions for the constants $a_{hor}$ and $f(+i0)$ are presented in propositions \ref{horizonprop} and \ref{f0f1prop} below. 

Moreover, define the $z$-dependent quantity $L$ by
\begin{align} \label{LLdef}
L =& -\frac{1}{2} \int_{\Gamma} d\kappa_1 \frac{dh}{dk}(\kappa_1) \int_{\Gamma}' h(\kappa_2) \omega_{\kappa_1^+\kappa_1^-}(\kappa_2^+) + \int_{\Gamma^+} h \omega_{-r_1^+, -r_1^-} - \text{\upshape sgn}(\zeta - r_1)\int_{\Gamma^+} h \omega_{r_1^+ r_1^-}
	\\ \nonumber
& + \frac{1}{2} \lim_{\epsilon \to 0} \left(\int_{\gamma_1(\epsilon)} \omega_{-r_1^+, -r_1^-}
- \text{\upshape sgn}(\zeta - r_1)\int_{\gamma_2(\epsilon)} \omega_{r_1^+ r_1^-} - 2\ln \epsilon \right),
\end{align}
where $\gamma_1(\epsilon)$ denotes the contour $\gamma$ with the segment covering $[-r_1, -r_1 + \epsilon]$ removed, $\gamma_2(\epsilon)$ denotes the contour $\gamma$ with the segment covering $[r_1 - \epsilon, r_1]$ removed, and the prime on the integral along $\Gamma$ indicates that the integration contour should be deformed slightly before evaluation so that the pole at $\kappa_2 = \kappa_1$ is avoided.\footnote{The result is indepedent of whether the contour is deformed to the right or to the left of the pole.}
Then the metric functions $e^{2U}$, $a$, $e^{2\kappa}$ of the line element (\ref{lineelement}) corresponding to the Ernst potential (\ref{ernstsolution}) are given for $z \in \mathcal{D}$ by
\begin{subequations}\label{solutionmetric}
\begin{align}\label{e2Uasolution}
 e^{2U(z)} = &\, \frac{Q(0)}{Q(u)} e^{I}, \qquad a(z) =  a_0 - \frac{\rho}{Q(0)}\left(\frac{\Theta(u + \int_{-iz}^{\infty^-} \omega  + \int_{i\bar{z}}^{\infty^-}\omega | B)}{Q(0)\Theta(u + \int_{-iz}^{i\bar{z}} \omega | B)} - Q(u)\right)e^{-I},
	\\ \label{e2ksolution}
e^{2\kappa(z)} = &\, K_0 \frac{\Theta(u|B)\Theta(u + \int_{-iz}^{i\bar{z}} \omega | B)}{\Theta(0 |B)\Theta(\int_{-iz}^{i\bar{z}} \omega | B)}
e^L.
\end{align}
\end{subequations}
where $Q(v)$ is defined by
\begin{equation}\label{Qvdef}
Q(v) = \frac{\Theta(v + \int_{-iz}^{\infty^-} \omega | B)\Theta(v + \int_{i\bar{z}}^{\infty^-} \omega| B)} {\Theta(v| B) \Theta(v + \int_{-iz}^{i\bar{z}} \omega| B)},\qquad v \in \C^4,
\end{equation}
and the two constants $a_0, K_0 \in \R$ are given explicitly by equations (\ref{a0solution}) and (\ref{K0solution}) below.
\end{theorem}

\begin{remark}\upshape\label{mainremark}
1. The function $h(k)$, $k \in \Gamma$, is well-defined by the right-hand side of (\ref{hdef}), because $w^2 + 1 \geq 0$ and $\sqrt{w^2 + 1} - w > 0$ for $k \in \Gamma$.

2. Unless stated otherwise, all integrals in this paper along paths on Riemann surfaces for which only the endpoints are specified are assumed to lie within the fundamental polygon obtained by cutting the Riemann surface along the given cut basis. This implies that some integrals will depend on the particular choice of the $a_j$'s and $b_j$'s within their respective homology classes. It is convenient to fix the $a_j$'s and $b_j$'s so that they are invariant under the involution $k^\pm \to k^\mp$. For $\zeta > \text{Re}\, k_4$, this is accomplished by fixing\footnote{For two complex numbers $z_1$ and $z_2$, $[z_1, z_2]^+$ and $[z_1, z_2]^-$ denote the covers of $[z_1, z_2]$ in the upper and lower sheets of $\Sigma_z$, respectively.}
$$b_4 = [i\bar{z}, k_4]^+ \cup [k_4, i\bar{z}]^-; \qquad
b_j = b_{j+1} \cup [\bar{k}_{j+1}, k_j]^+ \cup [k_j, \bar{k}_{j+1}]^-, \quad j = 1, 2, 3,$$
and by letting $a_j$, $j = 1,\dots, 4$, be the path in the homology class specified by Figure \ref{standardcutsall.pdf} which as a point set consists of the points of $\Sigma_z$ lying directly above $[k_j, \bar{k}_j]$. 
This implies the important symmetry $\omega(k^+) = -\omega(k^-)$.
For $\Sigma_z$ with $0 \leq \zeta < \text{Re}\, k_4$ and for other Riemann surfaces below, we will assume that an analogous fixing of the cut basis which assures $\omega(k^+) = -\omega(k^-)$ has been made.

3. The limit as $\epsilon \to 0$ of the expression within brackets on the right-hand side of (\ref{LLdef}) always exists and is finite because of the pole structure of $\omega_{-r_1^+, -r_1^-}$ and $\omega_{r_1^+ r_1^-}$.

4. The assumption that $r_1$ satisfies (\ref{r1betweencuts}) is made for simplicity and is not essential. The relevant formulas in the case when (\ref{r1betweencuts}) does not hold can be obtained from those presented here by analytic continuation.
\end{remark}

\subsection{Axis and horizon values}
Of particular interest are the values of the Ernst potential and of the metric functions on the regular axis $\{i\zeta\, |\, \zeta > r_1\}$ and on the black hole horizon $\{i\zeta \,| \, 0 < \zeta < r_1\}$.
In the limit $\rho \to 0$, the Riemann surface $\Sigma_z$ degenerates since the branch cut $[-iz, i\bar{z}]$ shrinks to a point. Thus, the values of $f$ on the $\zeta$-axis are given in terms of quantities defined on the Riemann surface $\Sigma'$ defined by the equation
\begin{equation}\label{Sigmapdef}  
  y'^2 = \prod_{j = 1}^4 (k - k_j)(k - \bar{k}_j),
\end{equation}
i.e., $\Sigma'$ is the Riemann surface $\Sigma_z$ with the cut $[-iz, i\bar{z}]$ removed.
\begin{figure}
\begin{center}
    \includegraphics[width=.4\textwidth]{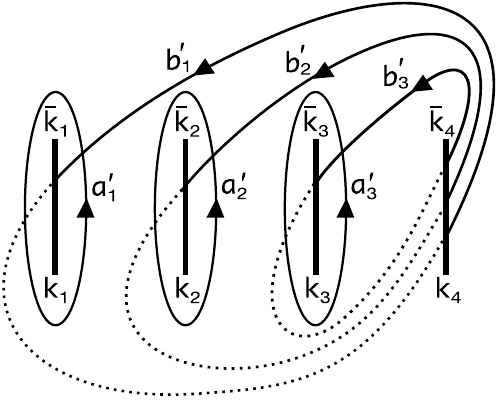} \quad
     \begin{figuretext}\label{degeneratedcuts.pdf}
     The cut system $\{a_j', b_j'\}_{j =1}^3$ on the degenerated Riemann surface $\Sigma'$ of genus $g = 3$.
     \end{figuretext}
 \end{center}
\end{figure}    
We introduce a canonical cut basis $\{a'_j, b'_j\}_1^3$ on $\Sigma'$ according to Figure \ref{degeneratedcuts.pdf} and let $\omega' = (\omega_1', \omega_2', \omega_3')^T$ denote the dual basis of holomorphic one-forms. We let $B'$ denote the associated period matrix and introduce the short-hand notation $\Theta'(v) := \Theta(v | B')$, $v \in \C^3$, for the associated theta function.

Let $\gamma'$ denote the $\zeta$-dependent contour on $\Sigma'$ which projects to the contour (\ref{gammaprojection}) in the complex $k$-plane and which lies in the upper sheet for $\text{Re}\, k < \zeta$ and in the lower sheet for $\text{Re}\,k > \zeta$. Define the $\zeta$-dependent quantities $u' \in \C^3$, $I' \in \R$,  and $K' \in \C$ by
\begin{equation}\label{upIpKpdef}
u' = \int_{\Gamma^+} h \omega' + \int_{\gamma'} \omega', \qquad
I' =  \int_{\Gamma^+} h \omega_{\infty^+\infty^-}' + \int_{\gamma'} \omega_{\infty^+\infty^-}', \qquad
K' =  \int_{k_4}^{\infty^-} \omega_{\zeta^+\zeta^-}'. 
\end{equation}
Let $\gamma^+$ denote the contour on $\Sigma'$ with the same projection onto the complex $k$-plane as $\gamma'$, but which lies entirely in the upper sheet, and define $J' \in \R$ by
\begin{equation}\label{Jprimedef}
J' = \begin{cases} \int_{\Gamma^+} h \omega_{\zeta^+\zeta^-}' +  \int_{\gamma'} \omega_{\zeta^+\zeta^-}', \qquad \zeta > r_1, 
\\
\int_{\Gamma^+} h \omega_{\zeta^+\zeta^-}'  + \left( \int_{r_1^-}^{r_1^+} + \int_{\gamma^+}' \right)\omega_{\zeta^+\zeta^-}', \qquad 0 < \zeta < r_1.
 \end{cases}
\end{equation}
The prime on the integral along $\gamma^+$ indicates that the path should be deformed slightly before evaluation, so that it avoids the pole of the integrand at $k = \zeta^+$. It is irrelevant for the formulas below if this deformation is performed so that the pole lies to the right or to the left of $\gamma^+$, since these two choices yield values of $J'$ which differ by a multiple of $2 \pi i$ and $J'$ only appears exponentiated. In the same way, the value of $J'$ changes by irrelevant multiples of $2\pi i$ if loops surrounding $\zeta^\pm$ are added to the contour from $r_1^-$ to $r_1^+$.

\begin{proposition}[Solution on the regular axis]\label{axisprop}
The behavior of the solution (\ref{ernstsolution}) near the regular axis $\{i\zeta\, |\, \zeta > r_1\}$ is given by
\begin{equation}\label{ernstsolutionnearaxis}
  f(\rho + i\zeta) = f(i\zeta) + O(\rho^2), \qquad \rho \to 0, \quad \zeta > r_1,
\end{equation}
where
\begin{equation}\label{ernstsolutionaxis}
f(i\zeta) = \frac{\Theta'(u' -  \int_{\zeta^-}^{\infty^-} \omega') - \Theta'(u' -  \int_{\zeta^+}^{\infty^-} \omega')e^{J' - K'}}{\Theta'(u' +  \int_{\zeta^-}^{\infty^-} \omega') -  \Theta'(u' +  \int_{\zeta^+}^{\infty^-} \omega')e^{-J'-K'}} e^{I'-J'}, \qquad \zeta > r_1.
\end{equation}
The behavior of the metric functions $e^{2U}$, $a$, and $e^{2\kappa}$ in (\ref{solutionmetric}) near the regular axis is given by
\begin{align} \nonumber
&e^{2U(\rho + i\zeta)} = e^{2U(i\zeta)} + O(\rho^2),  \qquad 
a(\rho + i\zeta) = O(\rho^2), 
	\\ \nonumber
& e^{2\kappa(\rho + i\zeta)} = 1 + O(\rho^2), \qquad& \rho \to 0, \quad \zeta > r_1,
\end{align}
where
\begin{align}\label{e2Uaxis}
e^{2U(i\zeta)} =  
\frac{\Theta'(u')^2\left[\Theta'(\int_{\zeta^-}^{\infty^-} \omega')^2 -  \Theta'(\int_{\zeta^+}^{\infty^-} \omega')^2e^{-2K'}\right]}
{\Theta'(0)^2\left[\Theta'(u' + \int_{\zeta^-}^{\infty^-} \omega')^2 -  \Theta'(u' + \int_{\zeta^+}^{\infty^-} \omega')^2e^{-2J' - 2K'}\right]}e^{I' - J'}, \qquad \zeta > r_1.
\end{align}
\end{proposition}
Define the constant $L_0$ by
\begin{align} \label{L0def}
L_0 =& -\frac{1}{2} \int_{\Gamma} d\kappa_1 \frac{dh}{dk}(\kappa_1) \int_{\Gamma}' h(\kappa_2)\omega'_{\kappa_1^+\kappa_1^-}(\kappa_2^+)
+ \int_{\Gamma^+} h \omega'_{-r_1^+, -r_1^-} - \int_{\Gamma^+} h \omega'_{r_1^+ r_1^-}
  	\\ \nonumber
& + \frac{1}{2} \lim_{\epsilon \to 0} \left(\int_{\gamma_1^+(\epsilon)} \omega'_{-r_1^+, -r_1^-} - \int_{\gamma_2^+(\epsilon)} \omega'_{r_1^+ r_1^-} - 2\ln \epsilon \right),
\end{align}
where $\gamma_1^+(\epsilon)$ denotes the contour $\gamma^+$ with the segment covering $[-r_1, -r_1 + \epsilon]$ removed, and $\gamma_2^+(\epsilon)$ denotes the contour $\gamma^+$ with the segment covering $[r_1 - \epsilon, r_1]$ removed.
Define the constants $a_0, K_0 \in \R$ by
\begin{align}\label{a0solution}
a_0 = &  - 2 i \frac{\Theta'(u' + 2\int_{k_4}^{\infty^-} \omega')}{\Theta'(u')} e^{-I'} \lim_{R \to \infty} \biggl(Re^{\int_{k_4}^{R^+}\omega_{\infty^+\infty^-}'}\biggr),
	\\ \label{K0solution}
K_0 = &\, \frac{\Theta'(0)^2}{\Theta'(u')^2}e^{-L_0},
\end{align}
where the right-hand sides are understood to be evaluated at some $\zeta > r_1$.\footnote{Their values are independent of the choice of $\zeta > r_1$.}
Define the $\zeta$-dependent quantities $L'$ and $M'$ by
\begin{align}\label{Lprimedef}
L' = &\, -\frac{1}{2} \int_{\Gamma^+} d\kappa_1 \frac{dh}{dk}(\kappa_1) \int_{\Gamma^+} h(\kappa_2) 
\omega'_{\kappa_1^+\kappa_1^-}(\kappa_2^+)
+ \int_{\Gamma^+} h \omega'_{-r_1^+ -r_1^-}
  + \int_{\Gamma^+} h \omega'_{r_1^+ r_1^-}
  	\\ \nonumber
  &\,+ \frac{1}{2}\lim_{\epsilon \to 0} \left(\int_{\gamma_1'(\epsilon)}\omega'_{-r_1^+ -r_1^-} + \int_ {\gamma_2'(\epsilon)} \omega'_{r_1^+ r_1^-} - 2\ln \epsilon \right)
- \int_{\Gamma^+} h \omega'_{\zeta^+\zeta^-}, \qquad 0 < \zeta < r_1,
\end{align}
and
\begin{equation}\label{Mprimedef}
M' = \frac{1}{2} \lim_{\epsilon \to 0} \left(\int_{(\zeta - \epsilon)^-}^{(\zeta - \epsilon)^+} \omega_{\zeta^+ \zeta^-}'  - 2\ln \epsilon - \ln 4 - \pi i\right), \qquad 0 < \zeta < r_1,  
\end{equation}
where $\gamma_1'(\epsilon)$ denotes the contour $\gamma'$ with the segment covering $[-r_1, -r_1 + \epsilon]$ removed, and $\gamma_2'(\epsilon)$ denotes the contour $\gamma'$ with the segment covering $[r_1 - \epsilon, r_1]$ removed.

\begin{proposition}[Solution on the horizon]\label{horizonprop}
The behavior of the solution (\ref{ernstsolution}) near the black hole horizon $\{i\zeta \, |\, 0 < \zeta < r_1\}$ is given by
\begin{equation}\label{ernstsolutionnearhorizon}
  f(\rho + i\zeta) = f(i\zeta) + O(\rho^2), \qquad \rho \to 0, \quad 0 < \zeta < r_1,
\end{equation}
where
\begin{equation}\label{ernstsolutionhorizon}
f(i\zeta) = - \frac{\Theta'(u' - \int_{\zeta^+}^{\infty^-} \omega')  - \Theta'(u' - \int_{\zeta^+}^{\infty^-}\omega' + \int_{\zeta^-}^{\zeta^+}\omega')e^{J' - K'}}{\Theta'(u' + \int_{\zeta^-}^{\infty^-}\omega') - \Theta'(u' + \int_{\zeta^-}^{\infty^-}\omega' + \int_{\zeta^-}^{\zeta^+}\omega')e^{J' + K'}} e^{I'-K'}, \qquad 0 < \zeta < r_1.
\end{equation}

The behavior of the metric functions $e^{2U}$, $a$, and $e^{2\kappa}$ in (\ref{solutionmetric}) near the black hole horizon is given by
\begin{align}\nonumber
e^{2U(\rho + i\zeta)} = &\; e^{2U(i\zeta)} + O(\rho^2),  \qquad  a(\rho + i\zeta) = a_{hor} + O(\rho^2),
	\\ \label{e2kappahorizon}
e^{2\kappa(\rho + i\zeta)} = & \; e^{2\kappa_{hor}}  + O(\rho^2), \qquad \rho \to 0, \quad 0 < \zeta < r_1,
\end{align}
where
\begin{align}\label{e2Uhorizon}
e^{2U(i\zeta)} = & -\frac{\Theta'(u' + \int_{\zeta^-}^{\zeta^+} \omega')^2}{\Theta'(0)^2}
	\\ \nonumber
& \times  \frac{\Theta'(\int_{\zeta^-}^{\infty^-} \omega')^2 -  \Theta'(\int_{\zeta^+}^{\infty^-} \omega')^2e^{-2K'}}{\Theta'(u' + \int_{\zeta^-}^{\infty^-} \omega')^2 - \Theta'(u' + \int_{\zeta^-}^{\infty^-} \omega'+ \int_{\zeta^-}^{\zeta^+} \omega')^2e^{2J' + 2K'}} e^{I' + J'} ,
	\\ \label{ahorsolution}
a_{hor} =&\; a_0 + \frac{\Theta'(u' + 2\int_{\zeta^-}^{\infty^-}\omega') \Theta'(0)^4}{\Theta'(u' + \int_{\zeta^-}^{\zeta^+}\omega')\left(\Theta'(\int_{\zeta^-}^{\infty^-}\omega')^2 - \Theta'(\int_{\zeta^+}^{\infty^-}\omega')^2e^{-2K'}\right)^2}e^{-I' - M'},
	\\ \label{e2kappahorsolution}
e^{2\kappa_{hor}} =&\, - K_0 \frac{\Theta'(u' + \int_{\zeta^-}^{\zeta^+} \omega')^2 }{\Theta'(0)^2}
e^{J' + L'}.
\end{align}
Moreover, $a_{hor}$ and $e^{2\kappa_{hor}}$ are constants independent of $0 < \zeta < r_1$.
\end{proposition}

By taking limits in the above formulas we can find the values of $f$ at the origin $z = +i0$ and at the point $z = ir_1$ where the regular axis meets the horizon. 

\begin{proposition}[Solution at $z = +i0$ and $z = ir_1$]\label{f0f1prop}
The value of the solution (\ref{ernstsolution}) at $z = ir_1$ is given by
\begin{align} \label{f1solution}
  f(ir_1) = - \frac{\Theta'(u' - \int_{r_1^+}^{\infty^-} \omega')}{\Theta'(u' + \int_{r_1^-}^{\infty^-} \omega')} e^{I' - K'} \Bigg|_{\zeta = r_1}.
\end{align}  
Define the value of $J'$ at $\zeta = 0$ by
$$J'|_{\zeta = 0} =  \int_{\Gamma^+}' h \omega_{0^+0^-}'  + \left( \int_{r_1^-}^{r_1^+} + \int_{\gamma^+}' \right)\omega_{0^+0^-}',$$
where the primes on the integrals indicate that the contours should be deformed slightly before evaluation so that they pass to the left of the pole at $k = 0^+$. 
Then the value $f(+i0)$ of the solution (\ref{ernstsolution}) at $z = +i0$ is given by the right-hand side of (\ref{ernstsolutionhorizon}) evaluated at $\zeta = 0$.
\end{proposition}

\section{Example}\nequation\label{examplesec}
Before presenting the derivation of the solution presented in the previous section, we wish to consider a concrete example.
In this regard we note that all the quantities appearing in section \ref{diskblackholesec} can easily be computed explicitly using standard results for Riemann surfaces. In order to find the canonical basis $\{\omega_i\}_{1}^4$ and the period matrix $B$, we start with the holomorphic one-forms $\{\zeta_i\}_1^4$ defined by
\begin{equation}\label{zetadef}
  \zeta_i = \frac{k^{i-1}dk}{y}, \qquad i = 1, \dots, 4.
\end{equation}
The $\zeta_i$'s form a noncanonical basis of holomorphic differentials (see e.g. \cite{FK}). Defining the two $4 \times 4$ matrices $A$ and $Z$ by
\begin{equation}\label{AZdef}
(A^{-1})_{ij} = \int_{a_j} \zeta_i, \qquad Z_{ij} = \int_{b_j} \zeta_i, \qquad i,j = 1, \dots, 4,
\end{equation}
we find 
\begin{equation}\label{omegaAzetaBAZ}
  \omega = A \zeta, \qquad  B = AZ.
\end{equation}
The one-form $\omega_{\infty^+\infty^-}$ on $\Sigma_z$ is given explicitly by
\begin{equation}\label{omegainftyinfty}
  \omega_{\infty^+\infty^-} = -\frac{k^4 dk}{y} + \sum_{j = 1}^4 \left( \int_{a_j} \frac{k^4 dk}{y}\right)\omega_j,
\end{equation}
whereas the one-form $\omega_{\zeta^+\zeta^-}'$ on $\Sigma'$ is given by
\begin{equation}\label{omegazetazeta}
\omega_{\zeta^+\zeta^-}' = \frac{y'(\zeta^+)dk}{(k - \zeta)y'(k)} - \sum_{j = 1}^3 \left( \int_{a_j'} \frac{y'(\zeta^+)dk}{(k - \zeta)y'(k)}\right)\omega'_j.
\end{equation}
For $k$'s which lie some distance away from the branch cuts, the value of $y$ in (\ref{Sigmazdef}) can be evaluated according to
$$y(z,k^+) = (k + iz) \sqrt{\frac{k - i\bar{z}}{k + iz}} \prod_{j = 1}^4 (k - k_j) \sqrt{\frac{k - \bar{k}_j}{k - k_j}}, \qquad k \in \hat{\C},$$
where the branches with strictly positive real part are chosen for the square roots. For $k \in [-iz, i\bar{z}]$ and $\epsilon > 0$ infinitesimally small, we have
$$y(z, (k + \epsilon)^+) = (k + iz) \left(-i \sqrt{\left|\frac{k - i\bar{z}}{k + iz}\right|}\right) \prod_{j = 1}^4 (k - k_j) \sqrt{\frac{k - \bar{k}_j}{k - k_j}}, \qquad k \in \hat{\C};$$
similar expressions are valid when $k \in [k_j, \bar{k}_j]$. Using formulas of this type, it is straightforward to numerically evaluate all the expressions presented in section \ref{diskblackholesec}. In fact, the theta functions are particularly suitable for numerical evaluation, because the strictly positive imaginary part of $B$ implies that only a small number of terms in the sum (\ref{Thetadef}) have to be included. In this way, we have verified for several examples all formulas of section \ref{diskblackholesec} to high precision.

\subsection{Numerical data}
\begin{figure}
\begin{center}
    \includegraphics[width=.48\textwidth]{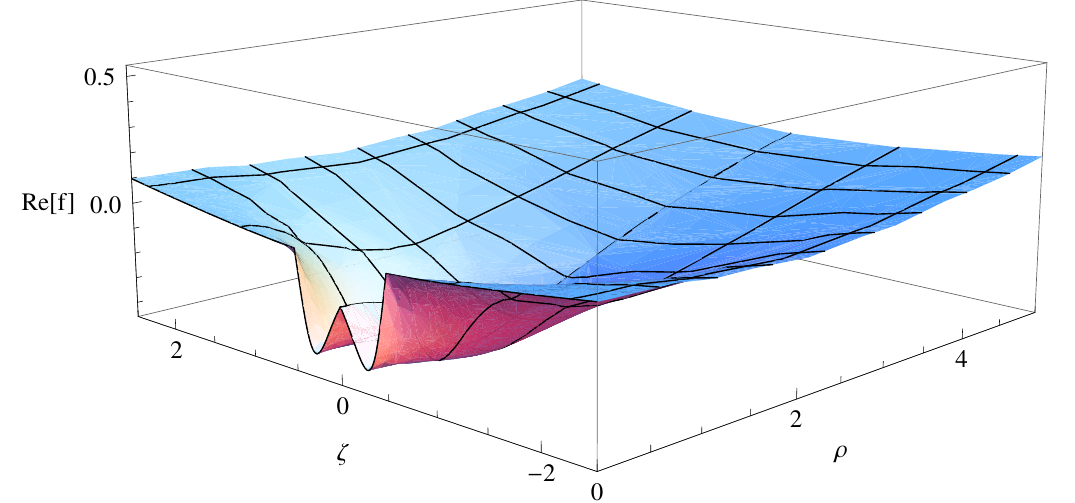} \quad
    \includegraphics[width=.48\textwidth]{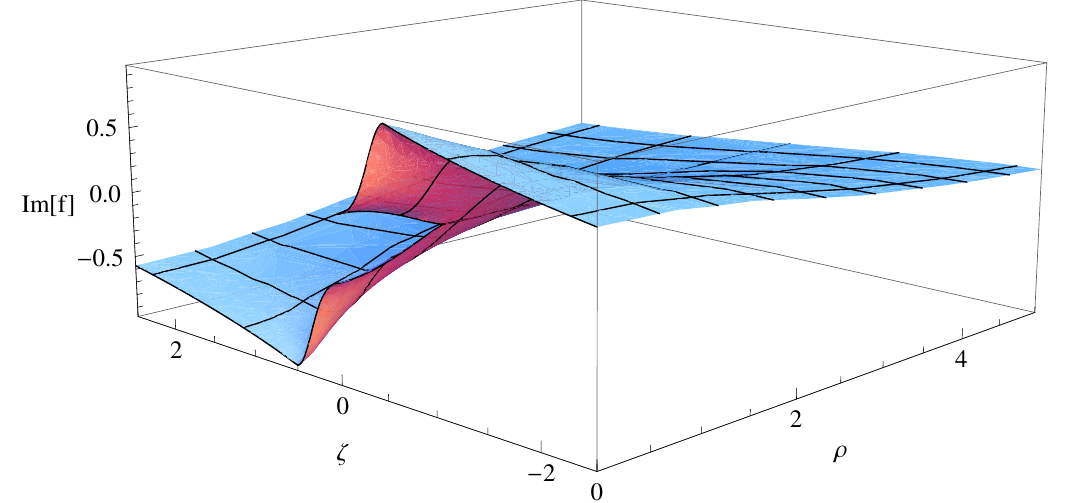} 
     \begin{figuretext}\label{f3Dfig}
       The real and imaginary parts of the solution (\ref{ernstsolution}) for the choice of parameters specified in (\ref{exampleparameters}).          
   \end{figuretext}
 \end{center}
\end{figure}

\begin{figure}
\begin{center}
    \includegraphics[width=.48\textwidth]{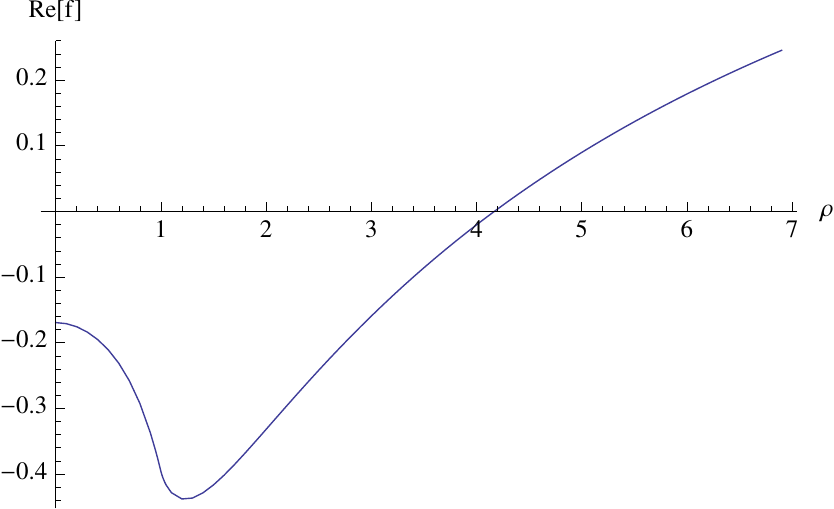} \quad
    \includegraphics[width=.48\textwidth]{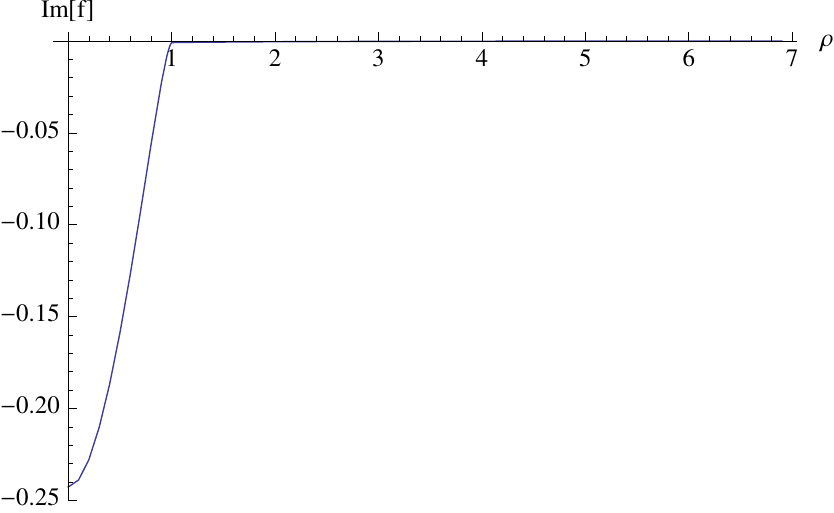} 
      \begin{figuretext}\label{fdiskfig}
       The values of $f$ in the equatorial plane $\zeta = 0^+$ for the choice of parameters specified in (\ref{exampleparameters}).       \end{figuretext}
 \end{center}
\end{figure}

Here we consider the particular example for which
\begin{equation}\label{exampleparameters}  
  \rho_0 = 1, \qquad r_1 = \frac{1}{2}, \qquad w_2 = 3, \qquad w_4 = \frac{1}{10}.
\end{equation}
Numerically, we find 
$$k_1 \approx -0.95 - 5.48i, \qquad k_2 \approx -0.21 - 0.95 i.$$
Thus, (\ref{r1betweencuts}) is satisfied and we compute
$$\Omega \approx 0.055, \qquad \Omega_h \approx 0.14, \qquad f(+i0) \approx -0.17, \qquad f(ir_1) \approx -0.94 i,$$
and
$$a_0 \approx -18.17, \qquad K_0 \approx 9.43, \qquad e^{2\kappa_{hor}} \approx -93.46.$$
The real and imaginary parts of $f$ are shown in Figure \ref{f3Dfig}. Note that the real part of $f$ is continuous but not smooth at the endpoints $\pm ir_1$ of the horizon. Moreover, as expected, the imaginary part of $f$ has a jump across the disk. The values of $f$ in the equatorial plane $\zeta = 0^+$ are shown in Figure \ref{fdiskfig}.

\begin{figure}
\begin{center}
    \includegraphics[width=.48\textwidth]{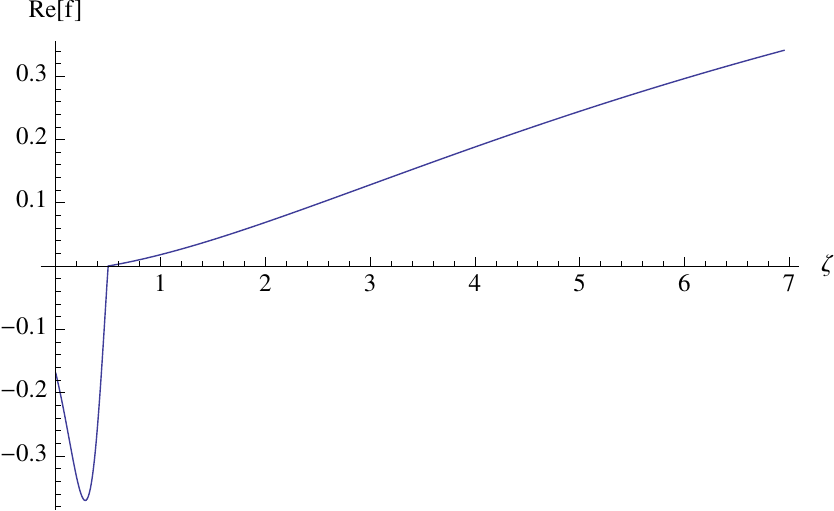} \quad
    \includegraphics[width=.48\textwidth]{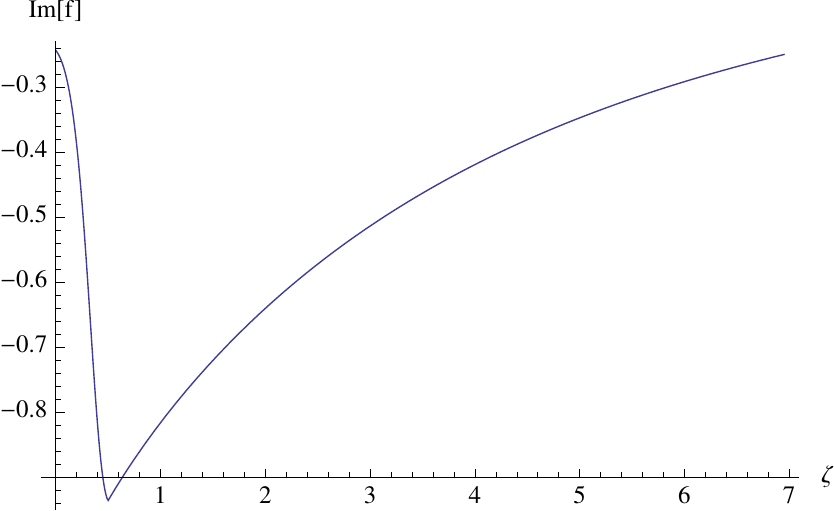} 
     \begin{figuretext}\label{faxisfig}
       The axis and horizon values of $f$ for the choice of parameters specified in (\ref{exampleparameters}).    \end{figuretext}
 \end{center}
\end{figure}

\begin{figure}
\begin{center}
    \includegraphics[width=.48\textwidth]{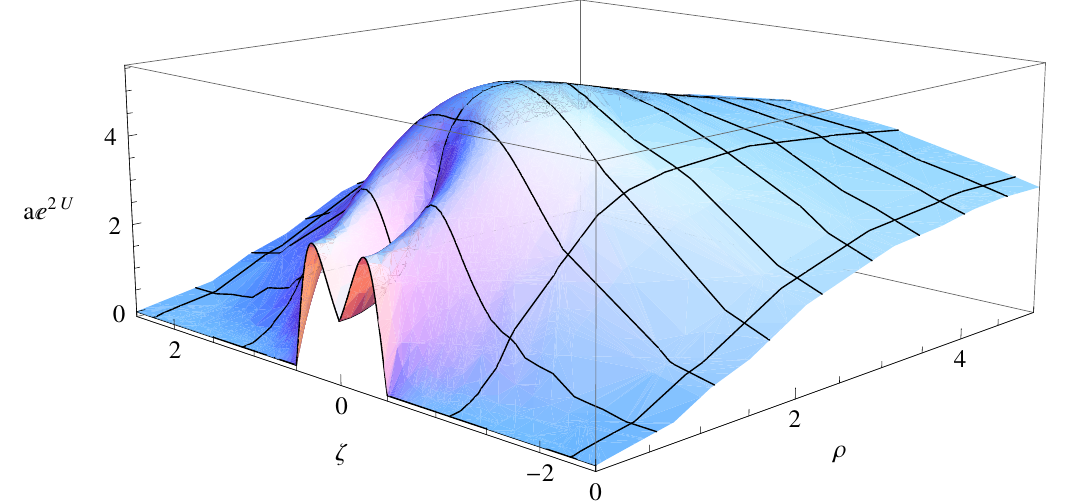} \quad
    \includegraphics[width=.48\textwidth]{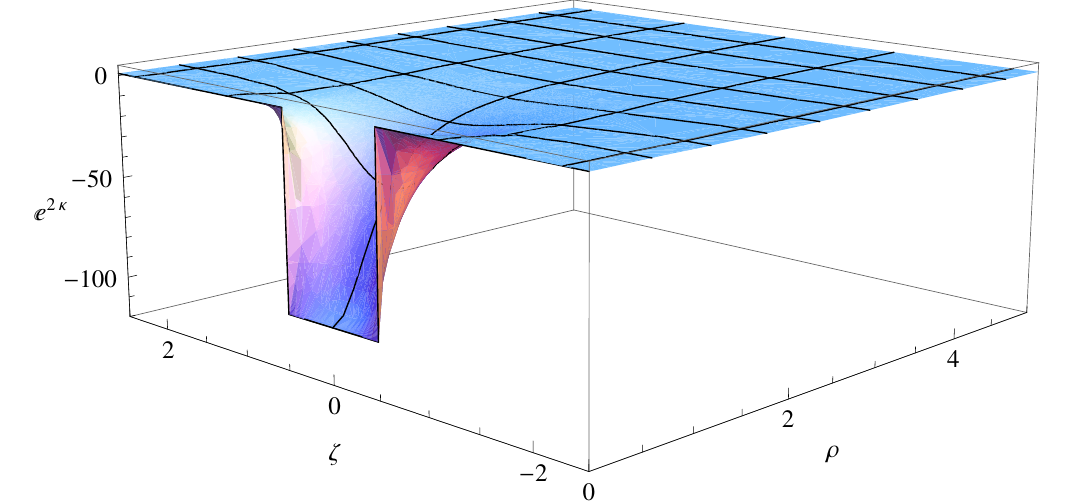} 
     \begin{figuretext}\label{ae2Ue2kappafig}
       The metric functions $ae^{2U}$ and $e^{2\kappa}$ for the choice of parameters specified in (\ref{exampleparameters}).          
      \end{figuretext}
 \end{center}
\end{figure}

\begin{figure}
\begin{center}
    \includegraphics[width=.48\textwidth]{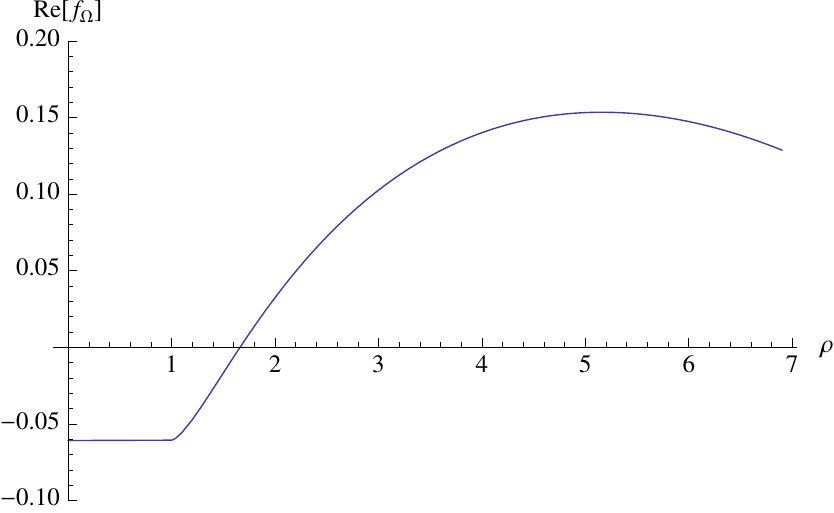} \quad
    \includegraphics[width=.48\textwidth]{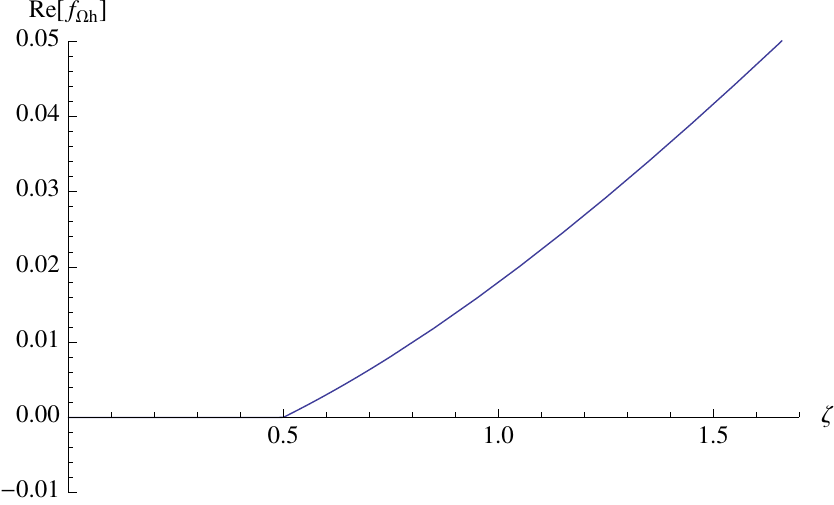} 
     \begin{figuretext}\label{e2UOmegasfig}
       The real parts of the corotating potentials $f_\Omega$ and $f_{\Omega_h}$ in the equatorial plane and along the $\zeta$-axis, respectively.   \end{figuretext}
 \end{center}
\end{figure}

The metric functions $ae^{2U}$ and $e^{2\kappa}$ are shown in Figure \ref{ae2Ue2kappafig}. Note that $ae^{2U} = 0$ on the regular part of the axis. Moreover, $e^{2\kappa} = 1$ on the regular axis and $e^{2\kappa} = e^{2\kappa_{hor}}$ on the horizon.
Figure \ref{e2UOmegasfig} shows the real parts of the corotating potentials $f_\Omega$ and $f_{\Omega_h}$ in the equatorial plane and along the $\zeta$-axis, respectively. In accordance with the boundary conditions (\ref{BVPdisk}) and (\ref{BVPhorizon}), $e^{2U_\Omega}$ is constant along the disk and $e^{2U_{\Omega_h}}$ vanishes along the horizon. 
It can also be verified to high accuracy that the metric functions $a$ and $\kappa$ defined by (\ref{solutionmetric}) satisfy the appropriate equations, i.e. (cf. \cite{KR2005})
\begin{equation}\label{kappaequation}
   a_z = \frac{i \rho}{e^{4U}} b_z  \qquad \hbox{and} \qquad \kappa_z = \frac{\rho}{2e^{4U}} f_z \bar{f}_z,
\end{equation}
and that 
\begin{equation}\label{daOmegadzeta}  
  \frac{da_{\Omega}}{d\zeta}\biggl|_{z = \rho + i0} = 0, \qquad 0 < \rho < \rho_0.
\end{equation}
Equation (\ref{daOmegadzeta}) implies that the imaginary part of $f_\Omega$ is constant along the disk in accordance with (\ref{BVPdisk}).

\section{Spectral theory}\nequation\label{specsec}
The now turn to the proof of the results of section \ref{diskblackholesec}. The proof will proceed through four main steps, presented in sections \ref{specsec}, \ref{scalarRHsec}, \ref{thetasec}, and \ref{axishorizonsec}, respectively. The first step consists of analyzing the Lax pair for equation (\ref{ernst}) and formulating two {\it matrix} Riemann-Hilbert (RH) problems: one main RH problem (which can be formulated for any choice of boundary conditions) and one auxiliary RH problem (which can be formulated because the boundary conditions of the BVP (\ref{BVP}) are linearizable). In the second step, we show that these two matrix RH problems can be combined into a single {\it scalar} RH problem. The third step consists of solving this scalar RH problem explicitly in terms of theta functions. In the final fourth step, we prove propositions \ref{axisprop}-\ref{f0f1prop} concerning the behavior of the solution near the $\zeta$-axis; this step will follow from a study of the theta function formulas of theorem \ref{mainth} in the limit $\rho \to 0$.

\subsection{A bounded and analytic eigenfunction}
The elliptic Ernst equation (\ref{ernst}) admits the Lax pair
\begin{equation}
\begin{cases}\label{ernstlax}
\Phi_z(z, k) = U(z, k) \Phi(z, k),
	\\
\Phi_{\bar{z}}(z, k) =V(z,k) \Phi(z,k),
\end{cases}
\end{equation}
where the $2 \times 2$-matrix valued function $\Phi(z, k)$ is an eigenfunction, $k$ is a spectral parameter, and the $2\times 2$-matrix valued functions $U$ and $V$ are defined by
$$U = \frac{1}{f + \bar{f}}\begin{pmatrix} \bar{f}_z & \lambda \bar{f}_z \\
\lambda f_z & f_z\end{pmatrix}, \qquad 
V =  \frac{1}{f + \bar{f}}\begin{pmatrix} \bar{f}_{\bar{z}} & \frac{1}{\lambda} \bar{f}_{\bar{z}} \\
\frac{1}{\lambda}  f_{\bar{z}}   & f_{\bar{z}}  \end{pmatrix}, 
\qquad \lambda(z,k) = \sqrt{\frac{k - i\bar{z}}{k + iz}}.$$
We write the Lax pair (\ref{ernstlax}) in the form
\begin{equation}\label{laxdiffform}  
  d\Phi = W\Phi \qquad \hbox{where} \qquad  W := Udz + Vd\bar{z}.
\end{equation}
Let
$$\sigma_1 = \begin{pmatrix} 0	&	1 \\ 1	& 0\end{pmatrix}, \qquad
\sigma_2 = \begin{pmatrix} 0	&	-i \\ i	& 0\end{pmatrix}, \qquad
\sigma_3 = \begin{pmatrix} 1	&	0 \\ 0	& -1\end{pmatrix}.$$
Suppose $f$ is a solution of the BVP (\ref{BVP}). Following the same procedure as in \cite{LF}, we define a solution $\Phi(z,k)$ of (\ref{laxdiffform}) with the following properties:
\begin{itemize}
\item For each $z$, $\Phi(z, \cdot)$ is a map from the Riemann surface $\mathcal{S}_z$ to the space of $2 \times 2$ matrices, where $\mathcal{S}_z$ is defined by the equation
$$\lambda^2 = \frac{k - i \bar{z}}{k + i z}.$$
We view $\mathcal{S}_z$ as a two-sheeted covering of the Riemann $k$-sphere endowed with a branch cut from $-iz$ to $i\bar{z}$; the upper (lower) sheet is characterized by $\lambda \to 1$ ($\lambda \to -1$) as $k \to \infty$. 

\item $\Phi$ satisfies the initial conditions
\begin{align}\label{phiinitial}
\lim_{z \to i\infty} [\Phi(z, k^-)]_1 = \begin{pmatrix} 1 \\ 1 \end{pmatrix}, \qquad
  \lim_{z \to i\infty} [\Phi(z, k^+)]_2 = \begin{pmatrix} 1 \\ -1 \end{pmatrix}, \qquad k \in \hat{\C},
\end{align}
where, for a $2 \times 2$-matrix $A$, $[A]_1$ and $[A]_2$ denote the first and second columns of $A$, respectively.

\item $\Phi$ obeys the symmetries
\begin{equation}\label{phisymmetriesk} 
\Phi(z, k^+) = \sigma_3 \Phi(z, k^-) \sigma_1, \qquad 
\Phi(z, k^+) = \sigma_1 \overline{\Phi(z, \bar{k}^+)} \sigma_3, \qquad k \in \hat{\C}.
\end{equation}

\item $\Phi$ is analytic for $k \in \mathcal{S}_z$ away from the set $\Gamma^+ \cup \Gamma^- \cup \{-r_1^\pm, r_1^\pm\}$, where $\Gamma^+$ and $\Gamma^-$ denote the coverings of $\Gamma = [-i\rho_0, i\rho_0]$ in the upper and lower sheets of $\mathcal{S}_z$, respectively.
\end{itemize}
We emphasize that $\Phi$, in general, has singularities (simple poles) at the points $k^\pm$ for $k = -r_1$ and $k = r_1$. 
These poles arise since the Lax pair (\ref{ernstlax}) is singular at points where $e^{2U}= 0$. Physically, the points at which $e^{2U}$ vanishes make up the {\it ergospheres} of the spacetime (within these surfaces there can be no static observer with respect to infinity). To see that $e^{2U} = 0$ at $z = ir_1$ and $z = -ir_1$, we note that the metric function $a$ vanishes identically on the regular axis (cf. \cite{SKMHH}). Thus, evaluating (\ref{UOmegaUrelation}) at $\rho = 0$, we find $U_\Omega(i\zeta) = U(i\zeta)$ for $\zeta > r_1$.
The boundary condition (\ref{BVPhorizon}) together with the continuity of $f_\Omega$ imply that $e^{2U(i\zeta)} \to 0$ as $\zeta \downarrow r_1$. 

In addition to the eigenfunction $\Phi(z,k)$, we will also need its corotating analog $\Phi_\Omega(z,k)$ . This eigenfunction satisfies the Lax pair equations (\ref{ernstlax}) with $f$ replaced by $f_\Omega$ and the initial conditions (\ref{phiinitial}) with $\Phi$ replaced by $\Phi_\Omega$. The eigenfunctions $\Phi$ and $\Phi_\Omega$ are related by
\cite{MAKNP}
\begin{equation}\label{Lambdadef}
  \Phi_\Omega(z, k) = \Lambda_\Omega(z, k)\Phi(z, k), \qquad k \in \mathcal{S}_z,
\end{equation}
where
\begin{equation}\label{Lambdadef2} 
  \Lambda_\Omega(z, k) =  (1 + \Omega a)\mathbb{I} - \Omega \rho e^{-2U}\sigma_3 + i(k + iz)\Omega e^{-2U}(\lambda(z, k)\sigma_1 - \mathbb{I})\sigma_3
\end{equation}  
and $\mathbb{I}$ denotes the $2 \times 2$ identity matrix.
The corotating eigenfunction $\Phi_{\Omega_h}(z,k)$ is defined analogously.

\subsection{The main Riemann-Hilbert problem}
Evaluation at $\rho = 0$ of equation (\ref{UOmegaUrelation}) with $\Omega$ replaced by $\Omega_h$ yields\begin{equation}\label{e2UOmegae2U}
  e^{2U_{\Omega_h}(i\zeta)} = e^{2U(i\zeta)}(1 + \Omega_h  a(i\zeta))^2, \qquad 0 < \zeta < r_1.
\end{equation}
Note that $e^{2U} < 0$ along the horizon, which lies inside the ergosphere. Thus, in view of the boundary condition (\ref{BVPhorizon}), we find that $a \equiv a_{hor}$ on the black hole horizon, where $a_{hor}$ is a constant given by
\begin{equation}\label{ahorOmegah}  
  a_{hor} = -1/\Omega_h.
\end{equation}

The next proposition expresses the values of $\Phi$ on the $\zeta$-axis in terms of two spectral functions $F(k)$ and $G(k)$. We let $f_1$ denote the value of $f$ at $z = ir_1$. 

\begin{proposition}\label{FGprop}
The values of $\Phi$ on the $\zeta$-axis can be expressed in terms of two spectral functions $F(k)$ and $G(k)$ as
\begin{subequations}\label{phionaxis}
\begin{align} \label{phionaxisa}
&\Phi(i\zeta, k^+) =  \begin{pmatrix} \overline{f(i\zeta)} & 1 \\ f(i\zeta) & -1 \end{pmatrix}A(k), & \zeta  > r_1,\quad k \in \hat{\C},
	\\ \label{phionaxisb}
&\Phi(i\zeta, k^+) =  \begin{pmatrix} \overline{f(i\zeta)} & 1 \\ f(i\zeta) & -1 \end{pmatrix}T_1(k)A(k), & 0 < \zeta  < r_1,\quad k \in \hat{\C},
	\\ \label{phionaxisc}
&\Phi(i\zeta, k^+) =  \begin{pmatrix} \overline{f(i\zeta)} & 1 \\ f(i\zeta) & -1 \end{pmatrix}T_2(k)\sigma_1A(k)\sigma_1, & -r_1 < \zeta  < 0,\quad k \in \hat{\C},
	\\ \label{phionaxisd}
&\Phi(i\zeta, k^+) = \begin{pmatrix} \overline{f(i\zeta)} & 1 \\ f(i\zeta) & -1 \end{pmatrix}\sigma_1A(k)\sigma_1, & \zeta  < -r_1,\quad k \in \hat{\C},
\end{align}
\end{subequations}
where the $2\times 2$-matrix valued functions $A(k)$, $T_1(k)$, and $T_2(k)$ are defined by
\begin{align} \label{Adef}
& A(k) = \begin{pmatrix} F(k) & 0 \\ G(k) & 1 \end{pmatrix}, \qquad k \in \hat{\C},
	\\ \label{T12def}
&T_1(k) = \frac{1}{2 (k - r_1) \Omega_h} \begin{pmatrix} 2(k - r_1) \Omega_h - i f_1 & i \\ - if_1^2 & 2(k - r_1) \Omega_h + i f_1 \end{pmatrix}, \qquad k \in \hat{\C}, 
	\\ \nonumber
&T_2(k) = \overline{T_1(-\bar{k})} , \qquad k \in \hat{\C}.
\end{align}
The functions $F(k)$ and $G(k)$ have the following properties:
\begin{itemize}
\item $F$ and $G$ are unique functions of $k \in \hat{\C}$, i.e. viewed as functions on $\mathcal{S}_z$ they satisfy
\begin{equation}\label{FGunique}
   F(k^+) = F(k^-), \qquad G(k^+) = G(k^-), \qquad k \in \hat{\C}.
\end{equation}

\item $F(k)$ and $G(k)$ are analytic for $k \in \hat{\C} \setminus (\Gamma \cup \{r_1, -r_1\})$.

\item Under the conjugation $k \mapsto \bar{k}$, $F$ andÊ $G$ obey the symmetries
\begin{equation}\label{FGsymm}  
 F(k) = \overline{F(\bar{k})}, \qquad G(k) = -\overline{G(\bar{k})}, \qquad k \in \hat{\C}.
\end{equation}

\item In the limit $k \to \infty$, 
\begin{equation}\label{FGlimit}  
  F(k) = 1 + O(1/k), \qquad G(k) = O(1/k), \qquad k \to \infty.
\end{equation}
\end{itemize}
\end{proposition}
\proofbegin
For $z = i \zeta$, $\lambda = 1$ for all $k$ on the upper sheet of $\mathcal{S}_z$. The axis values of $\Phi$ are thus determined by integration of the equation
\begin{equation}\label{dPhiWonaxis}  
  d\Phi = W(i\zeta, k^+) \Phi
\end{equation} 
where
\begin{equation}\label{Wlambda1}
 W(i\zeta, k^+) = \frac{1}{f + \bar{f}} \begin{pmatrix} d\bar{f} & d\bar{f} \\
 df & df\end{pmatrix}.
\end{equation}
Since the real part of $f$ vanishes at $z = \pm ir_1$, equation (\ref{dPhiWonaxis}) breaks down at $\zeta = \pm r_1$. 
Integration of (\ref{dPhiWonaxis}) for $\zeta$ in each of the four intervals $(-\infty, -r_1)$, $(-r_1, 0)$, $(0,r_1)$, and $(r_1, \infty)$ yields
\begin{align*}
&\Phi(i\zeta, k^+) =  \begin{pmatrix} \overline{f(i\zeta)} & 1 \\ f(i\zeta) & -1 \end{pmatrix}U_1(k), & \zeta  > r_1,\quad k \in \hat{\C},
	\\ 
&\Phi(i\zeta, k^+) =  \begin{pmatrix} \overline{f(i\zeta)} & 1 \\ f(i\zeta) & -1 \end{pmatrix}U_2(k), & 0 < \zeta  < r_1,\quad k \in \hat{\C},
	\\ 
&\Phi(i\zeta, k^+) =  \begin{pmatrix} \overline{f(i\zeta)} & 1 \\ f(i\zeta) & -1 \end{pmatrix}U_3(k), & -r_1 < \zeta  < 0,\quad k \in \hat{\C},
	\\ 
&\Phi(i\zeta, k^+) = \begin{pmatrix} \overline{f(i\zeta)} & 1 \\ f(i\zeta) & -1 \end{pmatrix}U_4(k), & \zeta  < -r_1,\quad k \in \hat{\C},
\end{align*}
where the matrices $U_j(k)$, $j = 1, \dots, 4$, are independent of $\zeta$. The initial conditions (\ref{phiinitial}) imply that $U_1 = A$ for some functions $F(k)$ and $G(k)$. This establishes (\ref{phionaxisa}). 
The value of $\Phi$ at $z = -i\infty$ is obtained from the value at $z = i\infty$ by integrating $W\Phi$ along a large semicircle at infinity. During this integration $k$ changes sheets. Thus, using (\ref{phionaxisa}) and the fact that $W\Phi$ vanishes for large $z$, we compute
\begin{equation}\label{limziinftPhi}
\lim_{z \to -i\infty} \Phi(z, k^+) = \lim_{z \to i\infty} \Phi(z, k^-) = \lim_{z \to i\infty} \sigma_3 \Phi(z, k^+)\sigma_1 = \begin{pmatrix} 1 & 1 \\ 1 & -1\end{pmatrix}\begin{pmatrix} 1 & G(k) \\ 0 & F(k) \end{pmatrix}.
\end{equation}
This shows that $U_4 = \sigma_1A(k)\sigma_1$ and proves (\ref{phionaxisd}).

We now use continuity of the matrices $\Phi$ and $\Phi_{\Omega_h}$ at the points $z = \pm i r_1$ to find $U_2$ and $U_3$. Let $f_2$ denote the value of $f$ at $z = -ir_1$.
The conditions that $\Phi(i\zeta, k^+)$ be continuous at $\zeta = r_1$ and $\zeta = -r_1$ are
\begin{subequations}\label{fourequations}
\begin{align}
  \begin{pmatrix} \bar{f}_1 & 1 \\ f_1 & -1 \end{pmatrix} ( A(k) - U_2(k)) = 0 \qquad \hbox{and} \qquad
  \begin{pmatrix} \bar{f}_2 & 1 \\ f_2 & -1 \end{pmatrix} ( \sigma_1A(k)\sigma_1 - U_3(k)) = 0,
\end{align}
respectively.
In view of equations (\ref{Lambdadef}), (\ref{Lambdadef2}), and (\ref{ahorOmegah}), the conditions that $\Phi_{\Omega_h}(i\zeta, k^+)$ be continuous at $\zeta = r_1$ and $\zeta = -r_1$ are
\begin{align}
 \left[ \begin{pmatrix} \bar{f}_1 & 1 \\ f_1 & -1 \end{pmatrix} + 2i\Omega_h(k - r_1)\begin{pmatrix} -1 & 0 \\ 1 & 0 \end{pmatrix}\right] A(k) = 2 i \Omega_h (k - r_1)\begin{pmatrix} -1 & 0 \\ 1 & 0 \end{pmatrix} U_2(k) ,
 \end{align}
 and
 \begin{align}
 \left[ \begin{pmatrix} \bar{f}_2 & 1 \\ f_2 & -1 \end{pmatrix} + 2i\Omega_h(k + r_1)\begin{pmatrix} -1 & 0 \\ 1 & 0 \end{pmatrix}\right] \sigma_1 A(k)\sigma_1 = 2 i \Omega_h (k + r_1)\begin{pmatrix} -1 & 0 \\ 1 & 0 \end{pmatrix} U_3(k),
 \end{align}
 \end{subequations}
respectively.
The top and bottom rows of each of the four matrix equations in (\ref{fourequations}) are linearly dependent since $f_1$ and $f_2$ are purely imaginary. Combining the four bottom rows into two matrix equations, we find
\begin{align}\nonumber
 & \begin{pmatrix} f_1 & -1 \\ f_1 + 2 i \Omega_h ( k - r_1) & -1 \end{pmatrix}A(k) = \begin{pmatrix} f_1 & -1 \\ 2 i \Omega (k - r_1) & 0 \end{pmatrix} U_2(k),
  	\\ \nonumber
 & \begin{pmatrix} f_2 & -1 \\ f_2 + 2 i \Omega_h ( k + r_1) & -1 \end{pmatrix}\sigma_1A(k)\sigma_1 = \begin{pmatrix} f_2 & -1 \\ 2 i \Omega_h (k + r_1) & 0 \end{pmatrix} U_3(k).
\end{align}
Using that $f_1 = -f_2$ in view of the equatorial symmetry, we deduce from these equations that $U_2(k) = T_1(k)A(k)$ and $U_3(k) = T_2(k)\sigma_1A(k)\sigma_1$, where $T_1$ and $T_2$ are given by (\ref{T12def}). This proves (\ref{phionaxisb}) and (\ref{phionaxisc}).

The properties of $F$ and $G$ are proved as in \cite{LF}.
\proofend

The functions $F(k)$ and $G(k)$ jump across $\Gamma = [-i\rho_0, i\rho_0]$. Let $F^+, G^+$ and $F^-, G^-$ denote the values of $F$ and $G$ for $k$ to the right and left of $\Gamma$, respectively. It follows as in \cite{MAKNP} (see also \cite{LF}) that 
\begin{equation}\label{Phijumps}
\Phi^-(z, k) = \Phi^+(z, k)D(k), \quad k \in \Gamma^+; \qquad \Phi^-(z, k) = \Phi^+(z, k)\sigma_1 D(k)\sigma_1, \quad k \in \Gamma^-,
\end{equation}
where the jump matrix $D$ is given in terms of $F^\pm$ and $G^\pm$ by
\begin{equation}\label{Dequation1}
  D(k) = \begin{pmatrix} F^+(k) & 0 \\ G^+(k) & 1 \end{pmatrix}^{-1}\begin{pmatrix} F^-(k) & 0 \\ G^-(k) & 1 \end{pmatrix}.
\end{equation}  

For a given $z$, equation (\ref{Phijumps}) provides the jump condition for a matrix RH problem on the Riemann surface $\mathcal{S}_z$ satisfied by $\Phi(z,k)$. We will refer to this as the {\it main} RH problem.\footnote{A complete formulation of this problem also involves specifying residue conditions at the four points $\pm r_1^\pm$ as well as a normalization condition.} 
In general, given {\it both} the Dirichlet and Neumann boundary values for a BVP for the Ernst equation, it is possible to determine the spectral functions $F$ and $G$, compute the jump matrix $D$, and then obtain the Ernst potential $f$ from the asymptotics of the solution of the main RH problem. However, for a well-posed problem only one of these boundary values is specified. In our analysis of (\ref{BVP}) we will therefore instead use the global relation and the symmetry of the boundary conditions to formulate an {\it auxiliary} RH problem from which $F$ and $G$ can be determined.

\subsection{The global relation}\label{eqsec}
The equatorial symmetry of the solution $f$ of (\ref{BVP}) implies that the spectral functions $F(k)$ and $G(k)$ satisfy an important relation. Recalling the axis values (\ref{phionaxis}) of $\Phi$, the following proposition is proved in the same way as proposition 4.3 in \cite{LF}. 

\begin{proposition}\label{eqprop}
The spectral functions $F(k)$ and $G(k)$ defined in proposition \ref{FGprop} satisfy
\begin{equation}\label{eqrelation}
  \overline{T_1A_+(k)\sigma_1A_+^{-1}(k) T_1^{-1}} = T_2 \sigma_1A_-(k) \sigma_1A_-^{-1}(k)\sigma_1 T_2^{-1}, \qquad k \in \Gamma,
\end{equation}
where $A(k)$ is defined in terms of $F(k)$ and $G(k)$ by equation (\ref{Adef}) and $A_\pm$ denote the values of $A$ to the right and left of $\Gamma$, respectively. 
\end{proposition}
Equation (\ref{eqrelation}) is referred to as the {\it global relation}. 

\subsection{Linearizable boundary conditions}
In general, the global relation alone is not sufficient for determining $F$ and $G$. However, for boundary conditions satisfying sufficient symmetry, so-called linearizable boundary conditions, there exist another algebraic relation satisfied by $F$ and $G$. The boundary conditions specified in (\ref{BVP}) turn out to be linearizable. Indeed, recalling the axis values (\ref{phionaxis}) of $\Phi$, the following proposition is proved in the same way as proposition 5.3 in \cite{LF}.

\begin{proposition}\label{fconstprop}
The spectral functions $F(k)$ and $G(k)$ satisfy the relation
\begin{equation}\label{fconstrelation}
  (B^{-1}\Lambda^{-1}\sigma_1\sigma_3\overline{\Lambda B})(\overline{T_1A_+ \sigma_1 A_+^{-1}T_1^{-1}}) = - (T_1 A_+\sigma_1A_+^{-1}T_1^{-1})(B^{-1} \Lambda^{-1}\sigma_1\sigma_3 \overline{\Lambda B}), \quad k \in \Gamma,
\end{equation}
where we use the short-hand notation $B$ and $\Lambda$ for
\begin{equation}\label{Bdef}
  B := \begin{pmatrix} \overline{f(+ i0)} &  1  \\	 f(+ i0) &   -1 \end{pmatrix}, \qquad \Lambda := \Lambda_\Omega(+i0, k^+) = \left(1 - \frac{\Omega}{\Omega_h}\right)\mathbb{I} + i k \Omega e^{-2U_0}(\sigma_1 - \mathbb{I}) \sigma_3.
\end{equation}\end{proposition}

\subsection{The auxiliary Riemann-Hilbert problem}
Combining the relations (\ref{eqrelation}) and (\ref{fconstrelation}), we can formulate a RH problem for the $2\times 2$-matrix valued function $\mathcal{M}(k)$ defined by
\begin{equation}\label{mathcalMdef}
\mathcal{M}(k) = A(k)\sigma_1A^{-1}(k)
= \left(
\begin{array}{cc}
 -G(k) & F(k) \\
 \frac{1-G(k)^2}{F(k)} & G(k)
\end{array}
\right), \qquad k \in \hat{\C}.
\end{equation}

\begin{proposition}\label{auxRHprop}
Suppose $f$ is a solution of the BVP (\ref{BVP}). Let $f_1 := f(ir_1) \in i\R$ denote the value of $f$ at $z = ir_1$.
Then the spectral functions $F(k)$ and $G(k)$ are given by
$$F(k) = \mathcal{M}_{12}(k), \qquad G(k) = \mathcal{M}_{22}(k), \qquad k \in \hat{\C},$$
where $\mathcal{M}$ is the unique solution of the following RH problem:
\begin{itemize}
\item \text{$\mathcal{M}(k)$ is analytic for $k \in \hat{\C} \setminus (\Gamma \cup \{-r_1, r_1\})$.} 

\item  Across $\Gamma$, $\mathcal{M}(k)$ satisfies the jump condition
\begin{equation}\label{auxiliaryRH}  
   \mathcal{S}(k)\mathcal{M}^-(k) = -\mathcal{M}^+(k)\mathcal{S}(k),\qquad k \in \Gamma, 
\end{equation}
where $\mathcal{M}^+$ and $\mathcal{M}^-$ denote the values of $\mathcal{M}$ to the right and left of $\Gamma$, respectively, and $\mathcal{S}(k)$ is defined by
\begin{equation}\label{mathcalSdef}
\mathcal{S}(k) = T_1^{-1}B^{-1}\Lambda^{-1}\sigma_1\sigma_3\overline{\Lambda B} T_2\sigma_1, \qquad k \in \Gamma.
\end{equation}

\item $\mathcal{M}$ has the asymptotic behavior
\begin{equation}\label{MtominusI}
  \mathcal{M}(k) = \sigma_1 + O(1/k), \qquad k \to \infty.
\end{equation}

\item The entries of $\mathcal{M}$ have simple poles at $k = r_1$ and $k = -r_1$. The associated residues are given by
\begin{equation}\label{Mresidues}
\underset{r_1}{\text{\upshape Res}} \, \mathcal{M}(k) = \frac{1}{\alpha}\begin{pmatrix} -f_1 & 1 \\ -f_1^2 & f_1 \end{pmatrix}, \qquad \underset{-r_1}{\text{\upshape Res}} \, \mathcal{M}(k) = \frac{1}{\alpha}\begin{pmatrix} f_1 & -|f_1|^2 \\ f_1/\bar{f}_1 & -f_1 \end{pmatrix},
\end{equation}
where 
\begin{equation}\label{alphadef}  
  \alpha = \frac{d^+}{d\zeta}\biggl|_{\zeta = r_1} e^{2U(i\zeta)}
\end{equation}
and $d^+/d\zeta$ denotes the right-sided derivative.
\end{itemize}
\end{proposition}
\proofbegin
We deduce from (\ref{eqrelation}) and (\ref{fconstrelation}) that the function $\mathcal{M}$ defined in (\ref{mathcalMdef}) satisfies the jump condition (\ref{auxiliaryRH}). The asymptotic behavior (\ref{MtominusI}) follows from the properties of $F$ and $G$. 

By evaluating the first symmetry in (\ref{phisymmetriesk}) at the branch point $i\bar{z}$ and taking the limit as $z$ approaches the regular axis, we find (cf. equations (2.63)-(2.64) in \cite{MAKNP})
\begin{subequations}\label{FGaxisvalues}
\begin{align}
& F(\zeta) = \frac{1}{\text{Re}\, f(i\zeta)}, \qquad G(\zeta) = \frac{i\text{Im}\, f(i\zeta)}{\text{Re}\, f(i\zeta)}, \qquad \zeta > r_1,
	\\
& F(\zeta) = \frac{|f(i\zeta)|^2}{\text{Re}\, f(i\zeta)}, \qquad G(\zeta) = \frac{-i\text{Im}\, f(i\zeta)}{\text{Re}\, f(i\zeta)}, \qquad \zeta <  -r_1.
\end{align}
\end{subequations}
The poles of $F$ and $G$ arise since $\text{Re}\, f(i\zeta) = 0$ at $\zeta = \pm r_1$. Equations (\ref{FGaxisvalues}) together with the equatorial symmetry of $f$ yield
$$\underset{r_1}{\text{\upshape Res}} \, F(k) = \frac{1}{\alpha}, \qquad
\underset{r_1}{\text{\upshape Res}} \, G(k) = \frac{f_1}{\alpha}, \qquad \underset{-r_1}{\text{\upshape Res}}\, F(k) = -\frac{|f_1|^2}{\alpha}, \qquad
\underset{-r_1}{\text{\upshape Res}} \, G(k) = -\frac{f_1}{\alpha}$$
where $\alpha$ is given by (\ref{alphadef}). The residue conditions (\ref{Mresidues}) follow immediately from these relations.

In the last step of the proof, we show that $\mathcal{M}_{21}(k)$ does {\it not} have poles at the possible zeros of $F$, despite the form of (\ref{mathcalMdef}). We first extend the definition (\ref{mathcalSdef}) of $\mathcal{S}(k)$ to all $k \in \hat{\C}$ by
\begin{equation}\label{Sextension}
  \mathcal{S}(k) = T_1^{-1}B^{-1}\Lambda_\Omega(+i0, k^+)^{-1}\sigma_1\sigma_3\Lambda_\Omega(-i0, k^+)\bar{B}T_2\sigma_1.
\end{equation}
This definition is consistent with (\ref{mathcalSdef}). Indeed, since $f$ is equatorially symmetric, 
\begin{equation}\label{Lambdasymm}
  \overline{\Lambda_\Omega(\bar{z}, k^+)} = \Lambda_\Omega(z, -\bar{k}^+),
\end{equation}
so that $\bar{\Lambda} = \overline{\Lambda_\Omega(+i 0, k^+)} = \Lambda_\Omega(-i0, k^+)$ for $k \in \Gamma$.
We claim that the matrices $\mathcal{S}$ and $\mathcal{M}$ satisfy
\begin{equation}\label{traceSM}
  \text{\upshape tr}(\mathcal{S}\mathcal{M}) = 0, \qquad k \in \hat{\C}.
\end{equation}
Indeed, the same type of argument used to prove proposition \ref{fconstprop} shows that the function
$$\mathcal{R} = \Phi_\Omega^{-1}(\rho + i0, k) \sigma_1 \sigma_3 \Phi_\Omega(\rho - i0, k)$$
is independent of $\rho$.
Evaluation at $\rho = 0$ using the axis values yields
$$\mathcal{R}(k^+) = A^{-1}T_1^{-1}B^{-1}\Lambda_\Omega^{-1} (+i0, k^+)\sigma_1\sigma_3\Lambda_\Omega(- i0, k^+) \bar{B}T_2\sigma_1A\sigma_1.$$
Evaluation at $\rho = \rho_0$ yields
$$\text{Tr}\,\mathcal{R} = 0.$$
The preceding two equations imply (\ref{traceSM}). 
It follows from (\ref{traceSM}) that $G^2 - 1$ must vanish whenever $F$ has a zero.
\proofend

The auxiliary RH problem presented in proposition \ref{auxRHprop} can be used to determine the spectral functions $F$ and $G$. These spectral functions can then be used to compute the jump matrix and to set up the main RH problem. However, in analogy with linearizable BVPs for other integrable PDEs, we expect that the jump condition of the auxiliary RH problem can also be substituted directly into the main RH problem with the result that the unknown quantities in the main RH problem disappear. In fact, an example of this mechanism was observed by Neugebauer and Meinel in the case of a rigidly rotating disk. They discovered that the analogs of the main and auxiliary RH problems can be combined into a single scalar RH problem from which the Ernst potential $f$ can be directly recovered, see \cite{MAKNP}. It turns out that a similar approach can be adopted in the present case---in the next section, we will combine the main and auxiliary RH problems with respective jump conditions (\ref{Phijumps}) and (\ref{auxiliaryRH}) into a scalar RH problem on the Riemann surface $\Sigma_z$ introduced in section \ref{diskblackholesec}.

\section{A scalar Riemann-Hilbert problem}\nequation\label{scalarRHsec}
We let the scalar-valued function $w(k)$ be defined by 
\begin{equation}\label{wdef2}  
  w(k) = -\frac{1}{2}\text{tr}(\mathcal{S}(k)), \qquad k \in \hat{\C},
\end{equation}
and define two $2\times 2$-matrix valued functions $\mathcal{L}$ and $\mathcal{Q}$ by
\begin{align}\label{Ldef}
\mathcal{L}(z, k) = \Phi(z, k)\sigma_1 \Phi^{-1}(z, k),  \qquad k \in \mathcal{S}_z,
\end{align}
and
\begin{align}\label{Qdef}
 \mathcal{Q}(z,k) = -\Phi(z,k) A(k)^{-1}\mathcal{S}(k)A(k) \Phi(z,k)^{-1} - w(k)\mathbb{I}, \qquad k \in \mathcal{S}_z.
\end{align}

\begin{lemma} The functions $\mathcal{L}$, $\mathcal{Q}$, and $w$ have the following properties:
\begin{itemize}
\item The traces and determinants of $\mathcal{Q}$ and $\mathcal{L}$ satisfy
\begin{equation}\label{QLtrdet}
\text{\upshape tr}\,\mathcal{Q} = 0, \qquad \text{\upshape tr}\,\mathcal{L} = 0, \qquad \text{\upshape det}\,\mathcal{L} = -1, \qquad \text{\upshape det}\,\mathcal{Q} = -1 - w^2.
\end{equation}

\item $\mathcal{Q}$ can be alternatively written as
\begin{equation}\label{Qalternative}  
  \mathcal{Q}(z,k) = \Phi(z,k) \sigma_1 A(k)^{-1} \mathcal{S}(k) A(k)\sigma_1 \Phi(z,k)^{-1} + w(k)\mathbb{I}.
\end{equation}

\item $\mathcal{Q}$ and $\mathcal{L}$ admit the symmetries
\begin{equation}\label{QLsymm}  
  \mathcal{Q}(z, k^-) = -\sigma_3\mathcal{Q}(z, k^+)\sigma_3, \qquad \mathcal{L}(z, k^-) = \sigma_3\mathcal{L}(z, k^+)\sigma_3.
\end{equation}

\item $\mathcal{Q}$ has no jump across $\Gamma^\pm$, whereas $\mathcal{L}$ satisfies the following jump conditions across $\Gamma^\pm$:
\begin{align} \label{QLjumps}
(\mathcal{Q} + w\mathbb{I})\mathcal{L}_- = -\mathcal{L}_+ (\mathcal{Q} + w\mathbb{I}), \qquad k \in \Gamma^+,
	\\ \nonumber
(\mathcal{Q} - w\mathbb{I})\mathcal{L}_- = -\mathcal{L}_+ (\mathcal{Q} - w\mathbb{I}), \qquad k \in \Gamma^-.
\end{align}

\item $\mathcal{Q}\mathcal{L} = -\mathcal{L}\mathcal{Q}$; in particular, $\text{\upshape tr}(\mathcal{Q}\mathcal{L}) = 0.$

\item Let $\hat{\mathcal{L}}_{22} = \mathcal{L}_{21}\mathcal{Q}_{11} + \mathcal{L}_{22}\mathcal{Q}_{21}$. Then
\begin{equation}\label{2.92analog}
  \hat{\mathcal{L}}_{22}^2 - \mathcal{L}_{21}^2(1 + w^2) = \mathcal{Q}_{21}^2.
\end{equation}

\item $w$ has the form 
$$w(k) = \frac{w_4 k^4 + w_2 k^2 + w_0}{k^2 - r_1^2}$$
where $w_4, w_2, w_0$ are real coefficients explicitly given by
\begin{subequations}\label{w0w2w4expressions}
\begin{align} \label{w4expression}
 w_4 = &\, -\frac{2 \Omega^2 \Omega_h^2}{e^{2U_0} (\Omega - \Omega_h)^2},
 	 \\ \label{w2expression}
 w_2 = & \,\frac{|f_0|^2 (\Omega -\Omega_h)^2+f_1^2\Omega  (\Omega -2 \Omega_h)+4 i f_1 r_1 \Omega^2 \Omega_h+\Omega_h^2 \left(4 r_1^2 \Omega^2-1\right)}{2 e^{2U_0} (\Omega -\Omega_h)^2},
	\\ \label{w0expression}
w_0 = & -\frac{1}{8 e^{2U_0}\Omega_h^2}\biggl(-2 i b_0 \left(f_1^3-2 i f_1^2 r_1 \Omega_h+f_1+2 i r_1 \Omega_h \right)
	\\ \nonumber
& - |f_0|^2 \left(f_1^2-4 i f_1 r_1\Omega_h-4 r_1^2 \Omega_h^2+1\right)+f_1^4+f_1^2+4 i f_1 r_1\Omega_h-4 r_1^2 \Omega_h^2\biggl),
\end{align}
\end{subequations}
where $f_0$ and $f_1$ denote the values of $f$ at $z = +i0$ and $z = ir_1$, respectively.

\end{itemize}
\end{lemma} 
\proofbegin
The first three properties in (\ref{QLtrdet}) are immediate from equations (\ref{wdef2})-(\ref{Qdef}). 
The fourth property follows since
\begin{equation}\nonumber
  \text{det}(\mathcal{S}) = -1
\end{equation}
and
$$\text{det}\,{\mathcal{Q}} =  \text{det}(-\Phi A^{-1}(\mathcal{S} + w\mathbb{I})A \Phi^{-1})
= \text{det}(\mathcal{S} + w\mathbb{I}) = \det{\mathcal{S}} - w^2.$$

Using (\ref{traceSM}) and the definitions of $A$ and $\mathcal{M}$, a computation shows that
$$A^{-1}(\mathcal{S} + w\mathbb{I})A = -\sigma_1A^{-1}(\mathcal{S} + w\mathbb{I})A\sigma_1.$$
Using this identity in the definition (\ref{Qdef}) of $\mathcal{Q}$, we find (\ref{Qalternative}).

The symmetries (\ref{QLsymm}) follow from (\ref{Ldef}), (\ref{Qdef}), and (\ref{Qalternative}) together with the first symmetry in (\ref{phisymmetriesk}).

By (\ref{Phijumps}) and (\ref{Dequation1}), $\Phi A^{-1}$ and $\Phi \sigma_1 A^{-1}$ do not jump across $\Gamma^+$ and $\Gamma^-$, respectively. It follows from the expressions (\ref{Qdef}) and (\ref{Qalternative}) that $\mathcal{Q}$ does not jump across $\Gamma^\pm$. 
For $k \in \Gamma^+$, the definitions (\ref{Ldef}) and (\ref{Qdef}) of $\mathcal{L}$ and $\mathcal{Q}$ show that
\begin{equation}\label{QwLminus}
(\mathcal{Q}_- + w\mathbb{I})\mathcal{L}_- = -\Phi_- A_-^{-1}\mathcal{S} A_-\sigma_1 \Phi_-^{-1}
= -\Phi_- A_-^{-1}\mathcal{S}\mathcal{M}_-A_- \Phi_-^{-1}
\end{equation}
and
\begin{equation}\label{LplusQw}
-\mathcal{L}_+(\mathcal{Q}_+ + w\mathbb{I}) = \Phi_+\sigma_1 A_+^{-1}\mathcal{S}A_+\Phi_+^{-1}
= \Phi_+A_+^{-1} \mathcal{M}_+ \mathcal{S}A_+\Phi_+^{-1}.
\end{equation}
Using that $\Phi A^{-1}$ does not jump across $\Gamma^+$ together with the jump condition (\ref{auxiliaryRH}), we see that the right-hand sides of (\ref{QwLminus}) and (\ref{LplusQw}) are equal.
Similarly, using (\ref{Qalternative}), the fact that  $\Phi \sigma_1 A^{-1}$ has no jump across $\Gamma^-$, and (\ref{auxiliaryRH}), we find the jump across $\Gamma^-$. This proves (\ref{QLjumps}). 

Since $\mathcal{M}$ is tracefree and $\text{tr}(\mathcal{S}\mathcal{M}) = 0$, we deduce that $MS + SM + 2w M = 0$. In view of the definitions of $\mathcal{Q}$ and $\mathcal{L}$, this implies $\mathcal{Q}\mathcal{L} = -\mathcal{L}\mathcal{Q}$.

Equation (\ref{2.92analog}) follows by direct computation using the identity $\text{tr}(\mathcal{Q}\mathcal{L}) = 0$ and the four properties in (\ref{QLtrdet}). 

The last statement concerning the form of $w$ follows from (\ref{Sextension}) and (\ref{wdef2}) by direct computation. 
\proofend

The condition that $\mathcal{M}$ does not jump at the endpoints of $\Gamma$ implies that $\text{tr}\, \mathcal{S}(\pm i\rho_0) = 0$, i.e.
\begin{equation}\label{rimcondition}
  w_0 = \rho_0^2(w_2 - w_4 \rho_0^2).
\end{equation}
In particular, the function $h(k)$ defined in (\ref{hdef}) vanishes at the endpoints of $\Gamma$.

\begin{lemma} There exist points $\{m_j\}_1^4 \subset \C$ such that
\begin{equation}\label{Q12asymptotics}
  \mathcal{Q}_{21}(k) = \frac{8 f \Omega^2 \Omega_h^2}{(f+\bar{f})
   (f_0 + \bar{f}_0) (\Omega -\Omega_h)^2} \frac{\prod_{j = 1}^4 ( k - m_j)}{(k^2 - r_1^2)}.
\end{equation}
\end{lemma}
\proofbegin
By (\ref{QLsymm}), $\mathcal{Q}_{21}$ is a unique function of $k$, i.e. $\mathcal{Q}_{21}(k^+) = \mathcal{Q}_{21}(k^-)$. Thus $(k^2 - r_1^2) \mathcal{Q}_{21}$ is an entire function of $k \in \C$. The existence of $\{m_j\}_1^4$ satisfying (\ref{Q12asymptotics}) therefore follows if we can show that
\begin{equation}\label{Q21atinfty}
  (k^2 - r_1^2) \mathcal{Q}_{21}(k) = \frac{8 f  \Omega^2 \Omega_h^2}{(f+\bar{f})
   (f_0 + \bar{f}_0) (\Omega -\Omega_h)^2}k^4 + O(k^3), \qquad k \to \infty.
\end{equation}
As $k \to \infty$, we have
\begin{equation}\label{Phiasymptotics}
\Phi(z, k^+) = \begin{pmatrix} \overline{f(z)} & 1 \\ f(z) & -1 \end{pmatrix} + O(1/k),\qquad 
A(k) = \mathbb{I} + O(1/k), \qquad k \to \infty.
\end{equation}
Thus, by (\ref{Qdef}) and (\ref{Sextension}),
\begin{align}\label{Qasymptotics}
(k^2 - r_1^2)&\mathcal{Q}(z,k^+) =  - \begin{pmatrix} \bar{f} & 1 \\ f & -1 \end{pmatrix}\begin{pmatrix} 0 &  0 \\ 0 & \frac{8 k^4 \Omega ^2 \Omega_h^2}{(f_0 + \bar{f}_0)(\Omega - \Omega_h)^2} \end{pmatrix}
\begin{pmatrix} \bar{f} & 1 \\ f & -1 \end{pmatrix}^{-1} 
- w_4 k^4 \mathbb{I} + O(k^3)
	\\ \nonumber
&=
\frac{4 \Omega ^2 \Omega_h^2}{(f+\bar{f})
   (f_0+\bar{f}_0) (\Omega -\Omega_h)^2}
\left(
\begin{array}{cc}
 \bar{f}-f & 2 \bar{f} \\
 2 f & f-\bar{f}
\end{array}
\right)
k^4 + O(k^3), \qquad k \to \infty.
\end{align}
The $(21)$-entry of this equation yields (\ref{Q21atinfty}).
\proofend

Define four points $\{k_j\}_{j = 1}^4 \subset \hat{\C}$ by
$$w^2 + 1 = \frac{w_4^2 \prod_{j = 1}^4 (k - k_j)(k - \bar{k}_j)}{(k^2 - r_1^2)^2}.$$
We assume that the $k_j$'s are ordered as in subsection \ref{solutionsubsec}.
Let $\hat{S}_z$ denote the double cover of the Riemann surface $\mathcal{S}_z$ defined by adding cuts $[k_j, \bar{k}_j]$, $j = 1,\dots,4$, both on the upper and lower sheets of $\mathcal{S}_z$. Thus a point $(k, \pm\lambda, \pm\mu)$ of $\hat{S}_z$ is specified by giving a point $k \in \hat{\C}$ together with a choice of sign of $\lambda$ and of
$$\mu = \sqrt{\prod_{j = 1}^4 (k - k_j)(k - \bar{k}_j)}.$$
We specify the sheets so that $\lambda \to 1$ ($\lambda \to -1$) as $k \to \infty$ on sheets 1 and 2 (sheets 3 and 4), and $\mu \sim k^4$ ($\mu \sim -k^4$) as $k \to \infty$ on sheets 1 and 3 (sheets 2 and 4). 
As $k$Ê crosses the cut $[-iz, i\bar{z}]$, $\lambda$ changes sign whereas the sign of $\mu$ remains unchanged.
As $k$Ê crosses any of the other cuts, $\mu$ changes sign whereas the sign of $\lambda$ remains unchanged.

Consider the function $H$ defined by
\begin{equation}\label{Hdef}  
  H(z,k) = \frac{\hat{\mathcal{L}}_{22} - \mathcal{L}_{21}\sqrt{w^2 + 1}} {\hat{\mathcal{L}}_{22} + \mathcal{L}_{21}\sqrt{w^2 + 1}}, \qquad k \in \hat{S}_z,
\end{equation}
where $\hat{\mathcal{L}}_{22} = \mathcal{L}_{21}\mathcal{Q}_{11} + \mathcal{L}_{22}\mathcal{Q}_{21}$. We fix the sign of the root $\sqrt{w^2 + 1}$ in (\ref{Hdef}) by requiring that $\sqrt{w^2 + 1} = -w_4 k^2 + O(k)$ as $k \to \infty^+$. Since $w_4 > 0$, this implies that $\sqrt{w^2 + 1} \geq 0$ for $k \in \Gamma^+$. 
The eigenfunction $\Phi$ of the Lax pair (\ref{ernstlax}) satisfies $\det \Phi = -2 e^{2U} F(k)$ (see equation (2.65) in \cite{MAKNP}), so that the entries of $\mathcal{L}$ may have poles at the points which project to the zeros of $F(k)$ in the Riemann $k$-sphere. However, $H$ has no singularities at these points. Therefore, in view of (\ref{2.92analog}), the possible zeros and poles of $H$ belong to the set in $\hat{S}_z$ which projects to $\{\pm r_1\} \cup \{m_j\}_1^4$, and if $H$ has a double pole at $(m_j, \lambda, \mu)$, then it has a double zero at $(m_j, \lambda, -\mu)$, $j = 1, \dots, 4$.

By the symmetries (\ref{QLsymm}), we have
$$\hat{\mathcal{L}}_{22}(k, \lambda, \mu) = \hat{\mathcal{L}}_{22}(k, -\lambda, \mu), \qquad
\mathcal{L}_{21}(k, \lambda, \mu) = -\mathcal{L}_{21}(k, -\lambda, \mu).$$
Therefore
$$H(k, \lambda, \mu) = \frac{1}{H(k, -\lambda, \mu)}.$$
Similarly, we have $H(k, \lambda, \mu) = 1/H(k, \lambda, -\mu).$
Consequently, $H \to 1/H $ whenever $k$ crosses one of the cuts of the two-sheeted Riemann surface $\Sigma_z$ defined by (\ref{Sigmazdef}).
We can therefore view $H$ as a single-valued function on $\Sigma_z$ with the values on the upper sheet given by the values of $H$ on sheet 1 of $\hat{S}_z$, and the values on the lower sheet given by the inverses of these values.

\subsection{Formulation of the scalar RH problem}\label{scalarrhsubsec}
We want to formulate a scalar RH problem in terms of the complex-valued function $\psi(z, k)$ defined by 
\begin{equation}\label{psidef}  
  \psi(z, k) = \frac{\log H(z,k)}{y}, \qquad k \in \Sigma_z.
\end{equation}
However, since $\log H$ is a multi-valued function on $\Sigma_z$, this definition of $\psi$ needs to be supplemented by a choice of branches for the logarithm. We will fix a single-valued representative of $\psi$ on $\Sigma_z$ by introducing cuts which connect the zeros and poles of $H$. Across these cuts $\psi$ will jump by multiples of $2\pi i/y$. 
The problem is that even though (\ref{2.92analog}) implies that all zeros and poles of $H$ lie in the cover of the set $\{\pm r_1\} \cup \{m_j\}_1^4$, the exact distribution of these zeros and poles is not known. It is therefore not clear at this stage how to make a consistent choice of branches.

We address this problem by considering the limit in which the solution $f$ approaches the Kerr solution. For a solution near the Kerr solution, we can utilize the Kerr expressions for $F$ and $G$ to compute $H$ explicitly to first order. This will give us the correct choice of branches in the Kerr limit and by continuity this choice extends also to more general solutions. 

The Ernst potential for the Kerr black hole rotating with angular velocity $\Omega_h$ and with a horizon stretching fromÊ$-i r_1$ to $ir_1$ is given by
$$f^{kerr} = \frac{R_+ e^{-i\delta} + R_- e^{i\delta} - 2r_1}{R_+ e^{-i\delta} + R_- e^{i\delta} + 2r_1},$$
where $R_\pm$ are defined by
$$R_\pm = \sqrt{(\pm r_1 - \zeta)^2 + \rho^2},$$
and the parameter $\delta \in (-\pi/2, 0)$ is related to $\Omega_h$ by
$$\Omega_h = \frac{if_1^{kerr}(1 + (f_1^{kerr})^2)}{2r_1(1 - (f_1^{kerr})^2)}, \qquad f_1^{kerr} = i\tan(\delta/2).$$
The value at the origin is given by
$$f_0^{kerr} = \frac{\cos(\delta) - 1}{\cos(\delta) + 1}.$$
We consider adding a slowly rotating disk to the Kerr solution. Using the Kerr values for $f_0$ and $f_1$, we compute $w_4,w_2,w_0$ according to (\ref{w0w2w4expressions}) with $\Omega \ll 1$. The branch points $\{k_j\}_1^4$ are found by solving the equation $w^2 + 1 = 0$. As $\Omega \to 0$, $k_1$ and $k_4$ tend to infinity, whereas $k_2$ and $k_3$ approach finite values. The spectral functions $F^{kerr}(k)$ and $G^{kerr}(k)$ are given explicitly by (cf. section 2.4 in \cite{MAKNP}; note that there is a misprint in equation (2.349) of \cite{MAKNP})
\begin{align}\nonumber
& F^{kerr}(k) = \frac{2\Omega_h^2(k^2 - r_1^2) + 2i\Omega_h f_1^{kerr} k - (f_1^{kerr})^2}{2\Omega_h^2(k^2 - r_1^2)},
	\\ \nonumber
& G^{kerr}(k) = \frac{(2i\Omega_h r_1 - f_1^{kerr})(f_1^{kerr})^2}{2\Omega_h^2(k^2 - r_1^2)}.
\end{align}
For definiteness, we consider the example of $r_1 = 1/2$ and $\delta = -1/2$. 
Assuming that $z = \rho + i \zeta$ with $\rho \ll 1$ and $\zeta \gg 1$, we compute $\mathcal{Q}$ and $\mathcal{L}$ to first order by substituting the axis values (\ref{phionaxisa}) for $\Phi$ together with the values of the Kerr solution into the right-hand sides of (\ref{Ldef}) and (\ref{Qdef}). 
We find that $H$ has double poles and double zeros at the points in the sets
$$\{m_1^-, m_2^-, -r_1^-, m_3^+, m_4^+, r_1^+\} \quad \hbox{and} \quad \{m_1^+, m_2^+, -r_1^+, m_3^-, m_4^-, r_1^-\},$$ 
respectively. As $\Omega \to 0$, $m_1 \to k_1$ and $m_4 \to k_4$, whereas $m_2$ and $m_3$ converge to values close (but not equal) to $k_2$ and $k_3$, respectively. Let $\{a_j, b_j\}_1^4$ be the particular cycles on $\Sigma_z$ specified in remark \ref{mainremark}. As $k$ traverses each of these cycles, the argument of $H$ changes by the amount $\Delta\arg H$ according to TableÊ \ref{argHtable}.
\begin{table}[htdp]
\begin{center}
\begin{tabular}{cc}
 cycle & $\Delta\arg H$  \\
\hline
  $a_1$ & $-4\pi$ \\
  $a_2$ & $-4\pi$  \\
  $a_3$ & $4\pi$ \\
  $a_4$ & $4\pi$ 
  \end{tabular}
 \begin{tabular}{cc}
 \qquad\qquad\qquad\qquad
  \end{tabular}
\begin{tabular}{cc}
 cycle & $\Delta\arg H$  \\
\hline
  $b_1$ & $-8\pi$ \\
  $b_2$ & $-8\pi$ \\
  $b_3$ & $-12\pi$ \\
  $b_4$ & $-4\pi$ 
  \end{tabular}
\end{center}
\caption{The change $\Delta\arg H$ in $\arg H$ as $k$ traverses each of the cycles in the cut basis $\{a_j, b_j\}_{j=1}^4$.}
\label{argHtable}
\end{table}%

A choice of branches for $\log H$ consistent with the above properties is obtained by introducing cuts on $\Sigma_z$ according to Figure \ref{logHcuts.pdf}. The introduced cuts run from the poles of $H$ to the zeros of $H$. Letting $(\log H)^+$ and $(\log H)^-$ denote the values of $\log H$ for $z$ just to the right and to the left of a cut, respectively, we have $(\log H)^+ + 4 \pi i = (\log H)^-$. An overall choice of branch is made by requiring that $\log H \to 2\log f$ as $k \to \infty^+$, where an appropriate branch is chosen for $\log f$, see equation (\ref{psiasymptotics}) below.

For this choice of branches, we have $\log H(k^+) = -\log H(k^-)$, $k \in \hat{\C}$. Consequently, $\psi(z, k)$ is a unique function of $k$, i.e.
$$\psi(z, k^+) = \psi(z, k^-), \qquad k \in \hat{\C}.$$
We therefore view $\psi(z, \cdot)$ as a function $\hat{\C} \to \hat{\C}$.

\begin{proposition}\label{scalarRHprop}
Let $\psi$ be defined by (\ref{psidef}) with the choice of branches for $\log H$ specified above. Then the function $\psi(z, \cdot):\hat{\C} \to \hat{\C}$ has the following properties:
\begin{itemize}
\item $\psi(z, k)$ is analytic for $k \in \hat{\C} \setminus \left[\Gamma \cup \left(\cup_{j=1}^4 [m_j, k_j]\right) \cup \{\pm r_1\}\right]$.

\item Across $\Gamma$, $\psi(z, k)$ satisfies the jump condition
\begin{equation}\label{Psihatjump}
\psi^-(z, k) = \psi^+(z,k) + \frac{2}{y(z, k^+)} \ln \left(\frac{\sqrt{1 + w^2} - w}{\sqrt{1 + w^2} + w}\right), \qquad k \in \Gamma.
\end{equation}

\item Across the directed intervals $[k_j, m_j]$, $j = 1,2$; $[m_j, k_j]$, $j = 3,4$; $[r_1,k_3]$,  $[\bar{k}_3, k_2]$, and $[\bar{k}_2, -r_1]$, $\psi(z, k)$ satisfies the jump condition
\begin{equation} \nonumber
\psi^-(z, k) = \psi^+(z,k) + \frac{4 \pi i}{y(z, k^+)},
\end{equation}
where $\psi^+$ and $\psi^-$ denote the values of $\psi$ to the right and left of the cut, respectively.

\begin{figure}
\begin{center}
    \includegraphics[width=.7\textwidth]{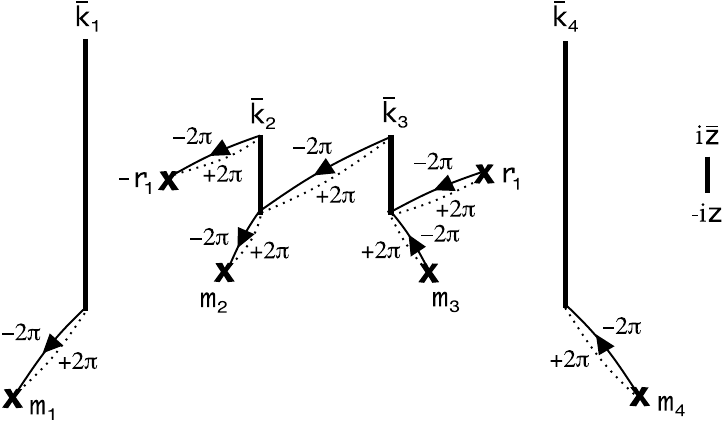} \quad
     \begin{figuretext}\label{logHcuts.pdf}
       The additional cuts introduced on $\Sigma_z$ in order to make $\log H$ a single-valued function.
     \end{figuretext}
 \end{center}
\end{figure}   

\item As $k \to m_j$, $j = 1,2$, $\psi(z, k)$ satisfies 
\begin{equation}\nonumber
  \psi(z, k) = \frac{2}{y(z, m_j^+)}\log(k - m_j), \qquad k \to m_j, \qquad j = 1, 2.
\end{equation}

\item As $k \to m_j$, $j = 3,4$, $\psi(z, k)$ satisfies 
\begin{equation} \nonumber
  \psi(z, k) = \frac{-2}{y(z, m_j^+)}\log(k - m_j), \qquad k \to m_j, \qquad j = 3, 4.
\end{equation}

\item As $k \to -r_1$, $\psi(z, k)$ satisfies 
\begin{equation} \nonumber
  \psi(z, k) = \frac{2}{y(z, -r_1^+)}\log(k + r_1), \qquad k \to -r_1.
\end{equation}

\item As $k \to r_1$, $\psi(z, k)$ satisfies 
\begin{equation} \nonumber
  \psi(z, k) = \frac{-2}{y(z, r_1^+)}\log(k - r_1), \qquad k \to r_1.
\end{equation}

\item As $k \to k_j$, $j = 1, \dots, 4$, $\psi(z, k)$ satisfies 
\begin{equation} \nonumber
\psi(z, k) = \frac{2\pi i}{y}, \qquad k \to k_j, \qquad j = 1, \dots, 4,
\end{equation}
where $y = y(z, k^+)$ for $k$ just to the left of the cut $[k_j, m_j]$ for $j = 1,2$, and just to the left of the cut $[m_j, k_j]$ for $j = 3,4$ and is analytically continued around the endpoint $k_j$ so that $y = y(z, k^-)$ to the right of the cut. 

\item As $k \to \infty$,
\begin{equation}\label{psiasymptotics}  
  \psi \to \frac{2\log(f)}{k^5} + O(1/k^6), \qquad k \to \infty.
\end{equation}

\end{itemize}
\end{proposition}
\proofbegin
We first show that $\psi$ satisfies the jump condition (\ref{Psihatjump}). Algebraic manipulation of (\ref{QLjumps}) using the identity $\text{\upshape tr}(\mathcal{Q}\mathcal{L}) = 0$ and the properties in (\ref{QLtrdet}) shows that the functions $\mathcal{L}_{21}$ and $\hat{\mathcal{L}}_{22} = \mathcal{L}_{21}\mathcal{Q}_{11} + \mathcal{L}_{22}\mathcal{Q}_{21}$ satisfy
\begin{align}
\begin{cases}
\mathcal{L}_{21+} = -2w\hat{\mathcal{L}}_{22-} + (1 + 2w^2)\mathcal{L}_{21-},
	\\
\hat{\mathcal{L}}_{22+} = (1 + 2w^2)\hat{\mathcal{L}}_{22-} - 2w(1 + w^2)\mathcal{L}_{21-},
\end{cases} \qquad k \in \Gamma^+,
	\\
\begin{cases}
\mathcal{L}_{21+} = 2w\hat{\mathcal{L}}_{22-} + (1 + 2w^2)\mathcal{L}_{21-},
	\\
\hat{\mathcal{L}}_{22+} = (1 + 2w^2)\hat{\mathcal{L}}_{22-} + 2w(1 + w^2)\mathcal{L}_{21-},
\end{cases} \qquad k \in \Gamma^-.
\end{align}
Thus
$$H_-(k) = \left(\frac{\sqrt{1 + w^2} - w}{\sqrt{1 + w^2} + w}\right)^{\pm2} H_+(k), \qquad k \in \Gamma^{\pm}.$$
Equation (\ref{Psihatjump}) follows from here since $1 + w^2 \geq 0$ and $\sqrt{1 + w^2} \pm w > 0$ for $k \in \Gamma$.

The behavior as $k \to \pm r_1$ and $k \to m_j$ follows since $H$ has double zeros at $-r_1$, $m_j^+$, $j = 1,2$, and double poles at $r_1$, $m_j^+$, $j = 3,4$.

In order to find the behavior of $\psi$ as $k \to \infty$, we note that, by (\ref{Phiasymptotics}),
$$\mathcal{L}(z, k^+) = \begin{pmatrix} \bar{f} & 1 \\ f & -1 \end{pmatrix} \sigma_1 \begin{pmatrix} \bar{f} & 1 \\ f & -1 \end{pmatrix}^{-1} + O(1/k), \qquad k \to \infty.$$
Thus, by (\ref{Qasymptotics}),
$$\hat{\mathcal{L}}_{22}(z,k^+) = -\frac{4 \left(1 + f^2\right) \Omega^2 \Omega_h^2}{(f+\bar{f}) (f_0+\bar{f}_0)
    (\Omega -\Omega_h)^2} k^2
   + O(k), \qquad k \to \infty.$$
SinceÊ $\sqrt{w^2 + 1} = -w_4 k^2 + O(k)$ as $k \to \infty^+$, we find
$$H(z, k^+) = f(z)^{2} + O(1/k), \qquad k \to \infty.$$
Therefore
$$\frac{\log H}{y} \to \frac{2\log f}{k^5} + O(1/k^6), \qquad k \to \infty.$$
\proofend


\subsection{Solution of the scalar RH problem}
The solution of the scalar RH problem presented in proposition \ref{scalarRHprop} is
\begin{align}\label{psisolution}
\psi(z,k) = &\; \frac{2}{\pi i} \int_{\Gamma} \frac{dk'}{y(z, k'^+) (k' - k)} \ln\bigl(\sqrt{1 + w(k')^2} - w(k')\bigr)
	\\ \nonumber
& + 2\sum_{j = 1}^2 \int_{[k_j, m_j]} \frac{dk'}{y(z,k'^+) (k' - k)}
- 2\sum_{j = 3}^4 \int_{[k_j, m_j]} \frac{dk'}{y(z,k'^+) (k' - k)}
	\\ \nonumber
&+ 2 \left(\int_{[r_1, k_3]}+\int_{[\bar{k}_3, k_2]}+\int_{[\bar{k}_2, -r_1]}\right) \frac{dk'}{y(z,k'^+) (k' - k)},
\qquad k \in \Sigma_z,
\end{align}
where we used that 
$$\ln\bigl(\sqrt{1 + w^2} + w\bigr) = -\ln\bigl(\sqrt{1 + w^2} - w\bigr), \qquad k \in \Gamma.$$
By deforming contours, we can replace the last three integrals on the right-hand side of (\ref{psisolution}) with an integral along $\gamma$, where $\gamma$ is the contour on $\Sigma_z$ defined in section \ref{diskblackholesec}. We define two divisors $\mathfrak{K}$ and $\mathfrak{M}$ on $\Sigma_z$. $\mathfrak{K}$ is defined by
\begin{equation}\label{divisorDdef}
  \mathfrak{K} = \sum_{j = 1}^4 k_j, 
\end{equation}  
whereas $\mathfrak{M}$ is defined as the sum of the points in $\Sigma_z$ which lie above the set $\{m_j\}_1^4$ and which are double poles of $H$, i.e.
\begin{equation}\label{divisorEdef}
  \mathfrak{M} = m_1^- + m_2^- + m_3^+ + m_3^+.
\end{equation}
We can then write (\ref{psisolution}) as
\begin{align}\nonumber
\psi(z,k) = & \, \frac{2}{\pi i} \int_{\Gamma^+} \frac{dk'}{y(z, k') (k' - k)} \ln\left(\sqrt{1 + w(k')^2} - w(k')\right)
	\\ \nonumber
&- 2\int_{\mathfrak{K}}^{'\mathfrak{M}} \frac{dk'}{y(z, k') (k' - k)} + 2 \int_\gamma \frac{dk'}{y(z, k') (k' - k)},
\end{align}
where the integrals are contour integrals on $\Sigma_z$ and the prime on the integral from $\mathfrak{K}$ to $\mathfrak{M}$ indicates that the paths of integration do not necessary lie in the complement of the cut basis $\{a_j, b_j\}$.
In view of (\ref{psiasymptotics}), this leads to
\begin{subequations}\label{logfboth}
\begin{align}\label{logf}
& \text{log}\, f = \int_{\mathfrak{K}}^{'\mathfrak{M}} \frac{k^4 dk}{y} - \int_{\Gamma^+} h(k) \frac{k^4 dk}{y} - \int_\gamma \frac{k^4 dk}{y},
	\\\label{intDEconditions}
 & \int_\mathfrak{K}^{'\mathfrak{M}} \frac{k^{n-1} dk}{y} 
 = \int_{\Gamma^+} h(k) \frac{k^{n-1} dk}{y} + \int_{\gamma} \frac{k^{n-1} dk}{y} , \qquad n = 1, \dots, 4,
\end{align}
\end{subequations}
where $h(k)$ is defined by (\ref{hdef}).

\begin{remark}\upshape
Although the equations (\ref{logfboth}) were derived under the assumption that the solution is a small perturbation of the Kerr solution, they are valid more generally. Indeed, the crucial facts used in the derivation were that $H$ has a double pole at $r_1^+$ and a double zero at $-r_1^+$, and these properties are preserved under a continuous deformation. It is conceivable that the double poles of $H$ that make up $\mathfrak{M}$ will change sheets under such a deformation so that (\ref{divisorEdef}) has to be modified, but the resulting equations (\ref{logfboth}) remain unchanged.
\end{remark}

\section{Theta functions}\nequation\label{thetasec}
In this section we derive explicit expressions for the Ernst potential $f$ and the metric functions $e^{2U}$ and $a$ in terms of theta functions.

\subsection{Explicit expression for the Ernst potential}
We will show that the right-hand side of (\ref{logf}) can be expressed in terms of theta function on $\Sigma_z$. We will first assume that the integration paths from $\mathfrak{K}$ to $\mathfrak{M}$ in (\ref{logfboth}) lie in the fundamental polygon determined by the cut basis; later we will see that the result is the same also when this is not the case.

Let $\{\zeta_j\}_{j=1}^4$ denote the noncanonical basis of holomorphic one-forms on $\Sigma_z$ defined in (\ref{zetadef}) and let $A$ be the matrix defined in (\ref{AZdef}). Then the canonical basis is $\omega = A\zeta$. 
Let $u, I \in \C^4$ be defined by (\ref{uIdef}). Applying $A$ to (\ref{intDEconditions}), we find
\begin{equation}\label{uDEomega}  
  u = \int_\mathfrak{K}^{\mathfrak{M}} \omega.
\end{equation}
Using that 
$$\omega_{\infty^+\infty^-} = -\frac{k^4 dk}{y} + \gamma^T \zeta,$$
for some vector $\gamma \in \C^4$, equation (\ref{logf}) yields
\begin{equation}\label{fDEomegaI}  
  f = e^{-\int_{\mathfrak{K}}^{\mathfrak{M}} \omega_{\infty^+\infty^-} + I}.
\end{equation}
where the terms involving $\gamma$ cancelled because of (\ref{intDEconditions}). Formula (\ref{ernstsolution}) for $f$ will follow if we can prove that 
\begin{equation}\label{expthetaquotient}
e^{-\int_{\mathfrak{K}}^{\mathfrak{M}} \omega_{\infty^+\infty^-}} = \frac{\Theta\left(u - \int_{-iz}^{\infty^-} \omega \right)}{\Theta\left (u + \int_{-iz}^{\infty^-} \omega \right)},
\end{equation}
where $\Theta(v) := \Theta(v|B)$.
Let $e^{(j)}$ and $\pi^{(j)}$ denote the $j$th columns of the $4 \times 4$ identity matrix $I$ and the period matrix $B$, respectively.
Then
\begin{equation}\label{thetashifts}
\Theta(v + e^{(j)} | B) = \Theta(v | B), \qquad \Theta(v + \pi^{(j)} | B) = e^{-2\pi i(v_j + \frac{1}{2} B_{jj})}\Theta(v | B), \qquad v \in \C^4.
\end{equation}
The Jacobian $Jac(\Sigma_z)$ of $\Sigma_z$ is defined as the complex torus $\C^4/\mathbb{L}$, where $\mathbb{L}$ is the discrete lattice generated by the $e^{(j)}$'s and the $\pi^{(j)}$'s. 
We define the map $\varphi:\Sigma_z \to \C^4$ by
$$\varphi(k) = \int_{-iz}^k \omega, \qquad k \in \Sigma_z,$$
with the contour fixed to lie within the fundamental polygon. Then $\varphi$ composed with the projection $\C^4 \to Jac(\Sigma_z)$ is the Abel map with base point $-iz$.
We write $\mathfrak{M} = \sum_{j=1}^4 M_j$, where, for $j = 1,\dots,4$, $M_j = m_j^+$ or $M_j = m_j^-$. 
Let $\mathcal{K} = \varphi(\mathfrak{K}) \in \C^4$. An argument following pp. 322-325 in \cite{FK} shows that $\mathcal{K}$ projects to the vector of Riemann constants in $Jac(\Sigma_z)$.  
Thus the functions
\begin{equation}\label{fun1}
\frac{\Theta(\varphi(P) - \varphi(\mathfrak{K}) + \mathcal{K})}{\Theta(\varphi(P) - \varphi(\mathfrak{M}) + \mathcal{K})}, \qquad P \in \Sigma_z,
\end{equation}
and
\begin{equation}\label{fun2}
e^{-\sum_{j = 1}^4 \int_{\infty^-}^P \omega_{M_jk_j}},\qquad P \in \Sigma_z,
\end{equation}
both have simple poles at the points of $\mathfrak{M}$ and simple zeros at the points of $\mathfrak{K}$. Moreover, the general identity (\cite{FK}, p. 67)
\begin{equation}\label{bperiodthirdkind}
  \int_{b_j} \omega_{RS} = 2\pi i \int_{S}^R \omega_j, \qquad j = 1, \dots, 4, \quad R,S \in \Sigma_z,
\end{equation}
implies that as $a_j$ is traversed the functions (\ref{fun1}) and (\ref{fun2}) both get multiplied by $1$ and as $b_j$ is traversed they both get multiplied by $e^{-2\pi i\int_\mathfrak{K}^{\mathfrak{M}} \omega_j}$.
Hence their quotient is a constant and we deduce that
\begin{equation}\label{emsumomega}
e^{-\sum_{j = 1}^4 \int_{\infty^-}^P \omega_{M_jk_j}}
= \frac{\Theta(\varphi(P) - \varphi(\mathfrak{K}) + \mathcal{K})\Theta(\varphi(\infty^-) - \varphi(\mathfrak{M}) + \mathcal{K})}{\Theta(\varphi(P) - \varphi(\mathfrak{M}) + \mathcal{K})\Theta(\varphi(\infty^-) - \varphi(\mathfrak{K}) + \mathcal{K})}.
\end{equation}
Using the identity
\begin{equation}\label{thirdkindswitch}
  \int_S^R \omega_{\infty^+\infty^-} = \int_{\infty^-}^{\infty^+} \omega_{RS}, \qquad R,S \in \Sigma_z,
\end{equation}
and the fact that $\mathcal{K} = \varphi(\mathfrak{K})$, evaluation of (\ref{emsumomega}) at $P = \infty^+$ yields
\begin{equation}\label{expfourthetaquotient}
e^{- \int_{\mathfrak{K}}^{\mathfrak{M}} \omega_{\infty^+\infty^-}}
= \frac{\Theta(\varphi(\infty^+))\Theta\bigl(\varphi(\infty^-) - \int_\mathfrak{K}^{\mathfrak{M}}\omega\bigr)}{\Theta \bigl(\varphi(\infty^+) - \int_\mathfrak{K}^{\mathfrak{M}} \omega\bigr)\Theta(\varphi(\infty^-))}.
\end{equation}
Our choice of the cut basis $\{a_j, b_j\}$ implies that 
\begin{equation}\label{varphiinftypm}  
  \varphi(\infty^+) = -\varphi(\infty^-) \quad \hbox{modulo a-periods}.
\end{equation}
Thus, since $\Theta(v)$ is an even function, we arrive at (\ref{expthetaquotient}).

Now suppose the integration paths from $\mathfrak{K}$ to $\mathfrak{M}$ in (\ref{logfboth}) do not lie within the fundamental polygon. Then there exist integer vectors $p, q \in \Z^4$ such that equations (\ref{uDEomega}) and (\ref{fDEomegaI}) get replaced by
\begin{equation}\nonumber  
  u = \int_\mathfrak{K}^{\mathfrak{M}} \omega + Bp + q \qquad \hbox{and} \qquad  f = e^{-\int_{\mathfrak{K}}^{\mathfrak{M}} \omega_{\infty^+\infty^-} - \sum_{j=1}^4 p_j \int_{b_j} \omega_{\infty^+\infty^-} + I},
\end{equation}
respectively. However, a computation using (\ref{thetashifts}), (\ref{bperiodthirdkind}), and (\ref{expfourthetaquotient}) shows that the terms involving $p$ and $q$ cancel, so that $f$ is still given by (\ref{ernstsolution}).


We can now complete most of the proof of theorem \ref{mainth}; the derivation of the formula for $e^{2\kappa}$ will be postponed to the appendix. 
 We first establish the formulas in (\ref{OmegahOmegae2U}): The expression for $\Omega_h$ follows from (\ref{ahorOmegah}); the expression for $\Omega$ follows by solving (\ref{w4expression}) for $\Omega$ recalling that $e^{2U_0} < 0$ and $w_4 > 0$; and the expression for $e^{2U_\Omega(+i0)}$ follows by evaluating (\ref{UOmegaUrelation}) at $z = +i0$.

\subsection{The metric functions $e^{2U}$ and $a$}
Using formula (\ref{ernstsolution}), which was established in the previous subsection, the expression for the metric function $e^{2U}$ in (\ref{solutionmetric}) can be derived as follows cf. \cite{KKS}. 
Since the entries of $u$ are purely imaginary, $I \in \R$, and $\Theta(\bar{v}) = \overline{\Theta(v)}$ for $v \in \C^4$ (\cite{KR2005}, p. 203), equation (\ref{ernstsolution}) yields
\begin{equation}\label{fplusfbar}
 f + \bar{f} = \left(\frac{\Theta(u - \int_{-iz}^{\infty^-} \omega)}{\Theta(u + \int_{-iz}^{\infty^-} \omega)} + \frac{\Theta(u - \int_{i\bar{z}}^{\infty^-} \omega)}{\Theta (u  + \int_{i\bar{z}}^{\infty^-} \omega)}\right)e^I.
 \end{equation}
Let $E(P,Q)$ denote the prime form on $\Sigma_z$. 
Applying Fay's identity (\cite{KR2005}, p. 205)
\begin{align}\nonumber
&\frac{E(P_3, P_1)E(P_2, P_4)}{E(P_3, P_4)E(P_2, P_1)}\Theta\left(v + \int_{P_2}^{P_3}\omega \right)\Theta \biggl(v + \int_{P_1}^{P_4} \omega \biggr)
	\\ \nonumber
& \qquad \qquad \qquad + \frac{E(P_3, P_2)E(P_1, P_4)}{E(P_3, P_4)E(P_1, P_2)}\Theta \left(v + \int_{P_1}^{P_3} \omega \right)\Theta\left(v + \int_{P_2}^{P_4} \omega \right) 
	\\ \nonumber
& \qquad \qquad \qquad \qquad \qquad \qquad \qquad = \Theta(v)\Theta\left(v + \int_{P_2}^{P_3} \omega + \int_{P_1}^{P_4} \omega \right), \qquad v \in \C^4,
\end{align}
with $(P_1, P_2, P_3, P_4) = (i\bar{z}, \infty^+, -iz, \infty^-)$ to (\ref{fplusfbar}), we find
\begin{align}\label{2Qmanythetas}
2Q(0) \Theta\biggl(v + \int_{\infty^+}^{-iz} \omega \biggr)\Theta \biggl(v + \int_{i\bar{z}}^{\infty^-} \omega \biggr)
&- \Theta \biggl(v + \int_{i\bar{z}}^{-iz} \omega \biggr)\Theta \biggl(v + \int_{\infty^+}^{\infty^-} \omega\biggr)
	\\ \nonumber
& = \Theta(v)\Theta \biggl(v + \int_{\infty^+}^{-iz} \omega + \int_{i\bar{z}}^{\infty^-} \omega\biggr),
\end{align}
where 
$$Q(0) = \frac{1}{2}\frac{E(-iz, i\bar{z})E(\infty^+, \infty^-)}{E(-iz, \infty^-)E(\infty^+, i\bar{z})}$$ 
and we used that (lemma 3.12 in \cite{KR2005})
$$\frac{E(-iz, \infty^+)E(i\bar{z}, \infty^-)}{E(-iz, \infty^-)E(i\bar{z}, \infty^+)} = -1.$$
By proposition 3.11 in \cite{KR2005}, $Q(0)$ can be written as in (\ref{Qvdef}).
Letting $v = u + \int_{-iz}^{\infty^+}\omega = u - \int_{-iz}^{\infty^-} \omega $ and dividing by $\Theta(u + \int_{-iz}^{\infty^-})\Theta(u + \int_{i\bar{z}}^{\infty^-} \omega)$, equation (\ref{2Qmanythetas}) yields
\begin{align}\label{thetathetaquotient}
2Q(0) \frac{\Theta(u)\Theta(u + \int_{i\bar{z}}^{-iz} \omega)}{\Theta(u + \int_{-iz}^{\infty^-} \omega)\Theta(u + \int_{i\bar{z}}^{\infty^-} \omega)}
- \frac{\Theta(u + \int_{i\bar{z}}^{\infty^+} \omega)}{\Theta(u + \int_{i\bar{z}}^{\infty^-} \omega)}
= \frac{\Theta(u + \int_{-iz}^{\infty^+} \omega)}{\Theta(u + \int_{-iz}^{\infty^-} \omega)},
\end{align}
Equations (\ref{fplusfbar}) and (\ref{thetathetaquotient}) lead to the expression for $e^{2U}$ in (\ref{solutionmetric}).

By (5.7) in \cite{KR1998}, we have
\begin{equation}\label{aminusa0}
(a - a_0)e^{2U} = -\rho\left(\frac{\Theta(u + \int_{-iz}^{\infty^-} \omega + \int_{i\bar{z}}^{\infty^-} \omega)}{Q(0)Q(u) \Theta(u + \int_{-iz}^{i\bar{z}} \omega)} - 1\right),
\end{equation}
where $a_0 \in \R$ is a constant determined by the condition that $a = 0$ on the regular axis.
In view of the expression for $e^{2U}$, this yields the expression for the metric function $a$ given in (\ref{e2Uasolution}). An alternative derivation of equation (\ref{aminusa0}) is presented in the appendix.

\section{Axis and horizon values}
\begin{figure}
\begin{center}
    \includegraphics[width=.5\textwidth]{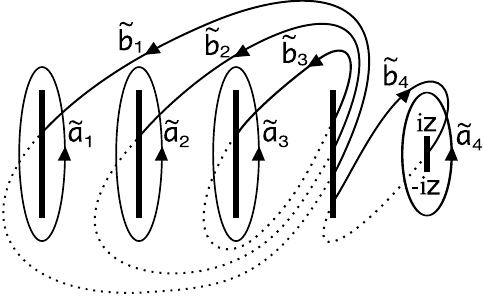} \quad
     \begin{figuretext}\label{axisadaptedcuts.pdf}
       The axis-adapted homology basis $\{\tilde{a}_j, \tilde{b}_j\}_{j = 1}^4$ on $\Sigma_z$.
     \end{figuretext}
 \end{center}
\end{figure}

\nequation\label{axishorizonsec}
In this section we consider the limits of the formulas in theorem \ref{mainth} as $z$ approaches a point on the $\zeta$-axis. As $\rho \downarrow 0$, the Riemann surface $\Sigma_z$ degenerates since the branch cut $[-iz, i\bar{z}]$ shrinks to a point. 
This type of degeneration of a Riemann surface is analyzed in Chapter III of \cite{Fay}. In order to utilize the results of \cite{Fay}, we introduce an axis-adapted cut basis $\{\tilde{a}_j, \tilde{b}_j\}_{j = 1}^4$ on $\Sigma_z$ by 
$$\begin{pmatrix} \tilde{a} \\ \tilde{b} \end{pmatrix} = \begin{pmatrix} \mathcal{A}^T & 0 \\ 0 &\mathcal{A}  \end{pmatrix}\begin{pmatrix} a \\ b \end{pmatrix}
\qquad \hbox{where} \qquad
\mathcal{A} = \begin{pmatrix} 1 & 0 & 0 & -1 \\ 0 & 1 & 0 & -1 \\ 0 & 0 & 1 & -1 \\ 0 & 0 & 0 & -1 \end{pmatrix}.$$
The axis-adapted cut basis is displayed in Figure \ref{axisadaptedcuts.pdf} in the case when $\zeta > \text{Re}\, k_4$. Note that $\tilde{a}_4$ surrounds the collapsing cut $[-iz, i\bar{z}]$. 
According to the transformation formula for theta functions (Eq. (12) in \cite{Fay}), there exists a constant $c_0$ independent of $v$ and $B$ such that
\begin{equation}\label{thetatransformation}
  \Theta(\tilde{v} | \tilde{B}) = c_0 \Theta(v | B)
\end{equation}
whenever $\tilde{v} = \mathcal{A}^{-1}v = \mathcal{A}v$.
We define $\tilde{u}$ and $\tilde{I}$ as the analogs of $u$ and $I$ in the axis-adapted basis, i.e.
$$\tilde{u} = \int_{\Gamma^+} h \tilde{\omega} + \int_\gamma \tilde{\omega}, \qquad
\tilde{I} = \int_{\Gamma^+} h \tilde{\omega}_{\infty^+\infty^-} + \int_\gamma \tilde{\omega}_{\infty^+\infty^-}.$$
Since $\tilde{\omega} = \mathcal{A}\omega$, we have $\tilde{u} = \mathcal{A}u.$
Since
$$ \omega_{\infty^+\infty^-} = \tilde{\omega}_{\infty^+\infty^-} - 2\pi i \tilde{\omega}_4,$$
we have $I = \tilde{I} -2 \pi i \tilde{u}_4.$ Thus, introducing the shorthand notation $\tilde{\Theta}(v) := \Theta(v|\tilde{B})$,
the Ernst potential (\ref{ernstsolution}) can be expressed in terms of the axis-adapted basis as
\begin{equation}\label{ernstpotentialtilde}
f(z) = \frac{\tilde{\Theta}(\tilde{u} - \int_{-iz}^{\infty^-} \tilde{\omega})}{\tilde{\Theta} (\tilde{u} + \int_{-iz}^{\infty^-} \tilde{\omega})}e^{-2 \pi i \tilde{u}_4 + \tilde{I}}.
\end{equation}

Let $\Sigma'$ denote the degenerated Riemann surface defined in (\ref{Sigmapdef}). 
According to \cite{Fay}, we have the following expansions as $\rho \downarrow 0$:
\begin{subequations}\label{Faylimits}
\begin{align}\label{tildeomegalimits}
& \tilde{\omega}_j = \omega_j' + O(\rho^2), \quad j = 1, 2,3; \qquad \tilde{\omega}_4 \to \frac{1}{2\pi i} \omega_{\zeta^+\zeta^-}'  + O(\rho^2),
	\\ \label{tildeBijlimits}
& \tilde{B}_{ij} \to B_{ij}'  + O(\rho^2), \quad i,j = 1, 2,3; \qquad \tilde{B}_{i4} \to 
\int_{\zeta^-}^{\zeta^+} \omega_i' + O(\rho^2), \quad i = 1, 2,3,
	\\ \label{tildeB44limit}
& \tilde{B}_{44} = \frac{1}{\pi i}\ln \rho + \frac{M'}{\pi i} + O(\rho^2),
\end{align}
\end{subequations}
where $M' \in \C$ is a constant. The path of integration from $\zeta^-$ to $\zeta^+$ in (\ref{tildeBijlimits}) must be chosen as the limit of the cycle $\tilde{b}_4$. For example, for $\text{Re}\, k_3 < \zeta < \text{Re}\, k_4$, this path is $[\zeta, \bar{k}_4]^- \cup [\bar{k}_4, \zeta]^+$. Whether this path lies within the fundamental polygon on $\Sigma'$ depends on the particular representatives of the homology cycles $a_j'$ and $b_j'$. 

\subsection{Values near the regular axis}
We first consider the case when $z$ approaches a point on the regular axis. We define $c \in \C^4$ and $c' \in \C^3$ by
\begin{equation}\label{cdef}
c = \int_{k_4}^{\infty^-} \tilde{\omega}, \qquad c' = \int_{k_4}^{\infty^-} \omega'.
\end{equation}

\begin{lemma}\label{thetalimitslemma}
The following limits hold as $z$ approaches a point on the regular axis (i.e. as $\rho \to 0$ with $r_1 < \zeta$):
\begin{subequations}\label{thetalimits}
\begin{align}\label{tildeuclimits}
\tilde{u} = &\, \begin{pmatrix} u' \\ \frac{J'}{2\pi i} \end{pmatrix} + O(\rho^2), \qquad c = \begin{pmatrix} c' \\ \frac{K'}{2\pi i} \end{pmatrix} + O(\rho^2),
	\\ \label{Thetauminusintlimit}
\tilde{\Theta}\biggl(\tilde{u} - \int_{-iz}^{\infty^-}\tilde{\omega}\biggr) = &\; \Theta'\biggl(u' - \int_{\zeta^-}^{\infty^-} \omega'\biggr) - \Theta'\biggl(u' - \int_{\zeta^+}^{\infty^-} \omega'\biggr)e^{J' - K'} + O(\rho^2),
	\\ \label{Thetauintlimit}
\tilde{\Theta} \biggl(\tilde{u} + \int_{-iz}^{\infty^-}\tilde{\omega} \biggr) = &\;
\Theta'\biggl(u' + \int_{\zeta^-}^{\infty^-} \omega'\biggr) -  \Theta'\biggl(u' + \int_{\zeta^+}^{\infty^-} \omega'\biggr)e^{-J' - K'}+ O(\rho^2),
	\\
\tilde{\Theta} \biggl(\tilde{u} + \int_{i\bar{z}}^{\infty^-}\tilde{\omega} \biggr) 
= &\;\Theta'\biggl(u' + \int_{\zeta^-}^{\infty^-} \omega'\biggr) + \Theta'\biggl(u«+ \int_{\zeta^+}^{\infty^-} \omega'\biggr)e^{-J' - K'} + O(\rho^2),
	\\
\tilde{\Theta} \biggl(\tilde{u} + \int_{-iz}^{i\bar{z}}\tilde{\omega} \biggr) = &\; \Theta'(u')+  \beta \rho + O(\rho^2), \qquad
\tilde{\Theta}(\tilde{u}) =  \Theta'(u') -  \beta \rho+ O(\rho^2), 
	\\
\tilde{\Theta}\biggl(\tilde{u} + \int_{-iz}^{\infty^-}\tilde{\omega} \; +& \int_{i\bar{z}}^{\infty^-}\tilde{\omega}\biggr)
 = -\frac{1}{\rho} \Theta'(u' + 2c')e^{-J' - 2K' - M'} + O(1),
 \end{align}
\end{subequations}
where $\beta \in \C$ is a constant and $u'$, $K'$, $J'$, and $M'$ are defined in (\ref{upIpKpdef}), (\ref{Jprimedef}), and (\ref{Mprimedef}). 
The equations obtained from (\ref{thetalimits}) by replacing $\tilde{u}$, $u'$, and $J'$ by $0$ everywhere are also valid.
\end{lemma}
\proofbegin
The expansions in (\ref{tildeuclimits}) follow immediately from (\ref{tildeomegalimits}).
We will also show (\ref{Thetauminusintlimit}); the proofs of the other expansions are similar. 
We first assume that $\zeta > \text{Re}\, k_4$.
In order for the argument of the theta function to have a finite limit, we shift the integration limit from $-iz$ to $k_4$. Our choice of the cut system $\{\tilde{a}_j, \tilde{b}_j\}$ implies that
\begin{equation}\label{basepointshift}
  \int_{-iz}^{\infty^-} \tilde{\omega} = c +\tilde{B}r + s,
\end{equation}
where $r,s \in \R^4$ are defined by
\begin{equation}\label{rsdef}  
  r = (0,0,0,1/2)^T, \qquad s = (0,0,0, -1/2)^T.
\end{equation}
Therefore,
\begin{align}\label{tildeThetatildeuint}
\tilde{\Theta}\biggl(\tilde{u} - \int_{-iz}^{\infty^-}\tilde{\omega}\biggr) = \tilde{\Theta}(\tilde{u} - c - \tilde{B}r - s) = \sum_{N \in \Z^4} e^{2\pi i(\frac{1}{2}N^T \tilde{B} N + N^T(\tilde{u}-c- \tilde{B}r-s))}.
\end{align}
Using (\ref{rsdef}), we can write the right-hand side of (\ref{tildeThetatildeuint}) as
$$\sum_{N \in \Z^4} e^{2\pi i \left(\frac{1}{2}\sum_{i,j=1}^3 N_i \tilde{B}_{ij} N_j + \frac{1}{2}\tilde{B}_{44}N_4(N_4 -1) + \sum_{i =1}^3 \tilde{B}_{i4}N_i (N_4 - \frac{1}{2}) + \sum_{i = 1}^3 N_i(\tilde{u}_i - c_i) + N_4(\tilde{u}_4 - c_4) + \frac{1}{2}N_4\right)}.$$
In view of (\ref{tildeB44limit}), only the terms with $N_4 = 0$ and $N_4 = 1$ give nonzero contributions in the limit $\rho \to 0$. 
Equations (\ref{Faylimits}) and (\ref{tildeuclimits}) imply that the subleading terms, which also receive contributions from the terms with $N_4 = -1$ and $N_4 = 2$, are of $O(\rho^2)$. We find
\begin{align}  \label{thetaucintlimit}
\tilde{\Theta}\biggl(\tilde{u} - \int_{-iz}^{\infty^-}\tilde{\omega}\biggr) = &\, 
\sum_{N' \in \Z^3} e^{2\pi i\bigl(\frac{1}{2} N'^T B' N' - \frac{1}{2} N'^T \int_{\zeta^-}^{\zeta^+} \omega' + N'^T(u' - c')\bigr)}
	\\ \nonumber
&+ \sum_{N' \in \Z^3} e^{2\pi i\bigl(\frac{1}{2} N'^T B' N' + \frac{1}{2} N'^T \int_{\zeta^-}^{\zeta^+} \omega' + N'^T(u' - c')\bigr) + J' - K' + \pi i} + O(\rho^2)
	\\\nonumber
  = &\; \Theta'\biggl(u' - c' - \frac{1}{2}\int_{\zeta^-}^{\zeta^+}\biggr) - \Theta'\biggl(u' - c' + \frac{1}{2}\int_{\zeta^-}^{\zeta^+}\biggr)e^{J' - K'} + O(\rho^2).
\end{align}
Since
\begin{equation}\label{cplusintrightcut}
c' + \frac{1}{2}\int_{\zeta^-}^{\zeta^+} \omega' = \int_{\zeta^-}^{\infty^-} \omega', \qquad
c' - \frac{1}{2}\int_{\zeta^-}^{\zeta^+} \omega' = \int_{\zeta^+}^{\infty^-} \omega',
\end{equation}
this proves (\ref{Thetauminusintlimit}) in the case when $\zeta > \text{Re}\, k_4$.

Similar arguments apply when $\zeta < \text{Re}\, k_3$ or $\text{Re}\, k_3 < \zeta < \text{Re}\, k_4$.
For example, if $\text{Re}\, k_3 < \zeta < \text{Re}\, k_4$, then equation (\ref{basepointshift}) gets replaced by
\begin{equation}\nonumber
  \int_{-iz}^{\infty^-} \tilde{\omega} = c +\tilde{B}r + t,
\end{equation}
where 
\begin{equation}\label{rtmiddlecutdef}  
  r = (0,0,0,1/2)^T, \qquad t = \left(-\frac{1}{2},-\frac{1}{2},-\frac{1}{2}, -\frac{1}{2}\right)^T, 
\end{equation}
Letting $t' = (-1/2, -1/2, -1/2)$, this leads to the following analog of equation (\ref{thetaucintlimit}):
\begin{align}\nonumber
\tilde{\Theta}\biggl(\tilde{u} - \int_{-iz}^{\infty^-}\tilde{\omega}\biggr) = &\, \Theta'\left(u' - c' - \frac{1}{2}\int_{\zeta^-}^{\zeta^+} \omega' - t'\right) 
	\\
& - \Theta'\left(u' - c' + \frac{1}{2}\int_{\zeta^-}^{\zeta^+}\omega' - t'\right)e^{J' - K'} + O(\rho^2),
\end{align}
Taking into account that, for $\text{Re}\, k_3 < \zeta < \text{Re}\, k_4$,
\begin{align} \label{cintsmiddlecut}
 c' + \frac{1}{2}\int_{\zeta^-}^{\zeta^+}\omega' + t' = \int_{\zeta^-}^{\infty^-} \omega', \qquad
c' - \frac{1}{2}\int_{\zeta^-}^{\zeta^+}\omega' + t'= \int_{\zeta^+}^{\infty^-} \omega',
\end{align}
we again arrive at (\ref{Thetauminusintlimit}).
\proofend

\begin{lemma}\label{Qaxislemma}
Let $Q$ be given by (\ref{Qvdef}). As $\rho \to 0$ with $r_1 < \zeta$,
\begin{align}\label{Quaxis}
Q(u) = &\frac{1}{\Theta'(u')^2}\biggl[\Theta'\biggl(u' + \int_{\zeta^-}^{\infty^-} \omega'\biggr)^2 -  \Theta'\biggl(u' + \int_{\zeta^+}^{\infty^-} \omega'\biggr)^2e^{-2J' - 2K'}\biggr] + O(\rho^2).
\end{align}
The behavior of $Q(0)$ as $\rho \to 0$ is given by the expression obtained by replacing $u'$ and $J'$ with zero in the right-hand side of (\ref{Quaxis}).
\end{lemma}
\proofbegin
In view of (\ref{thetatransformation}), the expression for $Q(u)$ is invariant under the change of cut basis fromÊ$\{a_j, b_j\}$ to $\{\tilde{a}_j, \tilde{b}_j\}$, i.e. 
\begin{equation}\label{tildeQdef}
Q(u) = \frac{\tilde{\Theta}(\tilde{u} + \int_{-iz}^{\infty^-} \tilde{\omega})\tilde{\Theta}(\tilde{u} + \int_{i\bar{z}}^{\infty^-} \tilde{\omega})} {\tilde{\Theta}(\tilde{u}) \tilde{\Theta}(\tilde{u} + \int_{-iz}^{i\bar{z}} \tilde{\omega})}.
\end{equation}
Utilizing the limits of lemma \ref{thetalimitslemma}, we find (\ref{Quaxis}).
\proofend

By applying the results of lemma \ref{thetalimitslemma} to formula (\ref{ernstpotentialtilde}) and using that $e^{I} = e^{\tilde{I} -2\pi i \tilde{u}_4 } = e^{I' - J'} + O(\rho^2)$ as $\rho \to 0$, we find that $f$ is given by (\ref{ernstsolutionnearaxis}) near the regular axis. 
The expression (\ref{e2Uaxis}) for $e^{2U}$ on the regular axis follows by applying the results of lemma \ref{Qaxislemma} to the equation $e^{2U} = \frac{Q(0)}{Q(u)}e^I$.
The limiting behavior $a = O(\rho^2)$ follows from (\ref{aminusa0}); the fact that the terms of $O(\rho)$ vanish in the expansion of $a$ is most easily seen from (\ref{bdef}). The behavior $e^{2\kappa} = 1 + O(\rho^2)$ near the regular axis follows from (\ref{e2ksolution}) and the condition that $\kappa = 0$ on the regular axis; the fact that the terms of $O(\rho)$ vanish in the expansion of $e^{2\kappa}$ is most easily seen from (\ref{kappaequation}). This completes the proof of proposition \ref{axisprop}.

\subsection{Values of $a_0$ and $K_0$}
The constant $a_0$ is determined by (\ref{aminusa0}) and the condition that $a = 0$ on the regular axis. We find
$$a_0 = \lim_{\rho \to 0}
\frac{\rho e^{-2U}}{Q(0)} \frac{\Theta(u + \int_{-iz}^{\infty^-}\omega + \int_{i\bar{z}}^{\infty^-} \omega)}{Q(u) \Theta(u + \int_{-iz}^{i\bar{z}}\omega)}, \qquad \zeta > r_1.$$
Substituting into this equation the expression (\ref{e2Uasolution}) for $e^{2U}$ and passing to the axis-adapted basis, we find
$$a_0 = \lim_{\rho \to 0}
\frac{\rho }{Q(0)^2} \frac{\tilde{\Theta}(\tilde{u} + \int_{-iz}^{\infty^-}\tilde{\omega} + \int_{i\bar{z}}^{\infty^-}\tilde{\omega})}{\tilde{\Theta}(\tilde{u} + \int_{-iz}^{i\bar{z}}\tilde{\omega})}e^{2\pi i \tilde{u}_4 - \tilde{I}}.$$
By lemmas \ref{thetalimitslemma} and \ref{Qaxislemma}, this yields
\begin{equation}\label{a0halfway}
a_0 = - 
\frac{\Theta'(0)^4}{\left(\Theta'(\int_{\zeta^-}^{\infty^-} \omega')^2 -  \Theta'(\int_{\zeta^+}^{\infty^-} \omega')^2e^{-2K'}\right)^2} \frac{\Theta'(u' + 2c')}{\Theta'(u')}e^{-M' -I' - 2K'}.
\end{equation}
Expression (\ref{a0solution}) for $a_0$ is obtained by letting $\zeta \to \infty$ in (\ref{a0halfway}). 
Indeed, since
$$\lim_{\zeta \to \infty} \frac{\Theta'(0)^4}{\left(\Theta'(\int_{\zeta^-}^{\infty^-} \omega')^2 -  \Theta'(\int_{\zeta^+}^{\infty^-} \omega')^2e^{-2K'}\right)^2} = 1$$
and $u', c', I'$ are independent of $\zeta > r_1$, we find
\begin{equation}\label{a0almostthere}
a_0 = - \frac{\Theta'(u' + 2c')}{\Theta'(u')}e^{-I'} \left(\lim_{\zeta \to \infty} e^{-M' - 2K'}\right).
\end{equation}
The constant $M'$ is given by\footnote{Expressions of this type are considered in \cite{Y}.}
$$M' = \frac{1}{2} \lim_{x \to 0} \left(\int_{(\zeta - x)^-}^{(\zeta - x)^+} \omega_{\zeta^+ \zeta^-}'  - 2\ln x - \ln 4 - \pi i\right).$$
The combination $M' + 2K'$ remains finite in the limit $\zeta \to \infty$ and we find
\begin{equation}\label{Mp2Kplimit} 
 \lim_{\zeta \to \infty} (M' + 2K') = \frac{1}{2}\lim_{R \to \infty} \left[-\int_{R^-}^{R^+}\omega_{\infty^+\infty^-}' - 2\ln R - \ln 4 - \pi i\right].
\end{equation}
Equations (\ref{a0almostthere}) and (\ref{Mp2Kplimit}) imply (\ref{a0solution}).

The constant $K_0$ in (\ref{e2ksolution}) is determined by the condition that $e^{2\kappa} = 1$ on the regular part of the axis. 
In order to compute $K_0$ we first rewrite (\ref{e2ksolution}) in terms of the axis-adapted cut system. Note that
\begin{equation}\label{omegatildeomegarelations}
\omega_{-r_1^+, -r_1^-} = \tilde{\omega}_ {-r_1^+, -r_1^-} - 2 \pi i \tilde{\omega}_4, \quad
\omega_{r_1^+ r_1^-} = \tilde{\omega}_ {r_1^+ r_1^-} - 2 \pi i \tilde{\omega}_4, \quad
\omega_{\kappa_1^+\kappa_1^-} = \tilde{\omega}_{\kappa_1^+\kappa_1^-} - 2 \pi i \tilde{\omega}_4.
\end{equation}
Since $\int_{\Gamma^+} d\kappa_1 \frac{dh}{dk}(\kappa_1) = 0$, we find
\begin{align}\nonumber
e^{2\kappa} = &\; K_0 \frac{\tilde{\Theta}(\tilde{u})\tilde{\Theta}(\tilde{u} + \int_{-iz}^{i\bar{z}} \tilde{\omega})}{\tilde{\Theta}(0)\tilde{\Theta}(\int_\xi^{\bar{\xi}} \tilde{\omega})}
e^{-\frac{1}{2} \int_{\Gamma^+} d\kappa_1 \frac{dh}{dk}(\kappa_1) \int_{\Gamma^+} h(\kappa_2) 
\tilde{\omega}_{\kappa_1^+\kappa_1^-}(\kappa_2^+)
+ \int_{\Gamma^+} h \tilde{\omega}_{-r_1^+, -r_1^-}
  - \int_{\Gamma^+} h \tilde{\omega}_{r_1^+ r_1^-}}
  	\\ \nonumber
&  \times e^{\frac{1}{2} \lim_{\epsilon \to 0} \left(
\int_{\gamma_1(\epsilon)} 
\tilde{\omega}_{-r_1^+, -r_1^-}
- \int_{\gamma_2(\epsilon)} 
\tilde{\omega}_{r_1^+ r_1^-} - 2\ln \epsilon \right)}, \qquad \zeta > r_1.
\end{align}
Taking the limit as $\rho \to 0$ of this expression, we find (\ref{K0solution}).

\subsection{Values near the black hole horizon}
The limits as $z$ approaches the black hole horizon have a slightly different flavor than those considered in the previous subsection, because $\tilde{u}_4$ diverges as $\rho \downarrow 0$ with $0 < \zeta< r_1$. In fact,
\begin{equation}\label{tildeu4expansion}  
  \tilde{u}_4 = -\frac{1}{\pi i}\ln \rho + \frac{P'}{\pi i} + O(\rho^2), \qquad \rho \downarrow 0, \quad 0 < \zeta< r_1,
\end{equation}
where
$$P' = \frac{1}{2}\int_{\Gamma^+} h \omega'_{\zeta^+\zeta^-} + \frac{1}{2} \lim_{\epsilon \to 0}\left(\int_{\gamma'(\epsilon)} \omega_{\zeta^+\zeta^-}' + 2 \ln \epsilon + \ln 4\right)$$ 
and $\gamma'(\epsilon)$ denotes the contour $\gamma'$ with the segments which lie above the interval $[\zeta - \epsilon, \zeta + \epsilon]$ removed.\footnote{For $0 < \zeta< r_1$, the contour $\gamma'$ contains the covering in the upper sheet of $[\zeta - \epsilon, \zeta]$ and the covering in the lower sheet of $[\zeta, \zeta + \epsilon]$.}
Note that 
\begin{equation}\label{e2P2MeJ}
  e^{2P' + 2M'} = e^{J'},  \qquad 0 < \zeta< r_1.
\end{equation}

\begin{lemma}\label{thetalimitshorizonlemma}
The following limits hold as $z$ approaches a point on the black hole horizon (i.e. as $\rho \to 0$ with $0 < \zeta< r_1$):
\begin{align}\label{Thetauminusinthorizonlimit}
\tilde{\Theta} \biggl(\tilde{u} - \int_{-iz}^{\infty^-}\tilde{\omega}\biggr) = & \frac{1}{\rho^2}\biggl[-\Theta'\biggl(u' - \int_{\zeta^+}^{\infty^-} \omega'\biggr)e^{J' - 2M' - K'} 
	\\ \nonumber
& + \Theta'\biggl(u' - \int_{\zeta^+}^{\infty^-} \omega' + \int_{\zeta^-}^{\zeta^+} \omega'\biggr)e^{2J' - 2M' - 2K'}\biggr] + O(1),
	\\ \nonumber
\tilde{\Theta} \biggl(\tilde{u} + \int_{-iz}^{\infty^-}\tilde{\omega} \biggr) = &
\Theta'\biggl(u' + \int_{\zeta^-}^{\infty^-}\omega'\biggr) - \Theta'\biggl(u' + \int_{\zeta^-}^{\infty^-}\omega' + \int_{\zeta^-}^{\zeta^+}\omega'\biggr) e^{J' + K'}  +O(\rho^2),
	\\ \nonumber
\tilde{\Theta} \biggl(\tilde{u} + \int_{i\bar{z}}^{\infty^-}\tilde{\omega} \biggr) 
= & \Theta'\biggl(u' + \int_{\zeta^-}^{\infty^-}\omega'\biggr) + \Theta'\biggl(u' + \int_{\zeta^-}^{\infty^-}\omega' + \int_{\zeta^-}^{\zeta^+}\omega') e^{J' + K'}  +O(\rho^2),
	\\ \nonumber
\tilde{\Theta} \biggl(\tilde{u} + \int_{-iz}^{i\bar{z}}\tilde{\omega} \biggr) = & -\frac{1}{\rho} \Theta'\biggl(u' + \int_{\zeta^-}^{\zeta^+}\omega'\biggr) e^{J' - M'} + \delta + O(\rho),
	\\ \label{Thetauhorizonlimit}
\tilde{\Theta}(\tilde{u}) = & \frac{1}{\rho} \Theta'\biggl(u' + \int_{\zeta^-}^{\zeta^+}\omega'\biggr) e^{J' - M'} + \delta + O(\rho),
	\\ \nonumber
\tilde{\Theta} \biggl(\tilde{u} + \int_{-iz}^{\infty^-}\tilde{\omega} +&\, \int_{i\bar{z}}^{\infty^-}\tilde{\omega} \biggr) 
=  \Theta'\biggl(u' + 2\int_{\zeta^-}^{\infty^-} \omega'\biggr) + O(\rho),
\end{align}
where $\delta$ is a constant.
\end{lemma}
\proofbegin
We prove (\ref{Thetauminusinthorizonlimit}) in the case when $\text{Re}\, k_3 < \zeta < r_1$; the proofs of the other identities are similar. For $\text{Re}\, k_3 < \zeta < r_1$,
$$\tilde{\Theta}\biggl(\tilde{u} - \int_{-iz}^{\infty^-}\tilde{\omega}\biggr)
= \tilde{\Theta}(\tilde{u} - c - \tilde{B}r - t),$$
where $r$ and $t$ are defined in (\ref{rtmiddlecutdef}).
We can write the right-hand side as
\begin{align*}
\sum_{N \in \Z^4}  &e^{2\pi i\bigl(\sum_{i,j=1}^3 N_i \tilde{B}_{ij} N_j + \frac{1}{2}\tilde{B}_{44}N_4(N_4 -1)\bigr)} 
	\\
& \times e^{2\pi i\bigl(\sum_{i = 1}^3 \tilde{B}_{i4}N_i(N_4 - \frac{1}{2}) + \sum_{i =1}^3 N_i(\tilde{u}_i - c_i) 
+ N_4(\tilde{u}_4 - c_4) - \sum_{i = 1}^3 N_i t_i + \frac{1}{2}N_4\bigr)},
\end{align*}
The factor involving the divergent quantities $\tilde{B}_{44}$ and $\tilde{u}_4$ is
$$e^{2\pi i\left(\frac{1}{2}\tilde{B}_{44}N_4(N_4 -1) + N_4\tilde{u}_4\right)}
= \rho^{N_4(N_4 -3)} e^{N_4(N_4 -1)M' + 2N_4P'}(1 + O(\rho^2)).$$
Thus the diverging terms are of $O(\rho^{-2})$ and arise when $N_4 = 1,2$. We find
\begin{align}\nonumber
\tilde{\Theta}\biggl(\tilde{u} - \int_{-iz}^{\infty^-}&\tilde{\omega}\biggr)
= -\frac{1}{\rho^2}e^{2 P' - K'} \sum_{N'\in \Z^3} e^{2\pi i\left(\frac{1}{2}N'^TB'N' + \frac{1}{2}N'^T\int_{\zeta^-}^{\zeta^+} \omega'
 + N'^T(u' - c' - t')\right)}
 	\\ \nonumber
 & \, + \frac{1}{\rho^2}e^{2M' + 4P' - 2K'} \sum_{N' \in \Z^3} e^{2\pi i\left(\frac{1}{2}N'^TB'N' + \frac{3}{2}N'^T\int_{\zeta^-}^{\zeta^+} \omega' + N'^T(u' - c' - t')\right)} + O(1).
\end{align}
Using (\ref{cintsmiddlecut}) and (\ref{e2P2MeJ}), we find (\ref{Thetauminusinthorizonlimit}).
\proofend

\begin{lemma}\label{Qhorizonlemma}
Let $Q$ be given by (\ref{Qvdef}). As $\rho \to 0$ with $0 < \zeta< r_1$,
\begin{align}\label{Quaxishorizon}
  Q(u) = & -\rho^2\frac{\Theta'(u' + \int_{\zeta^-}^{\infty^-} \omega')^2 - \Theta'(u' + \int_{\zeta^-}^{\infty^-} \omega'+ \int_{\zeta^-}^{\zeta^+} \omega')^2e^{2J' + 2K'}}{\Theta'(u' + \int_{\zeta^-}^{\zeta^+} \omega')^2e^{2J' - 2M'}} + O(\rho^4),
  	\\
Q(0) = &\; \frac{\Theta'(\int_{\zeta^-}^{\infty^-} \omega')^2 -  \Theta'(\int_{\zeta^+}^{\infty^-} \omega')^2e^{-2K'}}{\Theta'(0)^2}
+ O(\rho^2).
\end{align}
\end{lemma}
\proofbegin
By applying the limits in lemma \ref{thetalimitshorizonlemma} to (\ref{tildeQdef}), we find the statement for $Q(u)$.
The limit of $Q(0)$ is obtained as in the case of the regular axis, since the diverging factors involving $\tilde{u}_4$ are not present.
\proofend

By applying the limits of lemma \ref{thetalimitshorizonlemma} to equation (\ref{ernstpotentialtilde}), we find that $f$ is given by (\ref{ernstsolutionnearhorizon}) near the horizon. 
Similarly, the behavior of $e^{2U}$ near the horizon follows by applying lemma \ref{Qhorizonlemma} and the expansion (\ref{tildeu4expansion}) to the equation $e^{2U} = \frac{Q(0)}{Q(u)}e^{-2\pi i \tilde{u}_4 + \tilde{I}}$.
The expression for $a_{hor}$ is established by applying the results of lemmas \ref{thetalimitshorizonlemma} and \ref{Qhorizonlemma} to the axis-adapted version of (\ref{aminusa0}).

We next show formula (\ref{e2kappahorizon}) for the behavior of $e^{2\kappa}$ near the horizon. 
Equation (\ref{kappaequation}) implies that the terms of $O(\rho)$ in the expansion of $e^{2\kappa}$ vanish, so we only have to determine the leading term. 
Suppose $0 < \zeta < r_1$. 
Then, recalling (\ref{omegatildeomegarelations}) and using that $\int_{\Gamma^+} d\kappa_1 \frac{dh}{dk}(\kappa_1) = 0$, we can write $L$ Êas defined in (\ref{LLdef}) in terms of axis-adapted quantities as
\begin{align}\nonumber
L = &\,
-\frac{1}{2} \int_{\Gamma} d\kappa_1 \frac{dh}{dk}(\kappa_1) \int_{\Gamma}' h(\kappa_2) 
\tilde{\omega}_{\kappa_1^+\kappa_1^-}(\kappa_2^+)
+ \int_{\Gamma^+} h \tilde{\omega}_{-r_1^+, -r_1^-}
  + \int_{\Gamma^+} h \tilde{\omega}_{r_1^+ r_1^-}
  	\\ \nonumber
& \; + \frac{1}{2}\lim_{\epsilon \to 0} \left(\int_{\gamma_1(\epsilon)}\tilde{\omega}_{-r_1^+, -r_1^-} + \int_ {\gamma_2(\epsilon)} \tilde{\omega}_{r_1^+ r_1^-} - 2\ln \epsilon \right)
- 2 \pi i \int_{\Gamma^+} h \tilde{\omega}_4 
- 2 \pi i \tilde{u}_4.
\end{align}
In view of (\ref{tildeu4expansion}) and (\ref{e2P2MeJ}), we find
$$e^{L} = \rho^2 e^{L' + 2M' - J'} + O(\rho), \qquad \rho \to 0, \quad 0 < \zeta < r_1,$$
where $L'$ is defined by (\ref{Lprimedef}).
Applying this expansion together with lemma \ref{thetalimitshorizonlemma} to the expression for $e^{2\kappa}$ in theorem \ref{mainth}, we arrive at (\ref{e2kappahorizon}). 
This completes the proof of proposition \ref{horizonprop}.

We conclude this section by proving proposition \ref{f0f1prop}.
As $\zeta \downarrow r_1$, we have $e^{-J'} \to 0$ whereas $e^{K'}$ tends to a bounded constant. The expression (\ref{f1solution}) for $f_1 := f(ir_1)$ therefore follows immediately from (\ref{ernstsolutionaxis}). Similarly, the statement in proposition \ref{f0f1prop} regarding $f_0 := f(+i0)$ follows by taking the limit $\rho \downarrow 0$ in (\ref{ernstsolutionhorizon}).

\section{Parameter ranges}\nequation\label{paramsec}
In this section we consider the singularity structure of the solution (\ref{ernstsolution}) and its dependence on the four parameters $\rho_0, r_1, w_2$, and $w_4$.
 
\subsection{Singularity structure}
The solution $f$ presented in (\ref{ernstsolution}) is continuous but not smooth at the point $z = \pm ir_1$ where the regular axis meets the horizon. Moreover, $\text{Im}\, f$ has a jump across the disk. Away from these points $f$ is smooth except possibly at points in the set where the denominator of (\ref{ernstsolution}) vanishes. 
Physically, we are interested in solutions which are singularity-free away from the disk and the horizon. A complete characterization of the singularity-free solutions involves determining for which choices of $\rho_0$, $r_1$, $w_2$, $w_4$ the set $\bigl\{z \in \mathcal{D} \, | \, \Theta (u + \int_{-iz}^{\infty^-} \omega) = 0 \bigr\}$ is empty. We will not complete this analysis here, but we will indicate how a large class of singularity-free solutions can be constructed starting with parameters corresponding to a Kerr background. 

In subsection \ref{scalarrhsubsec}, the Kerr solutions were parametrized in terms of the parameters $r_1 > 0$ and $-\pi/2 < \delta < 0$. However, since the map 
\begin{equation}\label{deltaw2relation} 
  \delta \mapsto w_2 = \frac{2}{\tan\delta \sin\delta}:\left(-\frac{\pi}{2}, 0\right) \to \R_{>0}
\end{equation}
is one-to-one, we may also adopt a parametrization in terms of $r_1 > 0$ and $w_2 = 2/(\tan \delta \sin\delta) > 0$. Let $f^{kerr}_{r_1w_2}$ denote the unique Kerr solution corresponding to the parameters $r_1 > 0$ and $w_2 > 0$.
Moreover, let $f$ denote the solution in (\ref{ernstsolution}) corresponding to some strictly positive parameters $\rho_0, r_1, w_2, w_4$. Then $f \to f^{kerr}_{r_1w_2}$ as $\rho_0 \downarrow 0$ and $w_4 \downarrow 0$ with $r_1, w_2$ held fixed. Indeed, consider perturbing a Kerr background solution $f^{kerr}$ by adding a small disk of radius $\rho_0$ rotating with angular velocity $\Omega$. In the limit $\rho_0 \downarrow 0$ and $\Omega \downarrow 0$, the jump contour $\Gamma^+$ in the RH problem disappears and $\Lambda_\Omega$ reduces to the identity matrix. Thus, the BVP (\ref{BVP}) reduces to the Kerr black hole BVP and the perturbed solution $f$ approaches $f^{kerr}$ in this limit. 
Substituting the Kerr values of $f_0$ and $f_1$ and letting $\Omega \downarrow 0$ in (\ref{w0w2w4expressions}), we find that 
\begin{equation}\label{w2w4kerr}  
  w_0 \to 0, \qquad w_2 \to \frac{(f_0^{kerr})^2 - 1}{2 f_0^{kerr}} = \frac{2}{\tan\delta \sin\delta}, \qquad \hbox{and} \qquad w_4 \to 0,
\end{equation}
as the Kerr background is approached. This leads to the relation (\ref{deltaw2relation}) between $\delta$ and $w_2$. In view of (\ref{rimcondition}), the vanishing limiting value of $w_0$ is achieved by letting $\rho_0 \downarrow 0$.

\begin{figure}
\begin{center}
 \begin{overpic}[width=115mm]{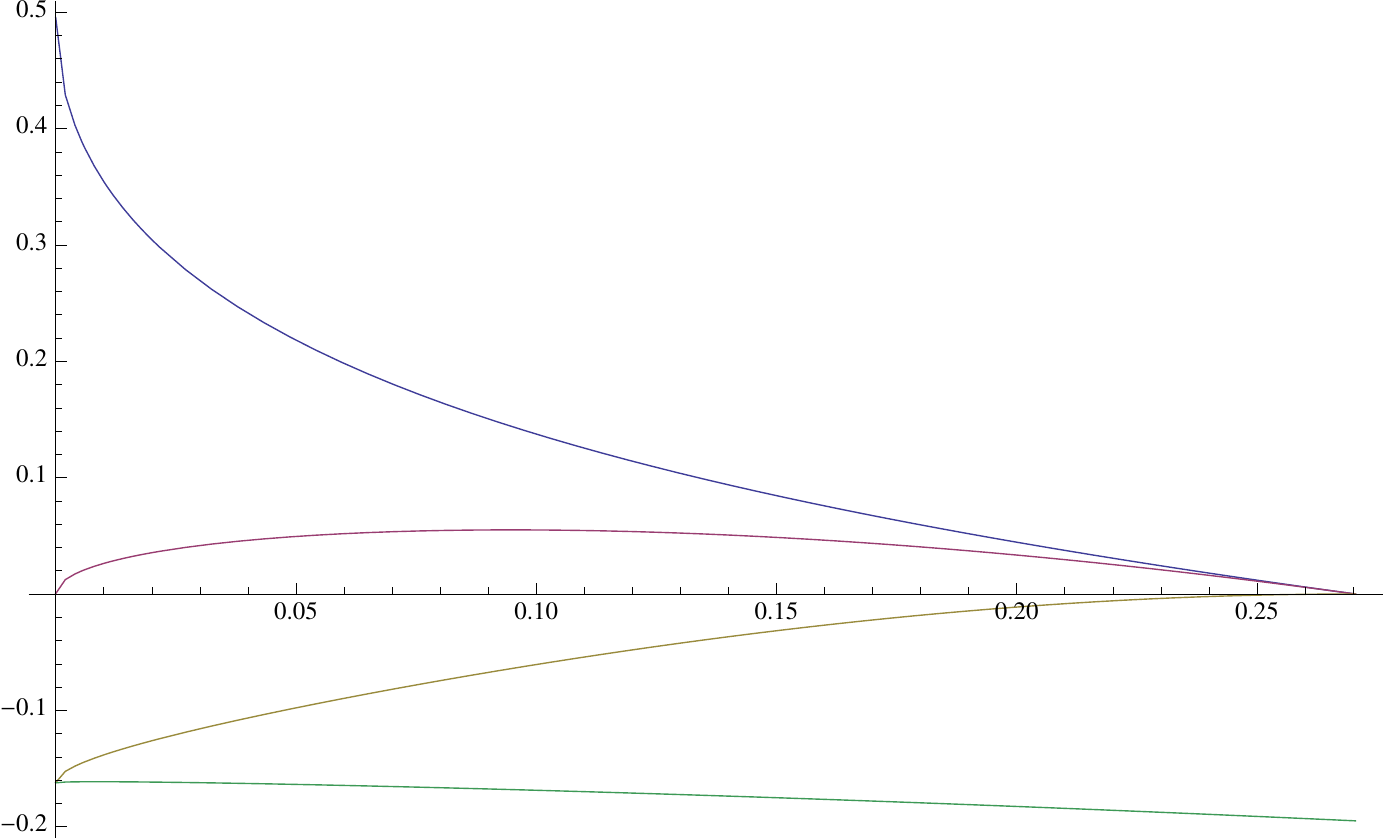}
      \put(70,123){$\Omega_h$}
      \put(70,76){$\Omega$}
      \put(45, 33){$e^{2U_{\Omega0}}$}
      \put(200,15){$e^{2U_0}$}
      \put(345,55){$w_4$}
      \put(312,40){$w_4^{max}$}
    \end{overpic}
      \begin{figuretext}\label{parametersw2equals3.pdf}
       The dependence on $w_4$ of the parameters $\Omega_h$, $\Omega$, $e^{2U_0} = \text{\upshape Re}\, f(+i0)$, and $e^{2U_{\Omega0}} = \text{\upshape Re}\, f_\Omega(+i0)$  for the example specified by (\ref{w4example}).
       \end{figuretext}
 \end{center}
\end{figure}   

Thus, for small values of $w_4 > 0$ and $\rho_0 > 0$, the solution $f$ corresponding to $\{\rho_0, r_1, w_2, w_4\}$ is a small perturbation of the Kerr background solution $f^{kerr}_{r_1w_2}$. In particular, $f$ is singularity-free for sufficiently small perturbations. By increasing $w_4$ and $\rho_0$, larger pertubarbations of the background are obtained until the construction eventually breaks down and the solutions become singular. In this way, a large class of singularity-free solutions can be constructed. Numerical data suggest that given strictly positive values of the parameters $\rho_0$, $r_1$, and $w_2$, there exists an interval $[0, w_4^{max}]$, $w_4^{max} > 0$, such that all solutions $f$ corresponding to $\{\rho_0, r_1, w_2, w_4\}$ with $w_4 \in (0, w_4^{max})$ are free of singularities. In the following subsection, we illustrate the general situation by considering a typical example.

\subsection{Dependence on $w_4$}
\begin{figure}
\begin{center}
 \begin{overpic}[width=80mm]{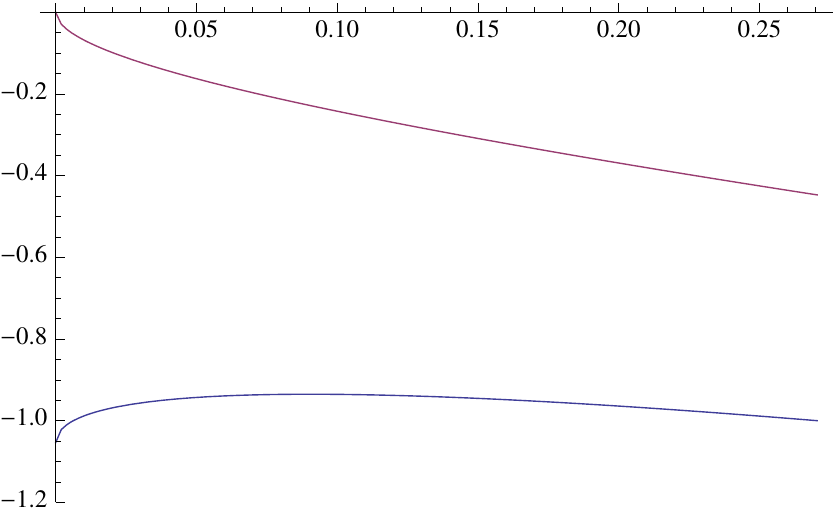}
      \put(70,102){$b_0$}
      \put(70,19){$\text{Im} \, f_1$}
      \put(247,135){$w_4$}
      \put(218,120){$w_4^{max}$}
    \end{overpic}
      \begin{figuretext}\label{imf0imf1w2equals3.pdf}
        The dependence on $w_4$ of the parameters $b_0 = \text{\upshape Im}\, f(+i0)$ and $f_1 = f(ir_1)$ for the example specified by (\ref{w4example}).         
           \end{figuretext}
 \end{center}
\end{figure}

\begin{figure}
\begin{center}
    \includegraphics[width=.6\textwidth]{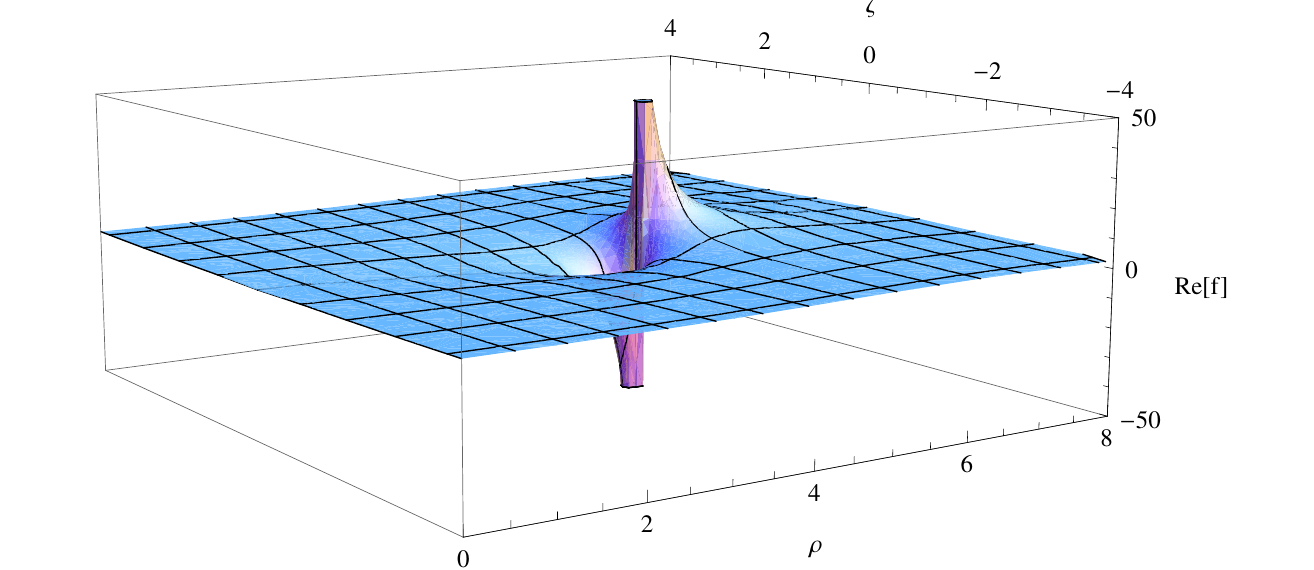}
     \begin{figuretext}\label{singularf3D.pdf}
       The graph of $\text{Re}\, f$ for $w_4 = 1/2 > w_4^{max}$. The disk and the black hole are too small to be visible.
     \end{figuretext}
 \end{center}
\end{figure}

We let 
\begin{equation}\label{w4example}  
  \rho_0 = 1, \qquad r_1 = \frac{1}{2}, \qquad w_2 = 3,
\end{equation}
and consider the solution $f$ given in (\ref{ernstsolution}) corresponding to $\{\rho_0, r_1, w_2, w_4\}$ as $w_4 > 0$ varies. We find that the solution is free of singularities for $0 < w_4 < w_4^{max}$ where $w_4^{max} \approx 0.27051$. The example presented in section \ref{examplesec} corresponds to taking $w_4 = 1/10$. The dependence on $w_4$ of several parameters is displayed in Figures \ref{parametersw2equals3.pdf} and \ref{imf0imf1w2equals3.pdf}. The parameter $w_4$ is analogous to the variable $\mu$ used in \cite{MAKNP} to parametrize the Neugebauer-Meinel disks and the Figures \ref{parametersw2equals3.pdf} and \ref{imf0imf1w2equals3.pdf} are the analogs of Figure 2.9 in \cite{MAKNP}.
To see how the solution $f$ becomes singular as $w_4$ increases beyond $w_4^{max}$, we note that as $w_4 < w_4^{max}$ increases, the ergosphere of the solution $f$ grows larger and larger until it eventually, in the limit $w_4 \uparrow w_4^{max}$, envelops all of spacetime. As $w_4$ increases beyond $w_4^{max}$ a singularity of $f$ enters the domain $\mathcal{D}$ at $z = +\infty$ and moves inward along the positive real axis. The graph of the singular function $\text{Re}\, f$ for $w_4 = 1/2 > w_4^{max}$ is shown in Figure \ref{singularf3D.pdf}.

\appendix
\section{Condensation of branch points}
\renewcommand{\theequation}{A.\arabic{equation}}\nequation
In this appendix we show that the Ernst potential (\ref{ernstsolution}) is related via a certain limiting procedure to the class of solutions of the Ernst equation studied in \cite{KM, KKS}. By applying this limiting procedure to the formula for the metric function $e^{2\kappa}$ given in \cite{KKS}, we will also establish the expression (\ref{e2ksolution}) for $e^{2\kappa}$ and so complete the proof of theorem \ref{mainth}. 
The limiting operation involves partially degenerating a Riemann surface by letting branch points coalesce along the curve $\Gamma^+$ and at the points $\pm r_1$. The construction of new solutions of the Ernst equation through this type of `condensation' of branch points along curves was first described in \cite{KM}.

Let $\hat{\Sigma}_z$ be a Riemann surface of genus $g > 4$ obtained by adding $g-4$ branch cuts $\{[E_{j}, F_{j}]\}_{j = 1}^{g-4}$ to $\Sigma_z$. Let $\xi = -iz$. Then $\hat{\Sigma}_z$ is defined by the equation
$$\hat{y}^2 = (k - \xi)(k - \bar{\xi})\prod_{i = 1}^{4} (k - k_i)(k - \bar{k}_i)\prod_{i = 1}^{g -4} (k - E_i)(k - F_i).$$
Let $\{\hat{a}_j, \hat{b}_j\}_{j = 1}^g$ be the cut basis on $\hat{\Sigma}_z$ which is the natural generalization of the basis $\{a_j,b_j\}_1^4$ on $\Sigma_z$, i.e. for $j = 1, \dots, 4$, $\hat{a}_j$ surrounds the cut $[k_j, \bar{k}_j]$; for $j = 5, \dots g$, $\hat{a}_j$ surrounds the cut $[E_{j - 4}, F_{j - 4}]$; the cycle $\hat{b}_j$ enters the upper sheet on the right side of $[-iz, i\bar{z}]$ and exits again on the right side of $[k_j, \bar{k}_j]$ for $j = 1, \dots, 4$ and on the right side of $[E_{j - 4}, F_{j -4}]$ for $j = 5, \dots, g$. For simplicity, we will assume that $\zeta > \text{Re}\, k_4$.
Let $\hat{\omega} = (\hat{\omega}_1, \dots, \hat{\omega}_g)^T$ denote the canonical dual basis and let $\hat{\Theta}(\hat{w}) := \Theta(\hat{w} | \hat{B})$ be the associated theta function.
Let $p, q \in \C^g$ be vectors which are indepedent of $z$ and which satisfy the reality condition $\hat{B}p + q \in \R^g$.
The theta function with characteristics $p, q \in \R^g$ is defined by
$$\Theta\begin{bmatrix} p \\ q \end{bmatrix}(\hat{v} | \hat{B}) = \Theta(\hat{v} + \hat{B}p + q | \hat{B}) e^{2\pi i\bigl(\frac{1}{2}p^T\hat{B}p + p^T(\hat{v} + q)\bigr)}, \qquad \hat{v} \in \C^g.$$
Then 
\begin{equation}\label{hatf}
\hat{f} = \frac{\hat{\Theta}\begin{bmatrix} p \\ q \end{bmatrix}(\int_{\xi}^{\infty^+} \hat{\omega})}{\hat{\Theta}\begin{bmatrix} p \\ q \end{bmatrix}(\int_{\xi}^{\infty^-} \hat{\omega})},
\end{equation}
is a solution of the Ernst equation (\ref{ernst}) and the corresponding metric functionÊ $e^{2\hat{\kappa}}$ is given by
\begin{equation}\label{pree2k}
e^{2\hat{\kappa}} = \hat{K}_0 \frac{\hat{\Theta}\begin{bmatrix} p \\ q \end{bmatrix}(0)\hat{\Theta}\begin{bmatrix} p \\ q \end{bmatrix}(\int_{\xi}^{\bar{\xi}} \hat{\omega})}{\hat{\Theta}(0) \hat{\Theta}(\int_{\xi}^{\bar{\xi}}\hat{\omega})}
\end{equation}
where $\hat{K}_0 \in \C$ is a constant determined by the condition that $e^{2\hat{\kappa}} = 1$ on the regular axis \cite{KM, KKS}. 

\begin{figure}
\begin{center}
    \includegraphics[width=.7\textwidth]{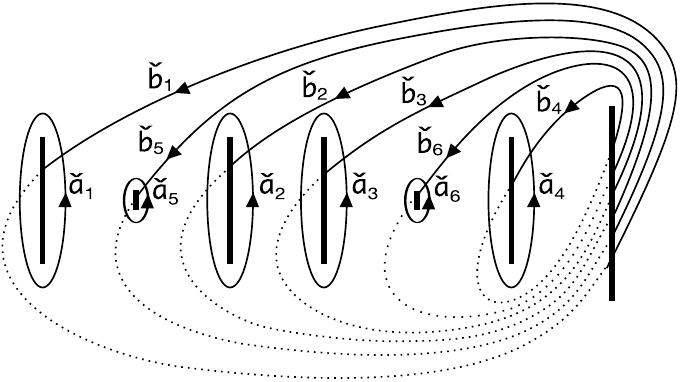}
     \begin{figuretext}\label{condensationcuts.pdf}
       The homology basis $\{\check{a}_j, \check{b}_j\}_{j=1}^6$ on the Riemann surface $\check{\Sigma}_z$.
     \end{figuretext}
 \end{center}
\end{figure}

We choose 
$$E_1 = -r_1 - i\epsilon, \qquad F_1 = \bar{E}_1, \qquad E_2 = r_1 - i\epsilon, \qquad F_2 = \bar{E}_2,$$
where $\epsilon > 0$ is a small number, and define the Riemann surface $\check{\Sigma}_z$ of genus $6$ by
$$\check{y}^2 = (k - \xi)(k - \bar{\xi})\prod_{i = 1}^{4} (k - k_i)(k - \bar{k}_i)\prod_{i = 1}^{2} (k - E_i)(k - F_i).$$
In other words, $\check{\Sigma}_z$ is obtained from $\Sigma_z$ by adding two short vertical cuts centered at $-r_1$ and $r_1$, respectively. The cut basis $\{\check{a}_j, \check{b}_j\}_{j=1}^6$ is shown in Figure \ref{condensationcuts.pdf}.

The condensation of branch points will now proceed in two steps: In the first step, we let the branch points $E_{j+2}, F_{j+2}$, $j = 1, \dots, g-6$ condense along the curve $\Gamma$. In doing this, the Riemann surface $\hat{\Sigma}_z$ degenerates to $\check{\Sigma}_z$ and the Ernst potential $\hat{f}$ approaches a solution $\check{f}$ defined in terms of theta functions on $\check{\Sigma}_z$. Intuitively $\check{f}$ has a disk, but no black hole. In the second step, we let $\epsilon \to 0$. Then $\check{\Sigma}_z$ degenerates to $\Sigma_z$ and we will find that $\check{f}$ approaches the solution $f$ in (\ref{ernstsolution}).

\subsection{The first degeneration}
Let
$$p = (\check{p}, m) \in \C^6 \times \R^{g - 6}, \qquad q = 0,$$
where the components of the vector $m \in \R^{g-6}$ satisfy $0 < m_j < 1/2$, $j = 1, \dots, g-6$.
We consider the limit $E_{j+2}, F_{j+2} \to \kappa_j$, $j = 1, \dots, g-6$, in which the branch cut $[E_{j+2}, F_{j+2}]$ shrinks to a point $\kappa_j \in \Gamma$. 
In this limit, (cf. Eq. (\ref{Faylimits}))
\begin{subequations} \nonumber
\begin{align}\nonumber
& \left(\hat{\omega}_1, \dots \hat{\omega}_g \right) \to
\left(\check{\omega}_1, \dots \check{\omega}_6, \frac{1}{2\pi i}\check{\omega}_{\kappa_1^+\kappa_1^-}, \dots, \frac{1}{2\pi i}\check{\omega}_{\kappa_{g - 6}^+\kappa_{g - 6}^-}\right);
	\\ \nonumber
& \hat{B}_{ij} \to \check{B}_{ij}, \quad i,j = 1, \dots, 6; \qquad \hat{B}_{i, j+6} \to \int_{\kappa_j^-}^{\kappa_j^+} \check{\omega}_i, \quad i = 1, \dots, 6, \quad j = 1, \dots, g-6;
	\\ \nonumber
& \hat{B}_{i+6, j+6} \to \frac{1}{2\pi i} \int_{\kappa_j^-}^{\kappa_j^+} \check{\omega}_{\kappa_i^+\kappa_i^-}, \qquad i, j = 1, \dots, g - 6, \quad i \neq j;
	\\ \nonumber
& \hat{B}_ {j+6, j+6} = \frac{1}{\pi i} \ln|E_{j +2} - F_{j+2}| + O(1), \qquad j = 1, \dots, g - 6.
\end{align}
\end{subequations}
For two points $P,Q \in \hat{\Sigma}_z$, we have
\begin{equation}\label{hatThetap0}
\hat{\Theta}\begin{bmatrix} p \\ 0 \end{bmatrix}\biggl(\int_P^Q \hat{\omega} \biggr) = 
\left(\sum_{N \in \Z^g} e^{2\pi i\left(\frac{1}{2} N^T \hat{B} N + N^T(\int_P^Q \hat{\omega} + Bp)\right)}\right)e^{2\pi i\left(\frac{1}{2} p^T \hat{B} p + p^T\int_P^Q\hat{\omega}\right)}.
\end{equation}
Letting $N = (\check{N}, n) \in \Z^6 \times \Z^{g-6}$ and using that $p = (\check{p}, m)$, we find that the factor in the sum on the right-hand side involving the diverging quantities $\hat{B}_{j+6, j + 6}$, $j = 1, \dots, g-6$, is
$$e^{\pi i \sum_{j =1}^{g-6} n_j(n_j + 2m_j)\hat{B}_{j+6, j+6}}.$$
Consequently, since $0 < m_j < 1/2$ by assumption, all terms in the sum in (\ref{hatThetap0}) approach zero except the ones with $n \equiv 0$. 
We infer that the sum on the right-hand side of (\ref{hatThetap0}) converges to
\begin{equation}\label{checkThetasumlimit}
\check{\Theta}\biggl(\int_P^Q \check{\omega} + \sum_{j =1}^{g-6}  m_j  \int_{\kappa_j^-}^{\kappa_j^+} \check{\omega} + \check{B}\check{p}\biggr).
\end{equation}
We let the $\kappa_j$'s condense onto the curve $\Gamma$ with a density determined by the measure $dm(\kappa)$ defined by
$$dm(\kappa) =  -\frac{1}{2}\frac{dh}{d\kappa}(\kappa)d\kappa, \qquad \kappa \in \Gamma,$$
where $h$ is the function defined in (\ref{hdef}).
Then, integrating by parts and using that $h$ vanishes at the endpoints of $\Gamma$, we find
\begin{equation}\label{condensetoucheck}
\sum_{j =1}^{g-6}  m_j  \int_{\kappa_j^-}^{\kappa_j^+} \check{\omega}
\to
\int_{\Gamma} dm(\kappa)  \int_{\kappa^-}^{\kappa^+} \check{\omega}
= \check{u},
\end{equation}
where $\check{u} \in \C^6$ is defined by
$$\check{u} = \int_{\Gamma^+} h \check{\omega}.$$
Combining (\ref{hatThetap0})-(\ref{condensetoucheck}), we find 
\begin{equation}\label{hatThetalimit}
\hat{\Theta}\begin{bmatrix} p \\ 0 \end{bmatrix}\biggl(\int_P^Q \hat{\omega} \biggr) \to 
\check{\Theta}\begin{bmatrix} \check{p} \\ 0 \end{bmatrix}\biggl(\check{u} + \int_P^Q \check{\omega} \biggr) 
e^{\check{L}/2 + \int_\Gamma dm(\kappa) \int_P^Q \check{\omega}_{\kappa^+\kappa^-}},
\end{equation}
where $\check{L}$ is defined by
$$\check{L} = -\frac{1}{2} \int_{\Gamma} d\kappa_1 \frac{dh}{dk}(\kappa_1) \int_{\Gamma}' h(\kappa_2) \omega_{\kappa_1^+\kappa_1^-}(\kappa_2^+),$$
and the prime on the integral along $\Gamma$ indicates that the integration contour should be deformed slightly before evaluation so that the pole at $\kappa_2 = \kappa_1$ is avoided.\footnote{The result is indepedent of whether the contour is deformed to the right or to the left of the pole.}

Applying this formula to (\ref{hatf}), we arrive at the following limit of $\hat{f}$:
\begin{equation}\label{checkf}
\hat{f} \to \check{f} = \frac{\check{\Theta}\begin{bmatrix} \check{p} \\ 0 \end{bmatrix}( \check{u} + \int_\xi^{\infty^+} \check{\omega})}{\check{\Theta}\begin{bmatrix} \check{p} \\ 0 \end{bmatrix}( \check{u} + \int_\xi^{\infty^-} \check{\omega})}
e^{\check{I}},
\end{equation}
where $\check{I} \in \R$ is defined by
$$\check{I} = \int_\Gamma dm(\kappa) \int_{\infty^-}^{\infty^+} \check{\omega}_{\kappa^+\kappa^-} =\int_{\Gamma^+} h \check{\omega}_{\infty^+ \infty ^-}.$$
Moreover, applying equation (\ref{hatThetalimit}) to the expression for $e^{2\hat{\kappa}}$ in (\ref{pree2k}), we find
\begin{equation}\label{e2checkkappa1}
e^{2\hat{\kappa}}
\to e^{2\check{\kappa}}
= \check{K}_0 \frac{\check{\Theta}(\check{u})\check{\Theta}(\check{u} + \int_\xi^{\bar{\xi}} \check{\omega})}{\check{\Theta}(0)\check{\Theta}(\int_\xi^{\bar{\xi}} \check{\omega})}
e^{\check{L}} e^{\int_\Gamma dm(\kappa) \int_\xi^{\bar{\xi}} \check{\omega}_{\kappa^+\kappa^-}},
\end{equation}
where $\check{K}_0$ is a constant independent of $z$. 
For some constant $C$, we have
$$e^{-\int_{P_0}^P \omega_{\xi\bar{\xi}}} = C \sqrt{\frac{P - \bar{\xi}}{P - \xi}}, \qquad P \in \check{\Sigma}_z,$$
because both sides have simple poles at $\xi$, simple zeros at $\bar{\xi}$, and are analytic elsewhere on $\check{\Sigma}_z$. Hence, 
$$\int_\xi^{\bar{\xi}} \check{\omega}_{\kappa^+\kappa^-}
= -\int_{\kappa^-}^{\kappa^+} \omega_{\xi\bar{\xi}} = \log \sqrt{\frac{P - \bar{\xi}}{P - \xi}}\Biggl|_{P = \kappa^-}^{\kappa^+} \in \pi i + 2\pi i \Z.$$
It follows that the last exponential factor in (\ref{e2checkkappa1}) is independent of $z$ and can be absorbed into $\check{K}_0$.
Thus,
\begin{equation}\label{e2checkkappa}
e^{2\check{\kappa}}
= \check{K}_0 \frac{\check{\Theta}(\check{u})\check{\Theta}( \check{u} + \int_\xi^{\bar{\xi}} \check{\omega})}{\check{\Theta}(0)\check{\Theta}(\int_\xi^{\bar{\xi}} \check{\omega})}
e^{\check{L}}.
\end{equation}

\subsection{The second degeneration}
We now consider the degeneration of $\check{\Sigma}_z$ as the cuts centered at $\pm r_1$ collapse. 
In the limit $\epsilon \to 0$,
\begin{align} \nonumber
& \left(\check{\omega}_1, \dots \check{\omega}_6 \right) \to
\left(\omega_1, \dots \omega_4, \frac{1}{2\pi i}\omega_{-r_1^+,-r_1^-}, \frac{1}{2\pi i}\omega_{r_1^+r_1^-}\right);
	\\ \nonumber
& \check{B} = \begin{pmatrix} B & B_1 \\ B_1^T & B_2 \end{pmatrix} + O(\epsilon^2),
\end{align}
where $B$ is the period matrix on $\Sigma_z$, the $4\times 2$ matrix $B_1$ is defined by
$$B_1 = \begin{pmatrix} \int_{-r_1^-}^{-r_1^+} \omega & \int_{r_1^-}^{r_1^+} \omega \end{pmatrix},$$
and the $2\times 2$ matrix $B_2$ is given by
$$B_2 = \begin{pmatrix} \frac{1}{\pi i }(\ln \epsilon + c_-) & 
\frac{1}{2\pi i} \int_{r_1^-}^{r_1^+} \omega_{-r_1^+, -r_1^-} \\
\frac{1}{2\pi i} \int_{-r_1^-}^{-r_1^+} \omega_{r_1^+ r_1^-} &
\frac{1}{\pi i }(\ln \epsilon + c_+)
\end{pmatrix}$$
where $c_+, c_- \in \C$ are constants.
Moreover,
$$\check{u} \to \begin{pmatrix} \int_{\Gamma^+} h \omega \\
\frac{1}{2\pi i}\int_{\Gamma^+} h \omega_{-r_1^+, -r_1^-} \\
\frac{1}{2\pi i}\int_{\Gamma^+} h \omega_{r_1^+ r_1^-} \end{pmatrix}.$$
We have
\begin{equation}\label{checkThetasum}
\check{\Theta}\begin{bmatrix} p \\ 0 \end{bmatrix}\biggl(\check{u} + \int_P^{Q} \hat{\omega}\biggr)
= \left(\sum_{\check{N} \in \Z^6} e^{2\pi i \left(\frac{1}{2} \check{N}^T \check{B} \check{N} + \check{N}^T(\check{u} + \int_P^{Q} \check{\omega} + \check{B}\check{p})\right)}\right)
e^{2\pi i\left(\frac{1}{2} \check{p}^T \check{B} \check{p} + \check{p}^T(\check{u} + \int_P^{Q} \check{\omega})\right)}.
\end{equation}
Letting $\check{N} = (N, n) \in \Z^4 \times \Z^2$ and choosing, for some $0 < \alpha < 1/2$,
$$\check{p} = (0,0,0,0, \alpha, -\alpha),$$
the same type of argument that led to (\ref{checkThetasumlimit}) shows that all terms in the sum approach $0$ as $\epsilon \to 0$ except those with $n \equiv 0$.
It follows that the sum in (\ref{checkThetasum}) converges to
$$\Theta\biggl(u + \int_P^{Q} \omega\biggr) \qquad \hbox{where} \qquad u = \int_{\Gamma^+} h \omega + \alpha \left(\int_{-r_1^-}^{-r_1^+} - \int_{r_1^-}^{r_1^+}\right)\omega.$$
On the other hand, the rightmost exponential factor in (\ref{checkThetasum}) can be written as
\begin{align}\nonumber
e^{2\pi i\left(\frac{1}{2} \check{p}^T \check{B} \check{p} + \check{p}^T(\check{u} + \int_P^{Q} \check{\omega})\right)}
= &\; e^{\pi i \alpha^2 (\check{B}_{55} - 2 \check{B}_{56} + \check{B}_{66})}
	\\ \nonumber
& \times e^{\alpha \int_{\Gamma^+} h \omega_{-r_1^+, -r_1^-}
- \alpha \int_{\Gamma^+} h \omega_{r_1^+ r_1^-}
+ \alpha\int_P^{Q}\omega_{-r_1^+,-r_1^-}
- \alpha\int_P^{Q}\omega_{r_1^+ r_1^-}}.
\end{align}
Applying these formulas to (\ref{checkf}), we find
\begin{equation}\label{checkftof}
\check{f} \to f = 
\frac{\Theta(u + \int_{\xi}^{\infty^+} \omega)}{\Theta(u + \int_{\xi}^{\infty^-} \omega)}e^{I},
\end{equation}
where
$$I = \int_{\Gamma^+} h \omega_{\infty^+ \infty^-} 
+ \alpha \left(\int_{-r_1^-}^{-r_1^+} - \int_{r_1^-}^{r_1^+}\right)\omega_{\infty^+ \infty^-}.$$
The contours in the integrals $\int_{-r_1^-}^{-r_1^+}$ and $\int_{r_1^-}^{r_1^+}$ are the limits of the cycles $\check{b}_5$ and $\check{b}_6$, respectively. By deforming these contours, we find that
$$\left(\int_{-r_1^-}^{-r_1^+} - \int_{r_1^-}^{r_1^+}\right)\omega_{\infty^+ \infty^-} = 
\int_\gamma \omega_{\infty^+ \infty^-}, \qquad 
\left(\int_{-r_1^-}^{-r_1^+}\omega - \int_{r_1^-}^{r_1^+}\right)\omega
= \int_\gamma \omega,$$
where $\gamma$ is the contour on $\Sigma_z$ defined in section \ref{diskblackholesec}.
Therefore, in the limit $\alpha \to 1/2$, $u$ and $I$ become exactly the $u$ and $I$ of theorem \ref{mainth} and the solution $f$ in (\ref{checkftof}) becomes the Ernst potential in (\ref{ernstsolution}). This provides the promised link between the solutions in \cite{KM} and the solution presented in this paper.

By applying the same limiting procedure to equation (\ref{e2checkkappa}) we will determine the corresponding metric function $e^{2\kappa}$.
We have
$$e^{\pi i \alpha^2 (\check{B}_{55} - 2 \check{B}_{56} + \check{B}_{66})} = \epsilon^{2\alpha^2} e^{\alpha^2\bigl(c_+ + c_- - \int_{-r_1^-}^{-r_1^+} \omega_{r_1^+r_1^-}\bigr)}( 1 + O(\epsilon)), \qquad \epsilon \to 0.$$
Hence, for the quotient in (\ref{e2checkkappa}) we find
\begin{align}\label{quotientexpansion}
\check{K}_0 \frac{\check{\Theta}(\check{u})\check{\Theta}( \check{u} + \int_\xi^{\bar{\xi}} \check{\omega})}{\check{\Theta}(0)\check{\Theta}(\int_\xi^{\bar{\xi}} \check{\omega})}
= & \; 
\frac{\Theta(u)\Theta(u + \int_\xi^{\bar{\xi}} \omega)}{\Theta(0)\Theta(\int_\xi^{\bar{\xi}} \omega)}
\epsilon^{4\alpha^2} e^{2 \alpha^2(c_+ + c_-) - 2 \alpha^2\int_{-r_1^-}^{-r_1^+} \omega_{r_1^+r_1^-}}
	\\ \nonumber
& \, \times e^{\int_{\Gamma^+} h \omega_{-r_1^+ -r_1^-} - \int_{\Gamma^+} h \omega_{r_1^+ r_1^-} }
e^{\frac{1}{2}\int_\xi^{\bar{\xi}} \omega_{-r_1^+ -r_1^-} - \frac{1}{2}\int_\xi^{\bar{\xi}} \omega_{r_1^+ r_1^-}}( 1 + O(\epsilon)).
\end{align}
This expression vanishes in the limit $\epsilon \to 0$. However, this behavior is compensated by the fact that the constant $\check{K}_0$ diverges as $\epsilon \to 0$, so that the limit of $e^{2\check{\kappa}}$ is finite and non-zero. The last exponential factor on the right-hand side of (\ref{quotientexpansion}) can be absorbed into $K_0$. Indeed, the same type of argument that we used to find (\ref{e2checkkappa}) shows that this factor is independent of $z$.
The constants $c_+$ and $c_-$ are given by the expressions obtained by replacing $\zeta$ in the right-hand side of (\ref{Mprimedef})
with $r_1$ and $-r_1$, respectively.
Letting $\alpha \to 1/2$ and using that
\begin{align*}
\lim_{\delta \to 0} \left(\int_{(-r_1 - \delta)^-}^{(-r_1 - \delta)^+} \omega_{-r_1^+, -r_1^-} 
+\int_{(r_1 - \delta)^-}^{(r_1 - \delta)^+} \omega_{r_1^+ r_1^-}  - 4\ln \delta \right)
- 2 \int_{r_1^-}^{r_1^+} \omega_{-r_1^+ -r_1^-}
	\\
= 2 \lim_{\delta \to 0} \left(\int_{\gamma_1(\delta)} \omega_{-r_1^+, -r_1^-}
- \int_{\gamma_2(\delta)}\omega_{r_1^+ r_1^-} - 2\ln \delta \right),
\end{align*}
we infer that the limit of $e^{2\check{\kappa}}$ is given by (\ref{e2ksolution}).

We finally point out that formula (\ref{aminusa0}) for the metric function $a$ can be derived in a similar way. Indeed, the metric function $\hat{a}$ corresponding to the solution $\hat{f}$ in (\ref{hatf}) is given by \cite{KKS}
$$(\hat{a}- \hat{a}_0)e^{2\hat{U}} = -\rho\left(\frac{\hat{\Theta}\begin{bmatrix} p \\ q \end{bmatrix}(0)\hat{\Theta}\begin{bmatrix} p \\ q \end{bmatrix}(\int_{\xi}^{\infty^-} \hat{\omega} +  \int_{\bar{\xi}}^{\infty^-} \hat{\omega})}{\hat{Q}(0)\hat{\Theta}\begin{bmatrix} p \\ q \end{bmatrix}(\int_{\xi}^{\infty^-} \hat{\omega} )\hat{\Theta}\begin{bmatrix} p \\ q \end{bmatrix}(\int_{\bar{\xi}}^{\infty^-} \hat{\omega})} - 1\right),$$
where
$$\hat{Q}(0) = \frac{\hat{\Theta}(\int_{\xi}^{\infty^-} \hat{\omega})\hat{\Theta}(\int_{\bar{\xi}}^{\infty^-} \hat{\omega})} {\hat{\Theta}(0) \hat{\Theta}(\int_{\xi}^{\bar{\xi}} \hat{\omega})}.$$
An application of the above limiting procedure to this expression yields (\ref{aminusa0}).

 \bigskip
\noindent
{\bf Acknowledgement} {\it The author thanks M. Ehrnstr\"om and A. S. Fokas for helpful remarks on a first version of the manuscript.}

\bibliography{is}

\begin{thebibliography}{99}
\small

\bibitem[A]{ABP}
A. Abramowicz et al, {\it Theory of black hole accretion discs}, Edited by M. A. Abramowicz, G. Bj\"ornsson, and J. E. Pringle, 309 p., Cambridge University Press, Cambridge, UK, 1999.

\bibitem[BW]{BW1969}
J. M. Bardeen and R. V. Wagoner, Uniformly rotating disks in general relativity,
{\it Astrophys. J.} {\bf 158} (1969), L65--L69.

\bibitem[BW2]{BW1971}
J. M. Bardeen and R. V. Wagoner, Relativistic disks. I. Uniform rotation,
{\it Astrophys. J.} {\bf 167} (1971), 359--423.

\bibitem[B]{Bicak}
J. Bi\v{c}\'ak, Selected solutions of Einstein's field equations: their role in general relativity and astrophysics, in {\it Einstein's field equations and their physical implications}, Edited by B. G. Schmidt, Lecture Notes in Physics, vol. 540, Springer-Verlag, Berlin, 2000, p.1--126.

\bibitem[C]{Chandra}
S. Chandrasekhar, {\it The mathematical theory of black holes}, Reprint of the 1992 edition, Oxford Classic Texts in the Physical Sciences, The Clarendon Press, Oxford University Press, New York, 1998. xxii+646 pp.

\bibitem[FK]{FK}
H. M. Farkas and I. Kra, {\it Riemann surfaces}, 2nd edition, 
Graduate Texts in Mathematics {\bf 71}, Springer-Verlag, New York, 1992, xvi+363 pp. 

\bibitem[Fa]{Fay}
J. D. Fay, {\it Theta functions on Riemann surfaces,} 
Lecture Notes in Mathematics {\bf 352}, Springer-Verlag, Berlin-New York, 1973, iv+137 pp.

\bibitem[F1]{F1997}
A. S. Fokas, A unified transform method for solving linear and certain nonlinear PDEs, 
{\it Proc. Roy. Soc. Lond.} A {\bf 453} (1997), 1411--1443.

\bibitem[F2]{F2002}
A. S. Fokas, Integrable nonlinear evolution equations on the half-line, 
{\it Comm. Math. Phys.} {\bf 230} (2002), 1--39.

\bibitem[F3]{Fbook}
A. S. Fokas, {\it A unified approach to boundary value problems}, CBMS-NSF regional conference series in applied mathematics, SIAM (2008).

\bibitem[GGK]{GGKM}
C.S. Gardner, J.M. Greene, M.D. Kruskal, and R.M. Miura, Method for solving the Korteweg-de Vries equation, {\it Phys. Rev. Lett.} {\bf 19} (1967), 1095--1097.

\bibitem[GH]{GH}
P. Griffiths and J. Harris, {\it Principles of algebraic geometry,} Wiley-Interscience [John Wiley \& Sons], New York, 1978, xii+813 pp. 

\bibitem[Ke]{Kerr}
R. P. Kerr, Gravitational field of a spinning mass as an example of algebraically special metrics, {\it Phys. Rev. Lett.} {\bf 11} (1963), 237--238.

\bibitem[KKS]{KKS}
C. Klein, D. Korotkin, and V. Shramchenko, Ernst equation, Fay identities and variational formulas on hyperelliptic curves, {\it Math. Res. Lett.} {\bf 9} (2002), 27--45. 

\bibitem[K]{K}
D. A. Korotkin, Finite-gap solutions of stationary axisymmetric Einstein equations in vacuum, Theoret. and Math. Phys., {\bf 77} (1989), 1018--1031.

\bibitem[KM]{KM}
D. A. Korotkin and V. B. Matveev, On theta-function solutions of the Schlesinger system and the Ernst equation. (Russian) Funktsional. Anal. i Prilozhen. 34 (2000), no. 4, 18--34, 96; translation in Funct. Anal. Appl. 34 (2000), no. 4, 252--264.

\bibitem[KR]{KR1998}
C. Klein and O. Richter, Physically realistic solutions to the Ernst equation on hyperelliptic Riemann surfaces, {\it Phys. Rev. D} {\bf 58} (1998), 124018, 18 pp.

\bibitem[KR2]{KR2005}
C. Klein and O. Richter, Ernst equation and Riemann surfaces. Analytical and numerical methods. Lecture Notes in Physics, 685. Springer-Verlag, Berlin, 2005.

\bibitem[LF]{LF}
J. Lenells and A. S. Fokas, Boundary value problems for the stationary axisymmetric Einstein equations: a rotating disk, arXiv:0911.1898.

\bibitem[NM1]{NM1993}
G. Neugebauer and R. Meinel, The Einsteinian gravitational field of the rigidly rotating disk of dust, 
{\it Astroph. J.} {\bf 414} (1993) L97--L99.

\bibitem[NM2]{NM1994}
G. Neugebauer and R. Meinel, General relativistic gravitational field of a rigidly rotating disk of dust: Axis potential, disk metric, and surface mass density,
{\it Phys. Rev. Lett.} {\bf 73} (1994), 2166--2168. 

\bibitem[NM3]{NM1995}
G. Neugebauer and R. Meinel, General relativistic gravitational field of a rigidly rotating disk of dust: Solution in terms of ultraelliptic functions,
{\it Phys. Rev. Lett.} {\bf 75} (1995), 3046--3047.

\bibitem[MAK]{MAKNP}
R. Meinel, M. Ansorg, A. Kleinw\"achter, G. Neugebauer, and D. Petroff,
{\it Relativistic figures of equilibrium}, Cambridge University Press, Cambridge, 2008.

\bibitem[P]{Pringle}
J. E. Pringle, Accretion discs in astrophysics, {\it Ann. Rev. Astron. Astrophys.} {\bf 19} (1981), 137--162.

\bibitem[SKM]{SKMHH}
H. Stephani, D. Kramer, M. MacCallum, C. Hoenselaers, and E. Herlt, {\it Exact solutions of Einstein's field equations}, Second Edition. 732 p., Cambridge University Press, Cambridge, 2003.

\bibitem[ZS]{ZS} V. E. Zakharov and A. B. Shabat, Exact theory of two-dimensional self-focussing and one- 
dimensional self-modulation in nonlinear media, {\it Soviet Physics-JETP} {\bf 34} (1972), 62--69. 

\bibitem[Y]{Y}
A. Yamada, Precise variational formulas for abelian differentials,
{\it Kodai Math. J.} {\bf 3} (1980), 114--143. 

\end{thebibliography}

\end{document}